\documentclass[10pt,onecolumn,english]{book}

\usepackage{geometry}
\usepackage{fancyhdr}
\usepackage{subfigure}
\usepackage{graphicx}
\usepackage{url}

\usepackage[english]{babel}
\usepackage{amsmath,amssymb,amscd,latexsym,dsfont}
\usepackage{cite}
\usepackage{textcomp}
\usepackage{psfrag}
\usepackage{rotating}
\usepackage{enumerate}
\usepackage{pdflscape}

\usepackage[final]{pdfpages}
\usepackage{xspace}
\usepackage{relsize}

\usepackage{url}
\usepackage{IEEEtrantools}

\usepackage{microtype}

\usepackage{todonotes}

\def\phdThesis{1}

\newcommand{\currentyear}{2021}
\newcommand{\authorname}{Mazen Mohamad}
\newcommand{\mytitle}{Towards Understanding and Applying Security Assurance Cases for Automotive Systems}
\newcommand{\division}{Interaction Design and Software Engineering}
\newcommand{\techReportNumber}{XXXX}

\usepackage[T1]{fontenc}
\usepackage{graphicx}
\usepackage{todonotes}
\usepackage{afterpage}
\usepackage{multirow}
\usepackage{booktabs}
\usepackage{graphicx,subfigure}
\newcommand*\ON[0]{$\surd$}
\usepackage{url}
\usepackage{pgfplots}
\usepackage{float}
\usepackage{longtable}
\usepackage{multicol}

\usepackage[T1]{fontenc}
\usepackage{graphicx}
\usepackage{afterpage}
\usepackage{multirow}
\usepackage{booktabs}

\usepackage{url}
\usepackage{pgfplots}
\pgfplotsset{compat=1.17}
\usepackage{float}
\usepackage{longtable}
\usepackage{multicol}
\usepackage{tabularx}
\newcolumntype{s}{>{\hsize=.4\hsize}X}
\newcolumntype{P}[1]{>{\raggedright\arraybackslash}p{#1}}
\usepackage{listings}
\lstdefinestyle{mystyle}{
    keepspaces=true,                 
    numbers=left,                    
    numbersep=5pt,                  
    showspaces=false,                
    showstringspaces=false,
    showtabs=false
}
\usepackage{longtable}
\usepackage{lscape}
\newcommand{\phdISBNNumber}{xxx-xx-xxxx-xxx-x}
\newcommand{\phdSeriesNumber}{xxxx}
\newenvironment{widequotation}{\list{}{\listparindent 1.5em \itemindent\listparindent
		\rightmargin 0pt \parsep 0pt plus 1pt}\item\relax}
{\endlist}
\usepackage{xspace}
\def\signed#1{{\leavevmode\unskip\nobreak\hfil\penalty50\hskip2em
		\hbox{}\nobreak\hfil\raise-3pt\hbox{#1}
		\parfillskip=0pt \finalhyphendemerits=0 \endgraf}}
\newsavebox\mybox
\newenvironment{aquote}[1]
{\savebox\mybox{(#1)}\begin{widequotation}\itshape``\ignorespaces}
	{\unskip"\signed{\usebox\mybox}\end{widequotation}}
	
\newcommand{\inlinequote}[1]{``\emph{#1}''}
\begin{document}
\frontmatter

%================================================================%
% Auxiliary pages
%================================================================%
%The first three pages. Regulated by the Chalmers/GU layouting guidelines.
%CHECK those to be sure the format is up to date and everything is included!
\ifx\phdThesis\undefined
\newcommand{\degreetitle}{Licentiate of Engineering}
\newcommand{\reportno}{TODO}
\newcommand{\reportNoText}{Technical Report No \techReportNumber \\
 ISSN 1652-876X\\ }
\else
\newcommand{\degreetitle}{Licentiate of Engineering}
\newcommand{\reportNoText}{ISBN \phdISBNNumber\\
Doktorsavhandlingar vid Chalmers tekniska h\"{o}gskola, Ny serie nr \phdSeriesNumber .\\
ISSN 0346-718X\\ 
\vspace{1cm}

\noindent
Technical Report No \techReportNumber \\
}
\fi

%================================================================%
%                          First page                            %
%================================================================%

\thispagestyle{empty} %\addtolength{\topmargin}{1cm}
\begin{center}
  \textsc{Thesis for The Degree of \degreetitle}\\
\end{center}

\vspace{6cm}
\begin{center} \Large \mytitle
\end{center}

\vspace{1cm}
\begin{center}
\textsc{\authorname} \\
\end{center}

\vspace{2cm}
\begin{figure}[h]
  \begin{center}
  %40mm from beginning
     %\includegraphics[width=30mm]{Fig/Avancez50PC.pdf}
    % \hspace{1cm}
     \includegraphics[width=30mm]{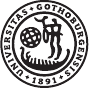}
  \end{center}
\end{figure}
\vspace{2cm}

\begin{center}
Division of \division \\
Department of Computer Science \& Engineering\\
Chalmers University of Technology and Gothenburg University\\
Gothenburg, Sweden, \currentyear \\
\end{center}

%================================================================%
%                       Printing page                            %
%================================================================%
\newpage
\thispagestyle{plain}

\vspace{2cm} \noindent \textbf
\mytitle\\

\noindent
\textsc{\authorname}\\

\vfill

\noindent
Copyright \copyright \currentyear \space \authorname \\
except where otherwise stated. \\
All rights reserved. \vspace{1cm}

%\noindent \reportNoText
Department of Computer Science \& Engineering\\
Division of \division \\
Chalmers University of Technology and Gothenburg University\\
Gothenburg, Sweden\\
\vspace{1cm}

\noindent This thesis has been prepared using \LaTeX.

\noindent
Printed by Chalmers Reproservice,\\
Gothenburg, Sweden \currentyear.

%The dedication.
%================================================================%
%                          Dedication                            %
%================================================================%

\newpage
\begin{table}
  \begin{flushright}
	\large{\emph{``Amateurs hack systems, professionals hack people.'' \\- Bruce Schneier}}  
\end{flushright}
\end{table}

\newpage
\thispagestyle{empty}

%Abstract. Word length regulated by Chalmers/GU layouting guidelines.
\cleardoublepage \addcontentsline{toc}{chapter}{Abstract}
\thispagestyle{empty}
\section*{Abstract}
%Context: What's the problem/context?

Security Assurance Cases (SAC) are structured bodies of arguments and evidence used to reason about security properties of a certain artefact.
SAC are gaining focus in the automotive domain as the need for security assurance is growing due to software becoming a main part of vehicles. Market demands for new services and products in the domain require connectivity, and hence, raise security concerns. Regulators and standardisation bodies started recently to require a structured for security assurance of products in the automotive domain, and automotive companies started, hence, to study ways to create and maintain these cases, as well as adopting them in their current way of working.

In order to facilitate the adoption of SAC in the automotive domain, we created CASCADE, an approach for creating SAC which have integrated quality assurance and are compliant with the requirements of ISO/SAE-21434, the upcoming cybersecurity standard for automotive systems. 

CASCADE was created by conducting design science research study in two iterative cycles. The design decisions of CASCADE are based on insights from a qualitative research study which includes a workshop, a survey, and one-to-one interviews, done in collaboration with our industrial partners about the needs and drivers of work in SAC in industry, and a systematic literature review in which we identified gaps between the industrial needs and the state of the art.

The evaluation of CASCADE was done with help of security experts from a large automotive OEM. It showed that CASCADE is suitable for integration in industrial product development processes. Additionally, our results show that the elements of CASCADE align well with respect to the way of working at the company, and has the potential to scale to cover the requirements
and needs of the company with its large organization and complex products

\vspace{5 mm} \noindent{\textbf{Keywords:}}  

\vspace{3 mm} \noindent{security, assurance case, automotive, automotive systems, arguments, evidence, security claims}

\newpage
\thispagestyle{empty}

%Acknowledgement
\cleardoublepage \addcontentsline{toc}{chapter}{Acknowledgement}
\chapter*{Acknowledgment}
\vspace{5 mm}

First of all, I would like to express my sincere gratitude to my supervisor and mentor Riccardo Scandariato for all the support, advice, patience, collaboration, and trust he provided me during this journey. I would also like to thank my co-supervisor Jan-Philipp Steghöfer for his great advice, feedback, collaboration and inspiring discussions. I also thank my examiner Ivica Crnkovic for trusting me and giving me the freedom to conduct my research.\\

Thank you to all my colleagues and friends at the Interaction Design and Software Engineering division for welcoming me in a very nice working environment and for all the nice social and sports activities. Special thanks to my friends whom I shared an office with: Rodi, Linda, Khaled, Joel, Mads, and Hamdi.
I would also like to extend my thanks to my industrial partners at Volvo and Volvo Cars for all their support.\\ 

Finally, I wish to express my deepest gratitude to my parents, friends, and my wife Rim, who did not spare a chance to motivate me and provide me with love and encouragement. I also wish to thank my son Bassam for inspiring me every single morning.\\

This work is partially supported by the CASUS research project funded by VINNOVA, a Swedish funding agency. 

\newpage
\thispagestyle{empty}

%Publication list. NOT the actual papers.
\cleardoublepage \addcontentsline{toc}{chapter}{List of Publications}
%================================================================%
%                       Appended papers                          %
%================================================================%
\chapter*{List of Publications}
\label{A:Papers}

\section*{Appended publications}
This thesis is based on the following publications:
\renewcommand{\labelenumi}{[\Alph{enumi}]}
\begin{enumerate}

\item M. Mohamad, A. Åström, Ö. Askerdal, J. Borg, R. Scandariato
\newblock ``Security Assurance Cases for Road Vehicles: an Industry Perspective''\\
\newblock {\em International Conference on Availability,Reliability and Security ARES, 2020} \cite{paperA,paperA_extended}. 

\item M. Mohamad, J.P. Steghöfer, R. Scandariato
\newblock ``Security Assurance Cases – State of the Art of an Emerging Approach''\\
\newblock {\em Empirical Software Engineering Journal 26, 70 (2021)} \cite{paperB}.

\item M. Mohamad, Ö. Askerdal, R. Jolak, J.P. Steghöfer, R. Scandariato
\newblock ``Asset-driven Security Assurance Cases with Built-in Quality Assurance''\\
\newblock {\em International Workshop on Engineering and Cybersecurity of Critical Systems (ENCYCRIS), 2021} \cite{paperC}.

\end{enumerate}

\newpage
%================================================================%
%                        Other papers                            %
%================================================================%
\section*{Other publications}
The following publications were published before or during my PhD studies, or are currently in submission/under revision.
However, they are not appended to this thesis, due to contents overlapping that of appended publications or contents not related to the thesis.

\renewcommand{\labelenumi}{[\alph{enumi}]}
\begin{enumerate}
 \item M. Mohamad, G. Liebel, E. Knauss
 \newblock ``LoCo CoCo: Automatically constructing coordination and communication networks from model-based systems engineering data''\\
 \newblock {\em Information and Software Technology Journal 92, 179-193, 2017} \cite{locococo}
 
  \item R. Jolak, T. Rosenstatter, M. Mohamad, K. Strandberg, B. Sangchoolie, N. Nowdehi, R. Scandariato
 \newblock ``CONSERVE: A Framework for the Selection of Techniques for Monitoring Containers Security''\\
 \newblock {\em In submission to The Journal of Systems \& Software} \cite{conserve}
 
   \item J.P. Steghöfer, B. Koopmann, J.S. Becker, M. Törnlund, Y. Ibrahim, M. Mohamad
 \newblock ``Design Decisions in the Construction of Traceability Information Models for Safe Automotive Systems''\\
 \newblock {\em In submission to the International Requirements Engineering Conference} \cite{designDes}
 \end{enumerate}

%Declaration of your own contribution to the included papers
\cleardoublepage \addcontentsline{toc}{chapter}{Personal Contribution}
\thispagestyle{empty}
\section*{Research Contribution}
My contribution in Paper A was mainly in the internal needs part. I, equally contributed in preparing the workshop, facilitating the brainstorming sessions and collecting the results. In the survey, I contributed the design, data collection, and analysis of results. I also equally contributed in leading the discussion in the interviews, and did the analysis of the outcome.
When it comes to writing the paper, I contributed the majority of all sections, except for RQ1 results and the introduction.

In Paper B, I contributed in running the search on the three digital repositories, collecting the results, filtering the results in three rounds, conducting the snowballing search, and analyzing and screening the included studies. I also equally contributed in the sessions for identifying the assessment and inclusion/exclusion criteria. I also did the majority of the writing in all sections but the introduction and discussion, in which I contributed.

In Paper C, I contributed equally in designing the CASCADE approach. I also created the SAC based on the example use case, conducted the evaluation of the approach, and analyzed the outcome of the evaluation. In terms of paper writing, I did the majority of the writing of all sections except for the background and validation, in which I contributed equally.

%Any notes, if needed.
% \cleardoublepage \addcontentsline{toc}{chapter}{Notes}
% \include{TexFiles/notes}

%================================================================%
% Table of contents, list of figures and tables
%================================================================%
\tableofcontents 

%If needed, lists of figures and tables
%\listoffigures
%\listoftables
 
\mainmatter

%================================================================%
% Chapters
%================================================================%
\chapter{Introduction}
\label{chap_Introduction}
%Context and related work.SAC, Approaches that are direct competition... safety critical systems.
%Research questions 
%Methodology and results: how each question is answered.
%Conclusion and future work
%One section per research question 
%In this chapter, we introduce a summary of the thesis, the context and related work, and research questions. We also introduce and discuss the research methodology, contribution, and the results of the thesis work. Finally, we conclude and discuss directions of future work.
%
%\section{Thesis at a glance}
\label{kappa:thesisAtGlance}
%turn into intro.. power summary.. give it a cool name "licentiate in a glance".
%Context and related work.SAC, Approaches that are direct competition... safety critical systems.
%Research questions 
%Methodology and results: how each question is answered.
%Conclusion and future work
%One section per research question 
Security is gaining more focus in safety-critical domains since more connectivity is needed in the services and products offered by companies in these domains. In automotive, software has become a main part of vehicles and the need for connectivity is essential to meet the market demands for functionalities and services the vehicles offer, e.g., mobile phone connectivity and navigation services.
This has raised an issue when it comes to security assurance, i.e., answering the question ``how do we make sure and prove that our product is secure?''. This becomes an even larger issue the more complex the systems become, consisting of multiple sub-systems with many stakeholders involved, e.g., different providers for different parts.

Regulators and standardisation bodies recently started to require a structured way of security assurance for automotive products and processes. For this reason, Security Assurance Cases (SAC) were specifically required in ISO/SAE-21434 \cite{iso21434} to prove security conformance. 
Automotive companies started, hence, to study ways to create and maintain these cases, as well as adopting them in their current way of working.
Assurance cases in general are not new to the automotive industry, as companies are already familiar with similar cases created for safety (safety cases), which are required by ISO-26262 \cite{iso26262} for functional safety for road vehicles. This opens up opportunities for knowledge transfer from the safety domain into the security domain. However, this knowledge transfer should be done cautiously and consider the differences between the two domains. 

In this work, we created CASCADE, an approach for creating SAC which are compliant with the requirements of ISO/SAE-21434 and have integrated quality assurance. 
CASCADE is based on insights from two studies: one done in collaboration with our industrial partners about the needs and drivers of work in SAC in industry and one systematic literature review in which we identified gaps between the industrial needs and the state of the art.

As a first step, we identified and studied different factors that would drive the work with security cases in the automotive domain.
We studied internal drivers, i.e., the requirements and needs from within an automotive company. We identified thirteen different scenarios in which SAC can be used. These scenarios spread over the entire life-cycle of automotive products and involve many different roles in automotive companies. These scenarios also imposed additional requirements on SAC. E.g., the quality assurance of a SAC is essential in order for it to be useful in industry.

External drivers that impose constraints of how SAC should look like were identified by our partners at industry. This was done analyzing how SAC were referenced in different documents (regulations, standards and best practices) in the three major automotive markets (EU, US, and China). Thirteen documents where SAC was either explicitly or implicitly required, or would assist to fulfill the requirements of the documents were identified. 

Based on what we learned about the internal and external needs for SAC in automotive, we conducted a systematic literature review to examine whether these needs are covered in literature or not. We systematically reviewed literature and looked for different characteristics, e.g., usage scenarios, approaches for creating SAC, and tool support.
In analysing our results, we made multiple observations. Most importantly, we saw a wide variety of approaches for creating SAC, but none of them considers actual constraints and needs from the automotive domain. We also observed a lack of quality assurance of SAC in the reviewed literature. Another observation we made is the wide range of potential benefits of SAC that was reported. However, there was a gap between the internal needs identified at the automotive company, and the usages suggested in literature. 

For all the reasons above, we designed our own approach for SAC creation based on what we learned from industry and literature. We focused on two main aspects: 
\begin{itemize}
    \item Align the requirements and work products of the upcoming standard ISO/SAE-21434 and the SAC (the outcome of the approach).
    \item Integrate quality within the cases themselves.
\end{itemize}
CASCADE, is an asset driven approach for creating security assurance cases with built-in quality assurance. We illustrated the approach using an example use case from ISO/SAE-21434 and evaluated it with help of security experts at a large automotive OEM.
The evaluation showed that CASCADE is suitable for integration in industrial product development processes. The elements of CASCADE align well with respect to the way of working at the company. Additionally, CASCADE has the potential to scale to cover the requirements and needs of the company with its large organization and complex products.
%get rid of the temporal terms
%get rid of: look into, digging,.. etc.
%connection between 2nd and third point... There is no approach that can be applied.
%be more assertive! 
%for all of the reasons above, we created our approach...

%in second point: none of the things are directly applicable. rethink argument.

\section{Research Focus}
\label{rq:ResGoal}
%In this section , we describe the goals of this thesis work, and present our research questions.
%\subsection{Research Goal}

This research work is motivated by the observation that SAC are becoming important in the safety critical domain, in particular, companies in the automotive industry.%, as mentioned in Section \ref{kappa:thesisAtGlance}. 
The main goal of this research is ``\emph{to support practitioners in the automotive domain to make go/no go decisions of the release of their products from a security point of view, with the help of security assurance cases}''.  
To achieve this overall goal, we addressed the following goals in this licentiate thesis:
\begin{itemize}
    \item \textbf{Goal 1:} to understand the specific needs concerning SAC in the automotive domain.
    \item \textbf{Goal 2:} to understand the state of the art about SAC in literature.
    \item \textbf{Goal 3:} to create and assess an approach for creating SAC taking into consideration the specific needs of the automotive domain.
\end{itemize}
%Discussion leaders: Barbara Gallina, 
%Robert Lagerström https://www.kth.se/profile/robertl

%industrial applicability / relevance of SAC in a specific domain (automotive / safety critical). 
%switch 1 and 2
% JP:
% A) Observed that SAC are becoming important. What is of importance to the industry?
% B) They struggle. Why so? SOTA appropriate?
% C) How do we contribute
%get rid of the sub-research questions - add this as a description instead
%\subsection{Research Questions}
To reach the goals of this thesis, we formulate the following research questions:
\begin{LaTeXdescription}
 \item [RQ1:] What are the drivers for working with security assurance cases in the automotive domain?
 
 This question addresses the emergence of several standards and regulations that are forcing the industry to develop a methodology for SAC in order to stay compliant and avoid legal risks. We call these the \emph{external drivers} that will impose constraints on what SAC should look like. 
  The need to develop a strategy for SAC is also perceived by the automotive companies as an opportunity to improve their cybersecurity development process. As such, the question also takes up the \emph{internal drivers} related to this aspect.
  
 \item[RQ2:] What are the gaps in the state of the art when it comes to the industrial applicability of SAC? 
 
 This question aims at identifying gaps in the state of the art with respect to the needs of companies in the automotive domain from two perspectives:
 \begin{itemize}
     \item Approaches for the creation of SAC
     \item Support to assist practitioner in creating SAC
 \end{itemize}
  
 %   \subitem RQ2.1: What approaches of security assurance cases' creation are reported in literature?
%    \subitem RQ2.2: What kind of support exists to assist practitioner in creating security assurance cases?
 \item [RQ3:] How can an approach for the construction of security assurance cases fulfill the needs of the automotive domain?
 
 The purpose of this question is to investigate how an approach for SAC creation can be built in order to fulfill both the external and internal needs of automotive companies, as well as closing the gaps between research and the industrial needs for SAC adoption.
% \item [RQ4:] How can machine learning techniques be leveraged to support the creation of security assurance cases?
%    \subitem RQ4.1: Can we reliably predict what requirements are security relevant and can be used in creating the arguments?
%    \subitem RQ4.2: Can we reliably predict what test cases are security relevant and can be used as evidence?
% \item [RQ5:] What artifacts exist at companies that can be used as evidence in security assurance cases?
% \item [RQ6:] How can security assurance cases be maintained throughout the life-cycle of products?
%    \subitem RQ6.1: Which sort of traceability information needs to be collected in order to support the maintenance of security assurance cases?
%    \subitem RQ6.2: How can an assurance case be transformed into a living document which is updated throughout the life-cycle of the corresponding product?
\end{LaTeXdescription}
\section{Context and related work}
In this section, we provide a background about security assurance cases, as well as a review of related work.
\subsection{Security Assurance Cases}
%\todo{talk about safety cases, and the differences to security (decide if we want to have a separate subsection before the approaches)}
\begin{figure}[htb]
        \centering
        \includegraphics[width=\textwidth]{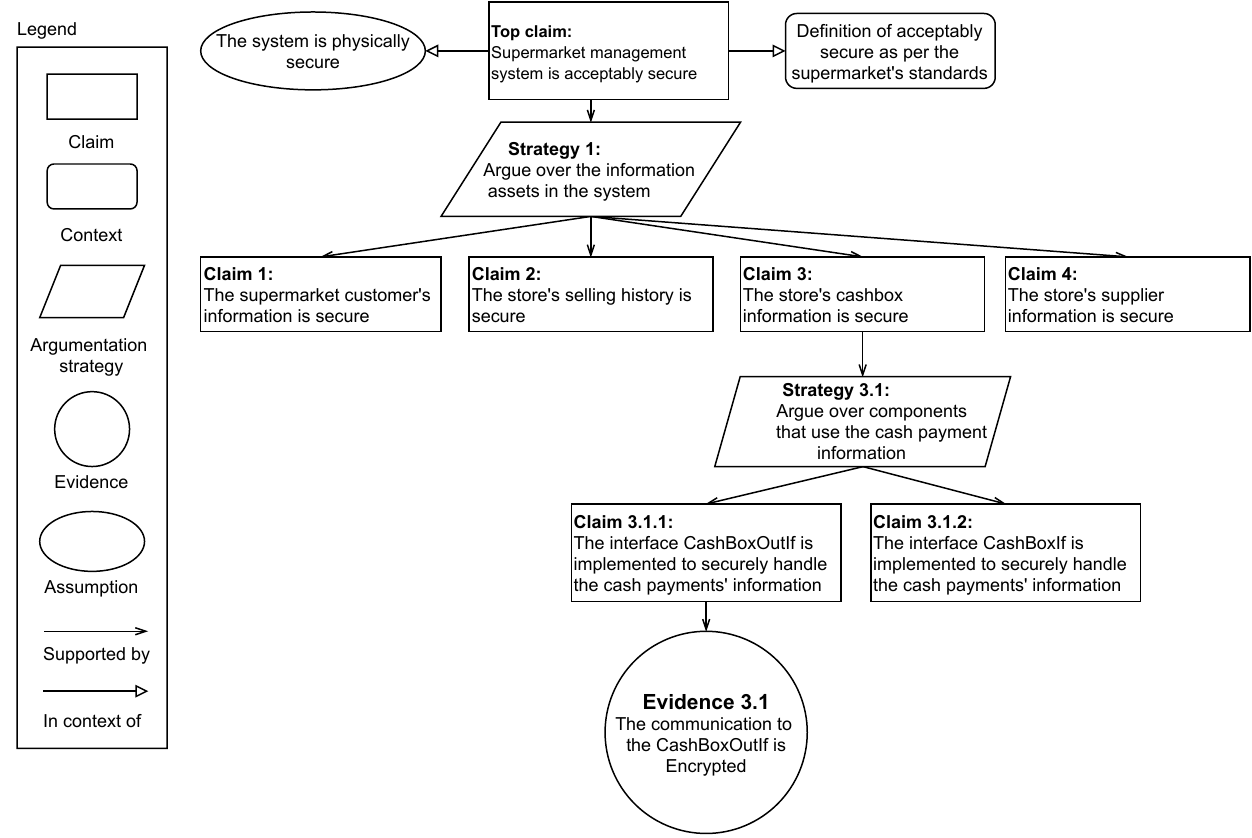}
        \caption{An example of a security assurance case}
        \label{fig:kappasacExample}
    \end{figure}
    
Assurance cases are defined by the GSN standard \cite{GSN_standard} as \inlinequote{A reasoned and compelling argument, supported by a body of evidence, that a system, service or organisation will operate as intended for a defined application in a defined environment.}

Assurance cases can be documented in either textual or graphical forms. Figure \ref{fig:kappasacExample} depicts an example of what an assurance case documented using the GSN notation looks like. The case in the example is a part of a larger case for a supermarket system.

Assurance cases consist of two main parts: the argument and the evidence. The case in the figure consists of the following nodes: claim (also called goal), context, strategy, assumption (also called justification), and evidence (also called solution). At the top of the case, there is usually a high level claim, which is broken down to sub-claims based on certain strategies. The claims specify the goals we want to assure in the case, e.g., that a certain  property is preserved. An example of a strategy is to break down a claim based on the information assets of the system as shown in \emph{Strategy 1} in Figure \ref{fig:kappasacExample}. 
Claims are broken down iteratively they reach a point where evidence can be assigned to justify them. Examples of evidence are test results, monitoring reports, and code review reports. The assumptions made while applying the strategies, e.g., that all relevant threats have been identified, are made explicit using the assumption nodes. Finally, the scope of a claim is set using the context nodes. An example of a context is the definition of an acceptably secure system. 

Assurance cases have been widely used for safety-critical systems in multiple domains \cite{bloomfield2010}. An example is the automotive industry, where safety cases have been used for demonstrating compliance with the functional safety standard ISO 26262~\cite{palin2011,birch2013,iso26262}. Another example is the medical domain, where safety cases where used to assure the safety of medical devices \cite{medicalSafetyExample}.
However, there is an increasing interest in using these cases for security as well. 
For instance, the upcoming automotive standard ISO 21434~\cite{iso21434} explicitly requires the creation of cyber-security arguments. 
SAC are a special type of assurance cases where the claims are about the security of the system in question, and the body of evidence justifies the security claims. 

\subsection{Related work}
In this section, we present the related work to the main contribution of this thesis, CASCADE.
It is an asset-driven approach for creating SAC with built-in quality assurance, inspired by ISO/SAE-21434 standard for cybersecurity in automotive. Hence, we introduce the main papers which use assets as argumentation strategies, are based on security standards, or are conducted in the automotive domain.
%In this section, we present related work to the main contribution of this thesis, CASCADE. We introduce the main papers about , which use assets t related to CASCADE. on approaches for creating SAC, as they relate to the main contribution of this thesis, which is the asset-driven approach for creating SAC with embedded quality assurance, CASCADE. 
%which focus on approaches for creating security assurance cases based on assets and security standards. We also review studies related to SAC that have been conducted in the automotive domain.
\subsubsection{Asset-based approaches}
Assets are artefacts of value to a certain organization, project, or system.
Researchers have been exploring several asset-based approaches for creating the argument part of SAC.
These approaches use assets and their decomposition as strategies to break down claims in SAC.

Biao et al.~\cite{20_xu2017} suggest dividing the argument into different layers, and using different patterns (one per layer) to create the part of the argument that corresponds to each layer. 
Assets are considered as one of these layers, and the pattern used to create it includes claims that the assets are ``under protection'', and strategies to break down critical assets. In contrast to our work, Biao et al.~\cite{20_xu2017}, however, do not consider the quality of the cases and only focus on creating arguments without touching upon the evidence part.

Luburic et al.~\cite{luburic2018} also present an asset-based approach for security assurance.
The info used in their approach is taken from: \emph{(i)} asset inventories; \emph{(ii)} Data Flow Diagrams (DFD) of particular assets and the components that manipulate them; and \emph{(iii)} the security policy that defines protective mechanisms for the components from the previous point.
They propose a domain model where assets are the center pieces. The assets are linked to security goals. The argument considers the protection of the assets throughout their life-cycles by arguing about protecting the components that store, process, and transmit those assets.
The SAC they provide is very high level and includes two strategies: ``reasonable protection for all sensitive assets'' and arguing over the data-flow of each related component.
The authors illustrate the approach with a conference management system example. They state that the main limitations of their are asset and data flow granularity.
In our work, we also consider the assets to be the driver of our approach, but we extend the argument to reach the level of concrete security requirements. We also derive our strategies from an industrial standard and validate our approach in collaboration with an OEM. Furthermore, we extend our approach to include case quality aspects.

\subsubsection{Standard-based approaches}
Using standards to extract requirements for creating the arguments of SAC has been done in multiple studies. However, none of these studies targets the upcoming standard ISO/SAE-21434 for cybersecurity in automotive.
Finnegan et al \cite{8_finnegan2014,45_finnegan2014} present a security case framework for the area of medical device security assurance. Their framework incorporates multiple standards and best practice documents as a guidance to develop a security argument pattern. The pattern provides a ``comprehensive matrix showing the link between the security risks, associated causes, the mitigating security controls and evidence of those controls being implemented to establish the security capability.''

Ankrum et al. \cite{9_ankum2005} studied how requirements from standards in safety-critical domains can be mapped to assurance cases using the most common notations for documenting assurance cases Goal Structuring Notation (GSN) and ASCAD (Claims – Arguments – Evidence).
One of the standards used in the study was the Common Criteria for Information Technology Security Evaluation, ISO/IEC 15408:1999 \cite{iso15408}, and the researchers describe challenges they encountered while conducting the mapping and lessons learned.

In our work, we have used the upcoming ISO/SAE 21434 standard to structure an approach for creating SAC, but also considered the industrial needs from the automotive domain.

\subsubsection{Studies in automotive}
Few studies about SAC have been conducted or evaluated in the automotive domain. 
Cheah et al. \cite{39_cheah2018} in their study ``Building an automotive security assurance case using systematic security evaluation''
review security engineering in the automotive industry, and the challenges to introducing a security engineering process in this domain, e.g., the overhead required to establish a security mechanism in general, and the diversity of the vehicles with many Parameters and configurations.
The authors presents a classification approach of security test results using security severity ratings. This classification can be included in the security evaluation, which may according to the study be used to improve the selection of future test cases, as well as evidence when creating security assurance cases.
The paper includes two case studies that demonstrate the method. The first case was done with a Bluetooth connection to the infotainment system of a vehicle, and the second was done on an aftermarket diagnostics tool. 
The results of both studies are severity rated evidences which could be used to prioritize countermeasure development, and to add evidence to security assurance cases.
No security assurance case is actually created, but rather severity rated evidences which the authors claim can be used in a security assurance case.
%No discussion of the contribution to the completeness
%and confidence of the security assurance case by the created evidences. It is not clear how to link these evidences to sub-claims and claims.

%common denominator in safety standards (transfer-ability among the domains...)
%is ISO-21434 based on something more generic?
%- suggestion: limit to automotive (future work :safety critical)
\section{Methodology}
This section summarizes the research methodology applied to answer the research questions of this thesis. 
RQ1 was answered through qualitative research methods. 
RQ2 was addressed using a systematic literature review and RQ3 was answered using the Design Science Research methodology. 
%\todo{Ask about the method of the third paper (CASCADE)..}

\subsection{Qualitative research methods}
\begin{figure}[htb]
        \centering
        \includegraphics[width=\textwidth,trim={0 5cm 0 2cm},clip]{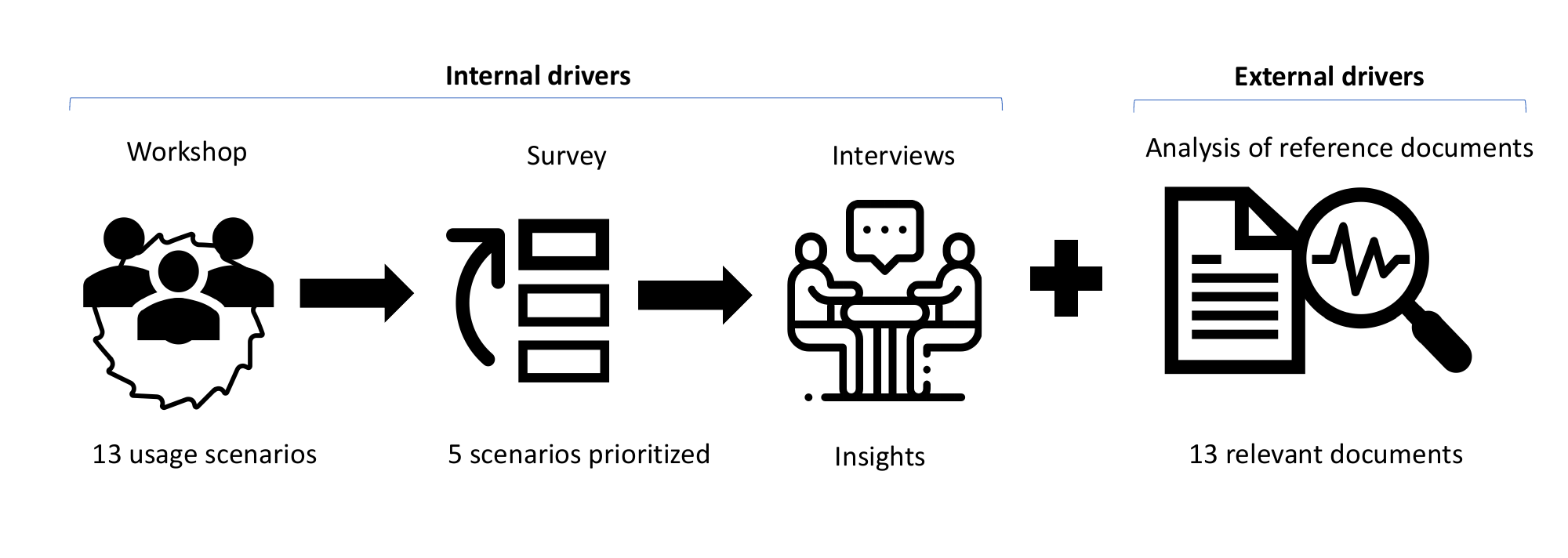}
        \caption{Qualitative research methods -- Paper A}
        \label{fig:ws}
    \end{figure}
    
We used various qualitative research methods in Paper A to answer the first research question \emph{``What are the drivers for working with security assurance cases in the automotive domain?''}, as shown in Figure \ref{fig:ws}. These include a workshop, a survey, and one-to-one interviews.
In Figure \ref{fig:ws}, The part marked as \emph{internal needs} was contributed by the author of this thesis, while the external needs part was contributed by co-authors of Paper A. 
%\subsubsection{Analysis of documents}
%To gain an understanding of the external drivers of SAC work, we used a knowledge base of documents relevant to cybersecurity, which was created and maintained by an industrial partner. This knowledge base consist of standards, regulations, guidelines, best practices,  etc applicable for various markets, and includes, among other things, information regarding the categorization of  requirements, their relevance, the parts of the organization that is affected, and which life-cycle phases of the products are impacted.
%
%We analyzed the documents for explicit references to security assurance cases or their parts.
%We also looked for implicit relationships to SAC, e.g., when the documents include requirements of processes for identification, assessment and mitigation of vulnerabilities. A SAC can then be used to show how this requirement is fulfilled listing the demanded processes and the evidence for them. 
%
%Concerning the internal needs at automotive companies, we used a three-steps method, as shown under \emph{internal needs} in Figure \ref{fig:ws}.

\subsubsection{Workshop}
We conducted a workshop at a large automotive OEM to elicit usage scenarios related to SAC. We invited stakeholders from different backgrounds and different parts of the organization. 
We had 12 participants and three moderators contributing. We divided the participants into three groups of 4, making sure to spread similar roles and competences among the groups. E.g., we had three participant who were familiar with safety cases, so we assigned them to different groups. 
We asked the groups to brainstorm for 45 minutes on usage scenarios for security assurance cases, and to describe them as user stories, like ``As a <<role>> I would use security assurance cases for <<usage>>''~\cite{cohn2004}. Each user story corresponds to one usage scenario. We explicitly asked the participants to come up with real-life scenarios in the context of their company. 
The participants shared their usage scenarios on a whiteboard, and we compiled a set of distinct scenarios as an outcome of this step and an input to the next step.

\subsubsection{Prioritization and interviews}
At this step, we wanted to dig deeper and get a better understanding of the most important scenarios. We also wanted to acquire the point of view of more diverse stakeholders. Hence we had to prioritize the usage scenarios and identify stakeholders to be interviewed for the top ones.

Concerning the prioritization, we aimed at getting expert opinions on which usage scenarios are of most value to the company, from a security perspective. 
We sent out the scenarios collected from the workshop to 10 security experts from an automotive OEM, and asked them to select the top five scenarios by assigning a rank from 1 to 5 to them, where 5 is assigned to the most valuable scenario for the company.

Afterwards, we selected the top five usage scenarios and identified a key stakeholder for each.
Finally, we conducted in-person interviews with these stakeholders to gain a deeper understanding of the usage scenarios. 
The interviewees were selected based on the relevance of their expertise to the actors of the user stories in the corresponding usage scenarios. For example, the actor of one of our top usage scenarios is a \emph{legal risk owner}. Hence, we selected an interviewee who has extensive experience in law and has the role \emph{senior legal counsel} in the company.

We organized each interview into four parts, according to the following themes:
\begin{enumerate}[i]
\item 
\textbf{Value} In the first part, we focus on the value that SAC might bring to the stakeholder in terms of, e.g., efficiency, and quality management. The objective of the discussion is to picture the ‘status quo’ (e.g., to understand how the level of security is currently appraised) and the expectations (i.e., how things should improve). 
\item \textbf{Content and structure} The focus of this part is to get the interviewees' technical opinions on how the content and structure of SAC should be, e.g., in terms of level of detail and types of claims. 
\item 
\textbf{Integration} This part is about understanding how SAC could be integrated with the current way of working, and whether it could fit in the current activities, or would require modifications to the process.
\item
\textbf{Challenges and opportunities} The last part of the interview is about understanding the challenges and opportunities that the stakeholders foresee in applying SAC.
\end{enumerate}

In each interview, there was an interviewer, an interviewee, and a security expert who acted as a discussion enabler. We recorded the interviews, and used the recordings to extract a transcript for each interview. To analyze the data, we used deductive coding using codes corresponding to our predefined themes. The analyzed data was then sent to the corresponding interviewees for validation and additional comments.

\subsubsection{Analysis of documents}
\label{section:AnalysisOfDocuments}
To gain an understanding of the external drivers of SAC work, a knowledge base of documents relevant to cybersecurity, which was created and maintained by an industrial partner was used. This knowledge base consist of standards, regulations, guidelines, best practices,  etc applicable for various markets, and includes, among other things, information regarding the categorization of  requirements, their relevance, the parts of the organization that is affected, and which life-cycle phases of the products are impacted.
Co-authors of Paper A analyzed the documents for explicit references to security assurance cases or their parts.
They also looked for implicit relationships to SAC, e.g., when the documents include requirements of processes for identification, assessment and mitigation of vulnerabilities. An SAC can then be used to show how this requirement is fulfilled listing the demanded processes and the evidence for them. 

\subsection{Systematic Literature Review(SLR)}
%\todo{Ready for review}
\begin{figure}[htb]
        \centering
        \includegraphics[width=\textwidth,trim={0 5cm 0 2cm},clip]{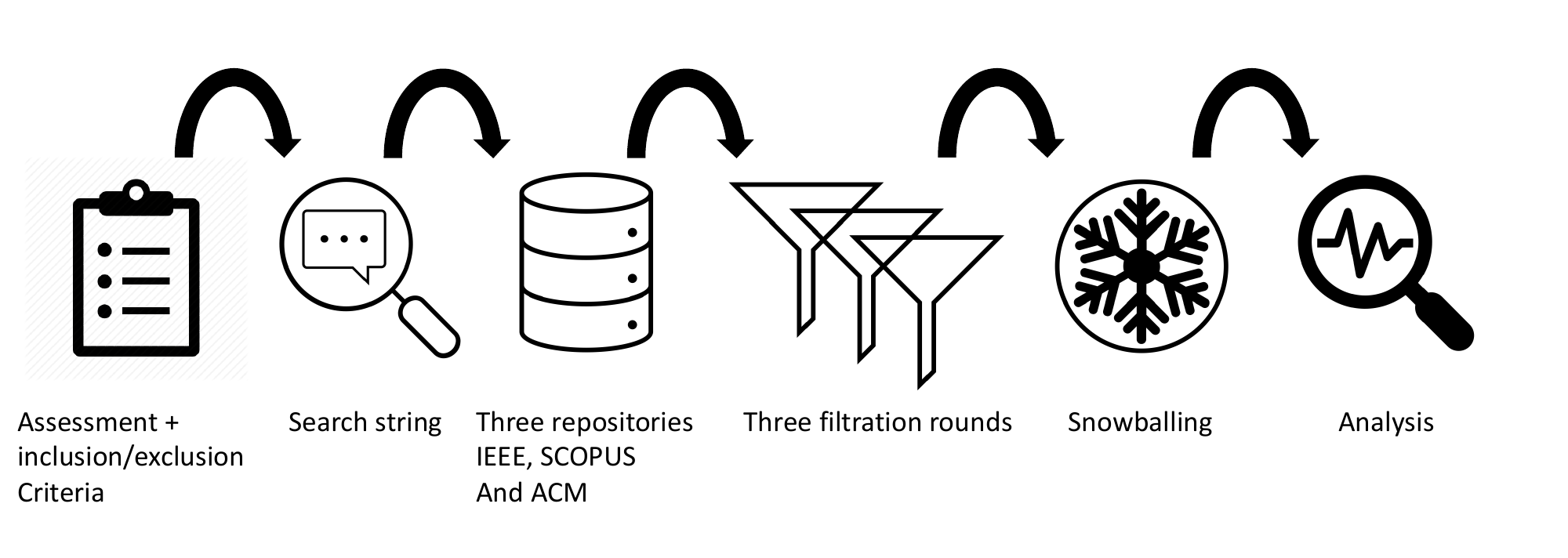}
        \caption{Systematic Literature Review steps -- Paper B}
        \label{fig:slr}
    \end{figure}

Systematic Literature Reviews (SLR) are conducted to collect and analyze data related to a specific research question \cite{kh2007guidelines}.
We conducted an SLR in Paper B to gain an understanding of existing work in security assurance. In particular, we looked for approaches for creating security assurance cases and evidence concerning their validity, support to facilitate the adoption of SAC, and rationale to support the adoption of SAC.
We followed the guidelines introduced by Kitchenham et al. \cite{kh2007guidelines}, and conducted the study in six steps as depicted in Figure \ref{fig:slr}.

In the first step, we carefully constructed the assessment criteria for each of our research questions and the inclusion/exclusion criteria for the retrieved papers. This was done in a series of brainstorming sessions including the three authors of the study.

The second step was creating the search string. In order to maximize the chance of obtaining all relevant papers in the field we familiarized ourselves with the specific terminology used by researchers in the field of security assurance. This was done by conducting a manual search for papers related to security assurance cases that were published in the past five years in multiple venues with high visibility in the security domain. 
We executed the query on three libraries (IEEE Xplore, ACM Digital Library,and Scopus) and got a total of 8440 results.

In the next step, we applied the inclusion/exclusion criteria on the results in three filtration rounds. In the first one, we filtered based on the title and keywords, which reduced the number of included studies to 211. In the second filtering round, we applied the inclusion and exclusion criteria to the abstracts and conclusions of the 211 remaining studies. After this step, the number of studies was reduced to 49. In the last filtering round, we fully read the remaining 49 papers, applied the inclusion and exclusion criteria on the whole text, and ended up with 44 included studies.
We also looked at the references in the included papers and performed  \emph{backward snowballing}~\cite{snowballing}. In this step, we did not restrict the search to only peer-reviewed studies in order to allow for potential gray literature to be included. This resulted in including additional 7 papers (including 2 technical reports) in our review. 

Finally, we analyzed the 51 included studies based on our defined assessment criteria to answer our research questions.

\subsection{Design Science Research (DSR)}
%\todo{Ready for review}
%- discuss the validation in the paper
%start from the asset, discussed with the company, improved the proposal (input from the company about wow), then we incorporated that and did a second cycle and we presented that to the experts.
%- present the larger story, and say that in the paper we focused on the results in the paper (make a story and put it in a picture "spiral")

    \begin{figure}[htb]
        \centering
        \includegraphics[width=19cm ,trim={1.3cm 2cm 0 1cm},clip]{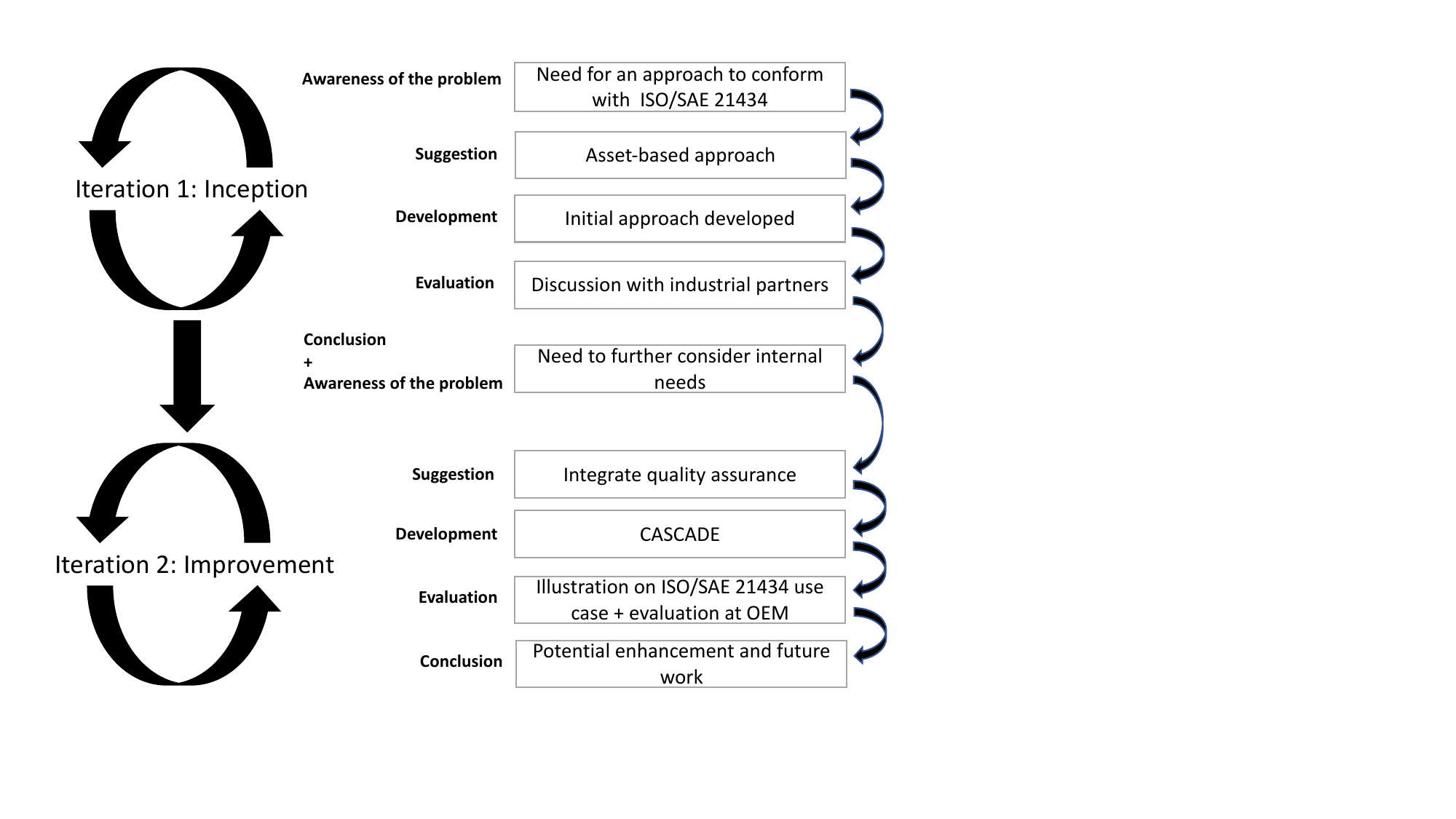}
        \caption{Two-iteration Design Science Research -- Paper C}
        \label{fig:dsr}
    \end{figure}
    
Design science research is a problem-solving methodology, which aims at developing artefacts to extend existing boundaries in a given context \cite{Hevner2004}.
% 
%\todo{Second iteration: need to consider quality aspects}
In paper C, we conducted two research iterations, following the design science guidelines proposed by Hevner et al.~\cite{Hevner2004} and the five-step process proposed by Vaishnavi and Kuechler~\cite{vaishnavi2004}, which consists of the \emph{awareness of the problem}, \emph{suggestion}, \emph{development}, \emph{evaluation} and \emph{conclusion} steps.
The two-iteration process is depicted in Figure~\ref{fig:dsr}.

The first iteration, \emph{initiation}, aimed at addressing the needs for security assurance cases which were identified in Paper A. Specifically, we aimed at investigating an asset-based approach for the creation of security assurance cases, in order to assist automotive companies to fulfill their needs to conform with the upcoming ISO/SAE-21434 standard.
We suggested an initial asset-based approach and used an online case for a supermarket system \cite{cocome} to illustrate the outcome of the approach.
The approach and the outcome of the illustration were discussed with security experts at two large automotive OEMs. The main input from the companies was focused on the need to align the structure of the approach with the internal way of working at the companies, which is also one of the internal needs identified in Paper A. Another aspect of improvement that emerged from the evaluation of the initiation iteration is the need for a mechanism to assure the quality of the approach's outcome.

In the second iteration \emph{improvement}, we  aimed at improving the artefact (asset-based approach) by incorporating the experience gathered in the first iteration.
We created CASCADE, an asset-based approach for SAC creation with built-in quality assurance. The structure of CASCADE is inspired by the requirements and work products of ISO/SAE-21434, and takes into consideration the need to quality assure the outcome, as well as the way of working at the automotive companies we consulted in the first iteration.
To evaluate CASCADE, we applied it on an example case of a headlamp item from ISO/SAE-21434, and presented the outcome to security experts at an OEM. 
As a conclusion, we identified areas for future enhancement of CASCADE to fulfill a wider range of the internal needs of the company.
%

%In Paper C, we focused on presenting the results of the the second iteration rather than extensively explaining the methodology. This is due to the page limit imposed by the venue.

\section{Contributions}
In this section, we provide a summary of the main contributions of each paper towards answering our research questions.

%\begin{figure}[t!]
%        \centering
%        \includegraphics[width=\textwidth,trim={2cm 2cm 0 2cm},clip]{figures/Kappa/contributions.pdf}
%        \caption{Contribution of each paper to answering the research questions}
%        \label{fig:kappa-contributions}
%    \end{figure}
%Figure \ref{fig:kappa-contributions} depicts the contribution of each paper included in this thesis to the corresponding research questions.

%\todo{revise the table with the drivers. Formulate it clearly as drivers.}
\subsection{RQ1: What are the drivers for working with security assurance cases in the automotive domain?}
To answer this research question we conducted the study presented in Paper A.
We investigated the internal drivers of SAC work in automotive and contextualized the results based on the external drivers which were identified by our co-authors, as explained in Section \ref{section:AnalysisOfDocuments}

Our results clearly indicate potential value of using SAC at the company. They show that SAC can be used by a variety of \emph{stakeholders}, e.g., product owners and compliance team members, for a variety of \emph{purposes}, e.g., quality assessment and communication with suppliers, in all the \emph{phases of an automotive product's life-cycle}, e.g., design and development.
%\subsubsection{External drivers of SAC in the automotive industry}
%Co-authors from volvo contributed the external and my contribution is... (start with my own contribution and then say that these are contextualized the results how does the part that is not mine relate to my part....).
%We identified 13 documents 
%\todo{add the table of external drivers and explain it}
\subsubsection{Internal drivers of SAC in the automotive industry}
What drives the SAC work in an automotive company is the value the cases can bring to people in different roles in the company. Hence, we identified 13 usage scenarios for SAC in an automotive company. We also prioritized these usage scenarios based on the potential added value to the company, and identified the top 5 scenarios, which are shown in Table \ref{tbl:rq1:usageScenarios}
\begin{table*}
\caption{Top 5 usage scenarios identified at an automotive company}
\label{tbl:rq1:usageScenarios}
\begin{tabular}{lP{10cm}}
\toprule
\textbf{US 2} & As a member of the compliance team, I would use detailed SAC to prove to authorities that the company has complied to a certain standard, legislation, etc., and show them evidence of my claim of compliance.  \\  
\midrule
\textbf{US 6} & As a product owner, I would use SAC to make an assessment of the quality of my product from a security perspective, and make a road-map for future security development. \\
\midrule    
\textbf{US 12} &  As a legal risk owner, I would use SAC in court if a legal case is raised against the company for security related issues. I would use the SAC to prove that sufficient preventive actions were taken.\\
\midrule 
\textbf{US 8} & As a member of the purchase team, I would include SAC as a part of the contracts made with suppliers, in order to have evidence of the fulfillment of security requirements at delivery time, and to track progress during development time.\\
\midrule 
\textbf{US 3} &  As a project manager, I would use SAC to make sure that a project is ready from a security point of view to be closed and shipped to production.\\
 
\bottomrule
\end{tabular}
\end{table*}

To gain a better understanding of the usage scenarios, we conducted interviews with corresponding roles and as a result, we extracted a set of drivers for companies wanting to adopt SAC in their work, as shown in Table \ref{tbl:rq1:recom}.

\begin{table*}
\small
\caption{Drivers of SAC work in automotive companies}
\label{tbl:rq1:recom}
\begin{tabular}{P{4cm}P{7cm}}
\toprule
\textbf{Driver} & \textbf{Description}\\
\midrule
The importance to cover both product and process to comply with regulations and standards & Several of the security-related standards/regulations contain both requirements on processes and the product. The processes include how to develop the product in a secure manner as well as keeping the product secure after its release. 
\\  
\midrule
The need for SAC on whole products over sub-projects & In industries producing complex products, e.g., automotive, it is common that the products are organized in multiple projects. Additionally, the changes to these products are also done using projects (commonly called delta projects). In this case, SAC should be created on a product level rather than a project level. \\
\midrule    
Essential that SAC work follows the development process &  It is possible to build SAC for existing products, but going forward, it is important to embed the work on SAC into the development process at the organization\\
\midrule 
The need to actively assess the quality of SAC & SAC are going to serve multiple purposes within the organization  with different levels of criticality. Therefore, it must be clear what the quality level of each SAC is, so that they are not used in the wrong context. \\
\midrule 
A common language is key to smooth collaboration with suppliers &  When it comes to working with suppliers, the SAC should be built using an exchangeable format. This is to enable the SAC created by the suppliers to be integrated with the SAC of the corresponding product.\\
\midrule 
The importance to plan for shared ownership with suppliers & The suppliers might require to keep parts of the SAC private (e.g., some evidence). In this case, it is important to have a mechanism to keep ensuring the overall quality of the SAC, e.g., by introducing a black-box with meta-information. Additionally, the ownership of the whole case has to be considered, as the complete SAC would not be in the hands of a single stakeholder. \\
\midrule 
The challenging nature of working with SAC & Working with SAC is not trivial and comes with many challenges. Traceability and change analysis were considered main challenges by the majority of the participants. Additionally, finding the right competences to carry out the SAC-related work, role identification and description, and acquiring the right tools and integrating them in the organizations tool chain were also considered major challenges.\\
\bottomrule
\end{tabular}
\end{table*}

%%%%%%%%%%%%%%%%%%%

\subsection{RQ2: What are the gaps in the state of the art when it comes to the industrial applicability of SAC?}
To answer this question, we need to know what the industrial needs are, which is covered in Paper A. Additionally, we need to know what exist in literature, and accordingly, we can identify the gaps. Paper B contributes with a Systematic Literature Review in which multiple research questions regarding the applicability of SAC in industry are studied. Specifically, we studied the motivations for creating and using SAC as reported in literature. We also studied different reported approaches for SAC creation, as well as their validations. Lastly, we studied the reported support for SAC creation as reported in literature.
%

%
%\subsubsection{Motivations and usage scenarios}
%ositively, the literature is full of motivations for using SAC, as well as suggestions for where to use them. In particular 73\% of the %included studies in the SLR mentioned at least one motivation for SAC. Additionally 28\% included at least one usage scenario where SAC %can be used for additional purposes beyond achieving security assurance.
%However, these motivations are on a high level and lack detailed studies to show how realistic and applicable they are.

%We believe that there is a substantial gap between the potential of SAC reported in literature and their application in industry. An obvious question to ask is: why are SACs not more widely adopted in industry even though there are so many motivations and usage scenarios for them in literature? It has already been shown that adopting SACs is non-trivial~\cite{industrialNeeds}. It requires a substantial amount of effort and time, which grows as the systems become more complex. It also comes with many challenges, such as finding the right expertise to create them. Furthermore, the challenges do not stop at the creation of SACs, but are extended to updating, maintaining, and making them accessible at the right level of abstraction to the right users. We believe that these matters need to be addressed in studies that suggest the usage of SACs in different domains.

\subsubsection{Wide variety of approaches, but not enough to cover industrial needs}
The literature includes a rich variety of studies which explore approaches for creating SAC, especially when it comes to the argumentation part. However, theses approaches do not consider the specific needs of companies in a specific industry, e.g., automotive.

The variety in approaches gives organizations the possibility to choose those that fit their way of working and the security artefacts they produce.
For example, a company that works according to an agile methodology could choose to adopt an SAC approach for iterative development~\cite{18_othmane2016}.
However, this choice has to consider constraints of the applicability of the approach, including benefits and challenges of its adoption, e.g., the impact on the the way of working. 
These aspects are not discussed in the literature and the burden is left to the adopter.

Another example is the question of conformance with different standards. While this has been discussed in literature, there is a lack of studies which systematically assess different approaches based on their ability to help achieving conformance with a certain standard. To generalize this, we observed that there is a lack of studies which compare different approaches in different contexts. In consequence, from an industrial perspective, organizations need to select suitable approaches in an exploratory way, which can be highly time and resource consuming. 

The studies presenting new approaches also lack the discussion of the granularity level that is possible or required to achieve using each approach. We believe that future studies should take into consideration the possible usages for SAC created using different approaches, and discuss the required granularity level based on that. For example, would an SAC created through the security assurance-driven software development approach \cite{34_vivas2011} be useful to companies which outsource parts of their development work to providers? In that case, on which level should these cases be created, e.g., on the feature level or on the level of the complete product?

\subsubsection{Lack of quality assurance}
Quality assurance is the weaker part of the literature reviewed in Paper B. We talk here about three main things. First is the quality of the outcomes when it comes to their applicability in practice. We have seen scarcity of industrial involvement. The reason might be a lack of interest, which contradict the reported motivations and usage scenarios, or simply because it is hard to get relevant data from industrial companies to validate the outcomes, as security-related data is considered to be sensitive (as we mentioned earlier). Furthermore, with the exception of a few cases, the creation and validation of SAC in literature is done by the authors of the studies. We believe that this contributes heavily to the lack of information addressing challenges and drawbacks of applying SAC in a practical context.

The second issue is the generalizability of the approaches with regards to the used argumentation strategies. The approaches we reviewed use a wide variety of argumentation strategies, e.g., based on threat analysis, requirements, or risk analysis. However, they lack validations and critical discussions as to whether the approaches work only with the used strategies or can use other strategies as well. We suggest to validate these approaches based on different types of strategies in future research.

The last point is the lack of mechanisms for including quality assurance within the SAC. We learned in RQ1 that it is essential for the argumentation provided in SAC to be complete in order for them to be useful. For that there needs to be a mechanism to actively assess the quality of the arguments to gain confidence in them. This is not addressed in literature apart from a few studies where it has been partially addressed, e.g., \cite{1_cyra2007,11_chindamaikul2014,25_rodes2014}.  
Similarly, the evidence part also needs to be assessed. e.g., by introducing metrics to assess the extension to which a certain evidence justifies the claim it is assigned to. 
The inter-relation between claims and evidence needs to be addressed to assess whether a claim is fully justified by the assigned evidence or not. %For example each claim can have a certain saturation level to be achieved, and each evidence provides a degree of saturation. Hence, it would be possible to assess whether the claim is fully satisfied or not by the assigned evidence.

\subsubsection{Imbalance in coverage}
\label{disc:coverage}
%\todo{Get to the point!}
%Our results in Paper B indicate a tendency towards covering the argumentation part more than the evidence part.
%The coverage of matters related to SAC in literature is imbalanced to a large extent. In consequence, 
Multiple needs and drivers, e.g., managing working with suppliers, quality assurance of SAC, and organization-related issues are not covered in literature.
 This indicates a weakness in the approaches, as elements of SAC cannot be evaluated in silos. For example, if we take an approach to create security arguments, how would we know which evidence to associate with these? Moreover, we will not be able to assess whether we actually reach an acceptable level of granularity for the claims to be justified by evidence. The same thing applies to the evidence part. If we only look at the evidence we will not be able to know which claims the suggested evidence can help justify. To be able to evaluate the evidence, they have to be put in context with the rest of the SAC. When reviewing the studies that focus on one element of SAC, we were not able to find any links to related studies focusing on the remaining elements, which indicates incompleteness of the approaches.

When it comes to other areas, the assessment and quality assurance of SAC is rarely covered, as we discussed in the previous sub-section. Furthermore, there is a lack of studies covering what comes after the creation of SAC. In particular, for SAC to be useful, they have to be updated and maintained throughout the life-cycles of the products and systems they target. Otherwise, they become obsolete, according to what we learned in Paper A. Particularly, there need to be traceability links between the created SAC and the artefacts of these products and systems. Many SAC approaches use GSN, which allows to reference external artefacts using the context and assumption nodes. However, these nodes are rarely exploited in the examples provided in the studies we reviewed. 
Moreover, there is a lack of studies targeting the organizational aspects of working with SAC, e.g., the ownership of SAC and how to handle sub-cases when working with suppliers. 

\subsection{RQ3: How can an approach for the construction of security assurance cases fulfill the needs of the automotive domain?}

The three papers included in this thesis contribute towards answering RQ3. We built on the gained knowledge when answering RQ1 and RQ2 to create an approach for SAC creation. The approach CASCADE is the main contribution of Paper C. It is an asset-driven approach with built-in quality assurance.
\subsubsection{Design goals of CASCADE}
CASCADE is inspired by the upcoming standard for cybersecurity in automotive SAE/ISO-21434 \cite{iso21434}. Conformance with this standard has been identified as one of the most important drivers for SAC work in Paper A, and it has not been covered in any of the papers included in the SLR of Paper B. 
CASCADE was designed to achieve the following goals:
\begin{itemize}
    \item Make assets the driving force of the SAC to allow creating security assurance based on what is valuable in the system. 
    \item Embed quality assurance in the approach to make sure the outcome satisfies the desired quality by the adopting entity.
    \item Divide the approach into different layers and blocks, so that different people can work on them in different development phases.
    \item Enable re-usability and scalability to prevent overhead and work repetition while creating SAC on lower-level items.
\end{itemize}
%% Splitting the work into different layers and blocks, so that it can be worked on by different people and in different development phases.
%% Make sure to preserve an area for scaleability and reuseability
%% Include quality assurance and build it in
%% Make assets the driving force! i.e., put it in the top of the arguments.
%%
%\todo{design goals (why it looks the way it looks), then structure, then discuss how the structure support these design goals."these are the high level goals, this is the structure, these are the activities needed. How the structure and activities help achieving the design goals."}
\subsubsection{Structure of the approach}
\begin{figure}[hbt!]
\begin{center}
  \includegraphics[width=0.9\linewidth]{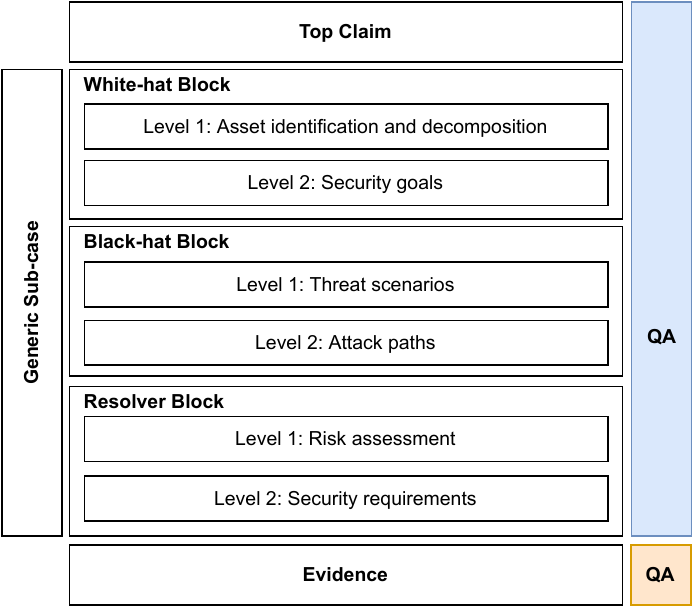}
  \caption{The CASCADE approach for creating security assurance cases} 
  \label{fig:kappa:assetApproach}
\end{center}
\end{figure}

CASCADE consists of blocks which correspond to the requirements and work products of SAE/ISO-21434. Figure \ref{fig:kappa:assetApproach} shows these blocks. 

\paragraph{Top Claim} consists of the top security claim of the artefact in question. It also includes the context of the claim and assumptions made to set the scope of the claim. 
\paragraph{Generic sub-case} helps achieve the goal of preserving re-usability and scalability. It contains a sub-case that is applicable not only to the artefact for which the SAC is being created, but instead to a larger context. For example, if a company defines a cybersecurity policy, enforced by cybersecurity rules and processes, then the policy can be used in security claims for all its products. 
\paragraph{White-hat}
This block starts with the identification of assets, which is the driver of our approach, as per our design goals.
Asset identification is done by conducting an analysis to find the artefacts of the system that are likely to be subject to an attack.
To link the assets to the main claim, we identify which assets exist and which components use or have access to these assets. 
To decompose assets, we look into the types of the identified assets. This gives an indication whether the asset would have implications on the local part of the vehicle (one electronic control unit/ECU), or on a bigger part of the vehicle (multiple ECUs). 

We also look into the relations among assets, e.g., dependability.
To link the asset to the lower level in the approach, i.e., the security goals, we identify the relevant security properties for the assets. Specifically, we look into the Confidentiality, Integrity, and Availability (CIA) triad.
When we have identified the relevant security properties for each asset, we create claims representing the security goals\footnote{A security goal is preserving a security concern (CIA) for an asset \cite{secGoals}}. 

\paragraph{Black-hat}
In this block, we aim to identify the scenarios that might lead to not fulfilling the identified security goals and hence cause harm to our identified assets. 
When we have identified the claims about the achievement of security goals, we proceed by identifying the threat scenarios and creating claims for negating the possibility of these scenarios. We connect these claims to the corresponding claims about achieving security goals. 
We then identify possible attack paths which can lead to the realization of a threat scenario. Each threat scenario might be associated with multiple attack paths. We then claim the opposite of these attack paths. 

\paragraph{Resolver}
This block is the last one in the argumentation part of the CASCADE approach. It links the claims derived from the attack paths to the evidence.
In this level, we assess the risk of the identified attack paths. Based on the risk level, the creators of the SAC create claims to treat the risk by, e.g., accepting, mitigating, or transferring it.

\paragraph{Requirements} At this point, requirements of risk treatments identified in the previous level are to be expressed as claims. This level may contain multiple decomposition of claims, based on the level of detail the creators of the SAC wish to achieve, which is driven by the potential usage of the SAC. For instance, if the SAC is to be used by a development team to assess the security level, this might require a fine grained requirement decomposition which might go all the way to the code level. In contrast, if the SAC is to be used to communicate security issues with outside parties, a higher level of granularity might be chosen. In either case, it is important to reach an \emph{``actionable''} level, meaning that the claims should reach a point where evidence can be assigned to justify them.

\paragraph{Evidence}
The evidence is a crucial part of an SAC. The quality of the argument does not matter if it cannot be justified by evidence. In our approach, evidence can be provided at any block of the argumentation.
For example, if it can be proven in the black-hat block that a certain asset is not subject to any threat scenario, then evidence can be provided and the corresponding claims can be considered as justified.
%\todo[inline]{I don't think the notion of ``branches'' has been introduced yet.} 
If the creators of the SAC cannot assign evidence to claims, this is an indication that either the argument did not reach an actionable point or that there is a need to go back and make development changes to satisfy the claims. For example, if we reach a claim which is not covered by any test report, then there might be a need to create test cases to cover that claim. 

\paragraph{Case Quality Assurance} assists the achievement of our design goal to embed quality assurance in CASCADE.
We consider two main aspects of quality assurance for SAC.
The first aspect is \emph{completeness} which refers to the level of coverage of the claims in each argumentation level of the SAC.
%We determine a level in the security assurance case by the strategy used to decompose the claims. 
Each level in CASCADE includes at least one strategy. 
%If a claim afterwards is decomposed using a different strategy, then we consider this to be a different level.
For each strategy, we add at least one completeness claim that refines it.
%In each level of argumentation, we add at least one completeness claim that refines the strategy used in that level. 
The role of this claim is to make sure that the strategy covers all and only the relevant claims on the argumentation level.
The completeness also relates to the context of the argumentation strategy. The context provides the information needed to determine if the completeness claim is fulfilled or not.

The second aspect is \emph{confidence} which indicates the level of certainty that a claim is fulfilled based on the provided evidence.
This is used in each level of a security assurance case where at least one claim is justified by evidence. The confidence aspect is expressed as a claim, which takes the form: ``The evidence provided for claim X achieves an acceptable level of confidence''. What makes an acceptable level of confidence is defined in the context of the strategy. The confidence claim itself must be justified by evidence.

\subsubsection{Evaluation of CASCADE}
In order to evaluate CASCADE, we collaborated with a security expert from the cybersecurity team at Volvo Trucks, which is a leading OEM that manufactures trucks in Sweden. We conducted several sessions during the development of CASCADE where we discussed the approach, its limitations and possible enhancements. When the approach was fully developed, we conducted a final evaluation session with the expert. 
We used the headlamp example from ISO/SAE-21434 as a context for this discussion. We then presented our approach and the example case for the headlamp item. The expert evaluated the approach by discussing what the overall structure of an SAC should look like from the company's perspective in order to satisfy the requirement for security cases in ISO/SAE-21434 and mapping the different elements of the example case to the internal way of working. The expert also provided insights on how to further enhance the approach.   

\begin{figure}
\begin{center}
  \includegraphics[width=1\linewidth]{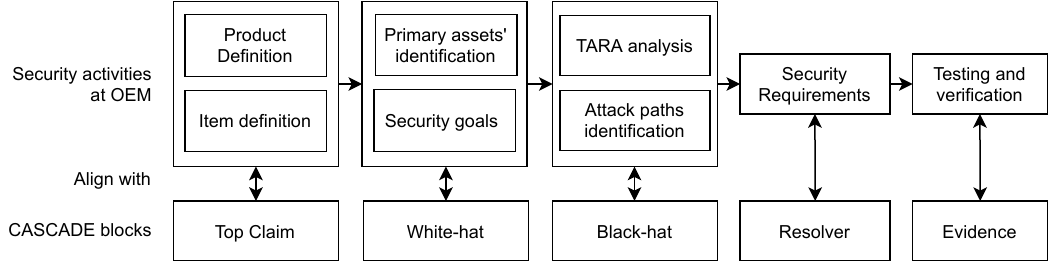}
\end{center}
        \caption{Mapping of the company's security activities to CASCADE blocks} 
        \label{fig:mappingKappa}
\end{figure}

Figure \ref{fig:mappingKappa} shows the different security activities at the company along with the corresponding CASCADE block. A link between an activity and a block indicates that the outcomes of the activity are used to create the SAC elements in the corresponding block. As shown in the figure, CASCADE aligns well with the way of working at the company.

\section{Threats to validity}
In this thesis, we consider the internal and external categories of validity threats as defined in \cite{ttv1}, and described in \cite{ttv3,kh2007guidelines}. 

In terms of \emph{external validity}, we are aware that the general validity of our results in Paper A and Paper C could be limited to the companies involved in the study. Also, the companies are from the same country. Therefore, the results might not directly translate to companies with a different culture.
However, the involved companies are of high profile, quite large and compete at the international level. Therefore, they are able to provide a quite broad perspective on the entire automotive industry.
In any case, the results presented in this paper are an important first important step towards a larger survey study involving more companies and professionals, internationally.

In terms of \emph{internal validity} we consider several aspects. In the the prioritization of the usage scenarios in Paper A, there is a risk that the selection of the top scenarios was biased by present market pressure towards compliance to the upcoming standards. Another limitation is the selection of the participants of the workshop and interviews of Paper A as well as the evaluation of Paper C, as it was based on expertise and availability (convenience sampling). 
%However, in this part of the study, we considered the knowledge of the business around the scenarios to be the most important factor. 
However, In Paper A, we have a balance mix of participants with different types of expertise: security, product development, business, and legal, and in Paper C we have an experienced security expert. This provides us with enough confidence that the results are representative of the expectations and needs across the studied companies. 

The work of conducting the SLR in Paper B was done by one researcher. This means that applying the inclusion / exclusion criteria in each of the four filtering rounds was done by one person. This imposes a risk of subjectivity, as well as a risk of missing results, which might have affected the internal validity of this study. To mitigate this, a preliminary list of known good papers was manually created and used for a sanity check of the selected and included papers. Additionally, a quality control was performed periodically by the other authors to check the included and excluded studies.

Another threat to validity in the SLR is publication bias \cite{kh2007guidelines}. This is due to the fact that studies with positive results are more likely to get published than those with negative results. This could compromise the conclusion validity of the SLR, as in our case we did not find any study that is, e.g., against using SAC, or which reported a failed validation of its outcome. 
In Paper B, we have partially mitigated this threat by also including a few technical reports (i.e., non peer-reviewed material). These papers have been identified as part of the snowballing, as we didn't restrict to peer-reviewed papers.

When it comes to the reliability of the SLR, we believe that any researcher with access to the used libraries will be able to reproduce the study, and get similar results plus additional results for the studies which get published after the work of the SLR is done.

In Paper C, we used an example from ISO/SAE-21434 to illustrate CASCADE. However, there is a risk that the example does not represent actual cases from industry. We believe that the structure of the example case is what is important for the evaluation rather than the actual content, as discussed and confirmed by the security expert who ran the evaluation at the OEM.
\section{Conclusion}
%\todo{main contributions described}
In this work, we have created CASCADE, an asset-driven approach for the creation of security assurance cases with built in quality assurance. CASCADE was inspired by the structure of the upcoming ISO/SAE-21434 standard for cybersecurity in automotive.
We have identified the drivers of security assurance case work in the automotive industry by investigating and analyzing requirements and usage scenarios for these cases in the industry.
We have also systematically reviewed literature of SAC and identified gaps between what is available in literature and what the industry needs. 
We utilized what we learned from the industry and the gaps we found in literature to design CASCADE. We evaluated this approach with a security expert at a large automotive OEM. An example case available in ISO/SAE-21434 was used to illustrate the approach, and the evaluation showed that the cyber-security activities at the company aligns well with the structure of CASCADE. 
\section{Future work}
In this section, we discuss the future work, which will build on the findings of this thesis to achieve the overall goal of my PhD thesis, as discussed in Section \ref{rq:ResGoal}.

\paragraph{Further development of CASCADE} CASCADE has been designed to close gaps between literature and industrial needs when it comes to adopting SAC. We will continue to develop CASCADE to further close these gaps. In particular, we the future development will target:
\begin{itemize}
    \item The maintenance of SAC, i.e., how to enable updating the cases following changes that are made to the system in question or to the artefacts used to build the case using traceability links.
    \item Organizational matters, i.e., how the work in SAC would affect the day to day work in an automotive company, and what impact it would have on the enterprise architecture of these companies.
    \item Work on the evidence part of SAC. In particular, we want to study the available evidence in automotive companies and assess what support is needed to continuously updated them in the SAC.
    \item CASCADE was inspired by the structure and requirements and work products of SAE/ISO-21434. However, there are other regulations and standards which require SAC, as we have seen in Paper A. We plan to study these requirements and reflect them in the structure of CASCADE.
\end{itemize}

\paragraph{Evaluation of CASCADE} To assess CASCADE, we plan to evaluate it by including a larger community of automotive companies and security experts. The plan to base the evaluation on the eventual added value of CASCADE to the company.
Additionally, we plan to study the application of CASCADE in other safety-critical domains, e.g., the medical domain, which includes different organizational structures to the automotive companies which we have been targeting. Moreover, we plan to reach out to stakeholders in the software engineering domain, e.g., architects with a questionnaire to evaluate CASCADE. This is mainly to eliminate the potential bias to the companies we collaborate with. %\todo{apply to other domains as well for instance the medical domain which include different organizational structure and context (smaller organizations..... This is biased to the companies.. so we can reach out with a questionnaire to ppl (architects for instance and other stakeholders in automotive and other companies)}
%\todo{get rid of the subsection and convert to bullet points}
%The direction of future work will take into consideration the answers to RQ1 and RQ2.
%\subsection{Further development and evaluation of CASCADE}
%\subsection{Evidence: close the gap to literature (imbalanced coverage)}
%\subsection{Addressing the main challenges of SAC work in industry (maintainability and tool support)}

%- how is CASCADE with respect to the list of recommendation \todo{think about it and maybe add something in the future work.}
%\todo{think about the weak spots and potential "nasty" questions. 10-15}

%The actual papers that are included in the thesis
\cleardoublepage 
%%===============================================================%
%%                     Appended papers                           %
%%===============================================================%
%Include each paper as shown below. Abstracts are in a separate file.

\chapter{Paper A}
\label{chap_paper_a}
\thispagestyle{empty}
\subsection*{Security Assurance Cases for Road Vehicles: an Industry Perspective}
\subsubsection*{M. Mohamad, A. Åström, Ö. Askerdal, J. Borg, R. Scandariato}
\subsubsection*{{\em In ARES ’20: International Conference on Availability,Reliability and Security. Article No.: 29 Pages 1–6.}} \cite{paperA}
\subsubsection*{Extended version available at: https://arxiv.org/abs/2003.14106} \cite{paperA_extended}
\newpage
\thispagestyle{empty}
\mbox{}
\newpage
\addtocounter{page}{-2}
\newpage
\section*{Abstract}
Assurance cases are structured arguments that are commonly used to reason about the safety of a product or service. Currently, there is an ongoing push towards using assurance cases for also cybersecurity, especially in safety critical domains, like automotive. While the industry is faced with the challenge of defining a sound methodology to build security assurance cases, the state of the art is rather immature. Therefore, we have conducted a thorough investigation of the (external) constraints and (internal) needs that security assurance cases have to satisfy when used in the automotive industry. This has been done in the context of two large automotive companies. The end result is a set of recommendations that automotive companies can apply in order to define security assurance cases that are (i) aligned with the constraints imposed by the existing and upcoming standards and regulations and (ii) harmonized with the internal product development processes and organizational practices. We expect the results to be also of interest for product companies in other safety critical domains, like healthcare, transportation, and so on.\\ 
\linebreak

\newpage
%Here you can just include the entire paper, starting from the first section.
%Skip all the meta Latex code, e.g., begin{document}, includes and so on.

% ARES 2020
% https://www.ares-conference.eu/conference-2020/cfp2020/

\section{Introduction}

\begin{figure}
    \centering
    \includegraphics[width=\columnwidth]{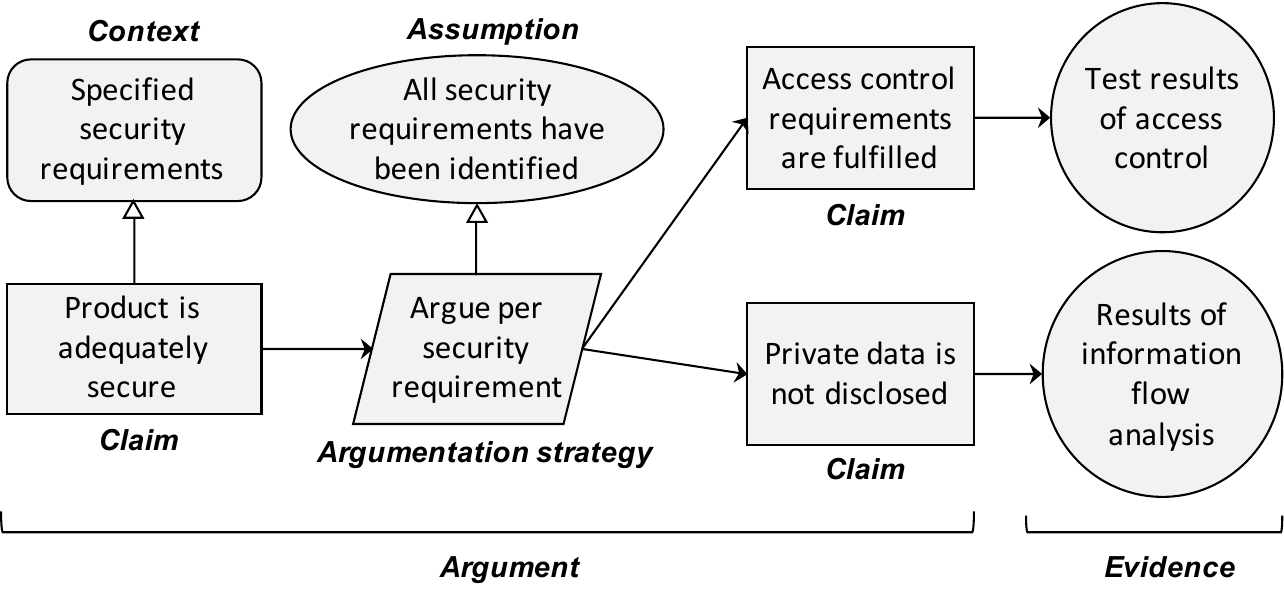}
    \caption{A simple example of a security assurance case}
    \label{fig:introSAC}
\end{figure}

An assurance case can be described as a structured set of arguments that are supported by evidence, e.g., collected from the results of the validation and verification activities \cite{goodenough2007}.
A simple example is given in Figure \ref{fig:introSAC} and the reader could recognize the resemblance with the logical argumentation of a legal case.
Assurance cases have been in use for several decades in order to argue for completeness and correctness of various dependability attributes in a wide range of industrial fields. 
In the automotive industry, assurance cases for functional safety (or safety cases) are a common practice since the release of the ISO 26262 standard on functional safety for road vehicles  in 2011 \cite{iso26262_1ed}. 
The necessity to adopt assurance cases also for cybersecurity is emerging in the automotive industry only now, especially because of the upcoming release of security standards that explicitly demand for them \cite{iso21434}.
The necessity is also felt from within the automotive industry.
The vehicle industry is going through a rapid transformation with features such as increased connectivity and automated driving as two of the major driving forces.
Features like these demands various external interfaces which exposes potential vulnerabilities in the connected devices of the vehicles and increases the risks in a way that was never seen before.
Therefore, a more systematic way to ``reason'' around security is desirable.

The push towards adopting security assurance cases represent a challenge that is both technical and organizational at the same time. 
For instance, the selection of a given argumentation strategy (i.e., the structuring of the security case) is not a purely technical choice, as it might require to re-organize the way the product is developed, e.g., in order to introduce extra activities and work products to create the necessary evidence.
Such organizational issues cannot be underestimated in large eco-systems, like the development environment in vehicle manufacturers (OEM).

In our analysis of the literature, we have found that the related work has not investigated the constraints and requirements around high-impact technical decisions such as (i) how to structure a security case, (ii) how to collaborate with suppliers on security cases, (iii) how to effectively update a security case, and so on.
Therefore, in a collaboration between one academic institution and two OEMs, we have performed a study of the industrial needs that pertain the technical choice of adopting (or defining) a methodology for security assurance cases in an automotive organization.

This paper has \emph{three main contributions}. 
First, we have performed a systematic study of the security-relevant regulations and standards in the automotive domain. 
In this analysis, we have identified the explicit and implicit constraints laid out by such documents with respect to security cases.
We call these the \emph{external forces} driving the adoption of security cases.
The analysis is performed by a pool of industrial security experts (working in two panels) who are also members of several standardization committees and, hence, know how to interpret these documents, which are, at times, somewhat fuzzy.

Second, we have performed an empirical study with a significant number of stakeholders (more than 20 people) that are affected either directly (e.g., as prospective producers) or indirectly (e.g., as prospective consumers) by security assurance cases.
By applying rigorous methods from the field qualitative research, we have systematically identified the \emph{internal organizational needs and opportunities} that must be taken in consideration for a successful adoption of a security case methodology.

Third, we have combined the observations collected in the two above-mentioned studies and translated them into a list of practical recommendations on security cases for the automotive industry.

The rest of the paper is structured as follows. 
In Section \ref{sec:rw}, we provide more background on assurance cases and discuss the related work. 
In Section \ref{sec:method}, we formulate the research questions and describe the research methodology. 
In Sections \ref{sec:rq1} and \ref{sec:rq2}, we present the results. 
In Section \ref{sec:threats}, we discuss the threats to the validity of our research.
In Section \ref{sec:discussion}, we discuss the results and provide our recommendations.
Finally, in Section \ref{sec:end}, we presents the concluding remarks.

\section{Background and Related work}\label{sec:rw}

\textbf{Security Assurance Cases.}
%A description of what they are and how they look like (from the generic papers).
%Assurance cases are bodies of evidence organized in structured arguments, used to justify that certain claims about a systems property hold \cite{goodenough2007}. 
The argumentation in a Security Assurance Case (SAC) consists of claims about security for the system in question, and the evidence to justify these security related claims.
A SAC consist of the following primary components: (i) security claims, (ii) the context in which the claims should hold, (iii) an argument about the security claim, (iv) the strategy used to build the argument, and (v) A body of evidence to prove the claims \cite{knight2015,alexander2011}
A SAC can be expressed in a textual or graphical format\cite{alexander2011}. The most common graphical formats are the Goal Structure Notation (GSN, \cite{gsn}), and the Claims Arguments, and Evidence notation (CAE, \cite{CAE}). 
%(focusing on security) Which aspects we want to highlight (Main decomposition strategy: security goals vs security threat) or other paper that argue over processes.
Researchers have been exploring several approaches for creating the argument part of SAC. Biao et al.~\cite{20_xu2017} suggest dividing the argument into different layers, and using different patterns to argue in each of these layers, which include an asset layer, and a threat layer. Another approach was used by Poreddy et al.~\cite{26_poreddy2011}, which uses different security properties as argumentation strategy. Other used approaches argue by development life-cycle phases \cite{16_ray2015}, standard recommendations \cite{4_he2012}, and product components \cite{15_hawkins2015}.

\textbf{Security Assurance Cases in Automotive.}
The review of literature shows very little use of assurance cases for security in the automotive industry. 
One of the very few related studies is the work by Cheah et al.~\cite{39_cheah2018}. The authors present a classification of security test results using security severity ratings. These security tests then form a body of evidence used as an input for constructing a security assurance case. The study suggests a bottom-up approach for constructing a security case, but does not provide a complete example case to show how the body of evidence is connected to claims. %Hence, it is not clear how this approach helps addressing specific claims in the automotive industry.
Another related study is the work by Fung et al.~\cite{35_fung2018} which studies maintaining assurance cases by using automated change impact analysis. A tool was created for that purpose and a case study was conducted in the automotive domain. The authors claim the tool can be used for both safety and security. However, the example case in the paper is only about safety. 
%2 papers from Mazen's SLR

%Functional Safety according to \cite{iso26262_1ed, iso26262_2ed}. Look at \cite{Birch2013} and  \cite{Palin2011}.
%Look for a PhD thesis by Fredrik Törner (safety cases)
%2/3 very relevant/referential works about automotive
%Surveys about safety cases in automotive?
 %Anything about industrial needs for safety cases (work similar to this, but for safety)?

\textbf{Safety Assurance Cases in Automotive.}
Safety cases have been in use in the automotive industry for several years.
%Palin et al.~\cite{Palin2011} argue that there were no consensus in the industry on the actual value that a safety case brings.
In the second edition of the ISO 26262 standard \cite{iso26262_2ed} it is stated that release for production shall only be approved if there is \emph{``sufficient evidence for confidence in the achievement of functional safety''} and that this could be provided by the safety case. This clearly shows the importance of a safety case. 
Birch et al.~\cite{birch2013} perform an industrial case study focusing on the product-related arguments of a safety case as opposed to the process-related arguments, which according to the authors often gets the overhand. The authors discuss the outcome of the case study listing challenges and advantages. They mention how the engineers that design the system may benefit from the safety case compilation throughout the project, especially the product-related part, and address issues in a timely manner.

\section{Research Methodology}\label{sec:method}

This study was conducted in two large automotive OEMs located in Europe. \emph{Company A} is a passenger car manufacturer, while \emph{Company B} is a truck manufacturer. 

%%%%%%%%%%%%%%%%%%%%%%%%%%%%%%%%%%%%%%%%%%%%%%%%%%%%%%%%%%%%%%
\subsection{Research Questions}

This work is motivated by the urgency that is currently perceived by the automotive industry with respect to implementing security assurance cases.
This is due to the emergence of several standards and regulations that are forcing the industry to develop a methodology for SAC in order to stay compliant and avoid legal risks.
We call these the \emph{external drivers} that will impose constraints on how SAC should look like.
Accordingly, we formulate the first research question as follows:\\
\textbf{RQ1}. What are the \textbf{constraints} for SAC coming from regulations and standards in the automotive markets of EU, US, and China?

The need to develop a strategy for SAC is also perceived by the automotive companies as an opportunity to improve their cybersecurity development process. Also, such methodology should integrate with the product lifecycle. 
As such, we have investigated these \emph{internal drivers}. Accordingly, we formulate the second research question as follows:\\
\textbf{RQ2}. What are the \textbf{needs and opportunities} related to security assurance cases in the automotive industry?

%%%%%%%%%%%%%%%%%%%%%%%%%%%%%%%%%%%%%%%%%%%%%%%%%%%%%%%%%%%%%%
\subsection{Methodology}\label{subsec:methodology}
%\todo{Mazen: Add regulations to standard column without destroy the format of the table.}
%\todo{Mazen: for some reason the footnote in the table is not visible, can you please fix it?}
\afterpage{% 
\begin{landscape}
\begin{table*}
\caption{List of the participants in all stages of the study}
\smaller
\label{tbl:participants}
%\begin{minipage}{\columnwidth}
%\begin{center}
\begin{tabular}{llllllll}
\toprule
               &               &                              & RQ1                   & RQ2        &            &                &            \\ 
           ID  & Company       & Role                         & Analysis of standards  & Pre-study  & Workshop   & Prioritization & Interviews  \\ 
           &&&and Regulations&&&&\\
\midrule
            1  & Company A     & Attribute Leaders %\footnote{Cybersecurity \& Privacy on product definition and complete vehicle level.} 
               &  \checkmark           & \checkmark &            &                &            \\
            2  & Company A     & Regulatory experts              &  \checkmark           & \checkmark &            &                &            \\
            3  & Company A     & Safety experts              &  \checkmark           & \checkmark &            &                &            \\
            4  & Company A     & Security R\&D experts              &  \checkmark           & \checkmark &            &                &            \\
            5  & Company A     & Product Owner Security              &  \checkmark           & \checkmark &            &                &            \\
            6  & Company A     & Security Engineers              &  \checkmark           & \checkmark &            &                &            \\
            7  & Company B     & Security expert              &                       &            & \checkmark & \checkmark     &            \\
            8  & Company B     & Security expert              &                       &            & \checkmark & \checkmark     &            \\
            9  & Company B     & Safety expert                &                       &            & \checkmark & \checkmark     &            \\
            10  & Company B     & Security Engineer            &                       &            & \checkmark &                &            \\
            11 & Company B     & Software architect           &                       &            & \checkmark &                &            \\ 
            12 & Company B     & Principal Engineer           &                       &            & \checkmark &                &            \\
            13 & Company B     & Software Engineer            &                       &            & \checkmark &                &            \\
            14 & Company B     & Security Engineer            &                       &            & \checkmark &                &            \\
            15 & Company B     & Security Engineer            &                       &            & \checkmark &                &            \\
            16 & Company B     & Security Engineer            &                       &            & \checkmark & \checkmark     &            \\
            17 & Company B     & Security Engineer            &                       &            & \checkmark & \checkmark     &            \\
            18 & Company B     & Security expert              &                       &            &            & \checkmark     & \checkmark (facilitator) \\
            19 & Company B     & Solution Train Engineer (STE)&                       &            &            &                & \checkmark (stakeholder) \\
            20 & Company B     & Solution Manager             &                       &            &            &                & \checkmark (stakeholder)\\
            21 & Company B     & Functional Safety Assessor   &                       &            & \checkmark &                & \checkmark (stakeholder)\\ 
            22 & Company B     & Component Owner              &                       &            &            &                & \checkmark (stakeholder)\\
            23 & Company B     & Senior Legal Counsel         &                       &            &            &                & \checkmark (stakeholder)\\
            24 & Company B     & Security expert              &                       & \checkmark &            & \checkmark     &            \\
            25 & Company B     & Security Manager             &                       & \checkmark &            & \checkmark     &            \\
            26 & Company B     & Security R\&D                &                       & \checkmark &            &                &            \\            
            27 & University C  & Researcher                   &                       &            &            & \checkmark     & \checkmark (interviewer) \\
            28 & University C  & Researcher                   &                       &            &            & \checkmark     &            \\ 
\bottomrule
\end{tabular}
%\end{center}
\footnotesize
%\end{minipage}
\end{table*}
\end{landscape}
}
As shown in Table \ref{tbl:participants}, this study involved a total of 28 participants.%\todo{Volvo Cars: update roles.}.

\textbf{RQ1.} Concerning the analysis of the standards and regulation, \emph{Company A} maintains a knowledge base of relevant documents as part of their security governance framework.
This knowledge base consist of standards, regulations, guidelines, best practices,  etc applicable for various markets.
Furthermore, this knowledge base covers both current and upcoming trends.
We assume that such knowledge base is fairly complete, at least for the most relevant markets (e.g., US, EU, China).
Furthermore, this knowledge base includes, among other things, information regarding the categorization of  requirements, their relevance, the parts of the organization that is affected, and which life-cycle phases of the products are impacted.
This knowledge base represented the pool of documents we have analyzed in order to answer RQ1.

In particular, in this study we prioritize new regulations and standards that will soon come into effect and focus on the markets mentioned above.
The filtered documents (listed in Section \ref{sec:rq1}) have been analyzed for explicit references to security assurance cases or their parts.
We also looked for implicit relationships to SAC.
For instance, in the SELF DRIVE Act \cite{SELFDRIVE} there is a demand that manufactures must have a \textit{Cybersecurity Plan} that includes, among other things, processes for identification, assessment and mitigation of vulnerabilities (that are reasonably foreseeable). A SAC can then be used to show how this requirement is fulfilled listing the demanded processes and the evidence for them. 

%Secondary input can be documents of lower prioritization but still giving valuable input to the area. Both the ISO/SAE 21434 \cite{iso21434} and the SAE J3061 \cite{J3061} puts direct requirements on SAC. 

\smallskip

\begin{figure}
    \centering
    \includegraphics[width=\columnwidth]{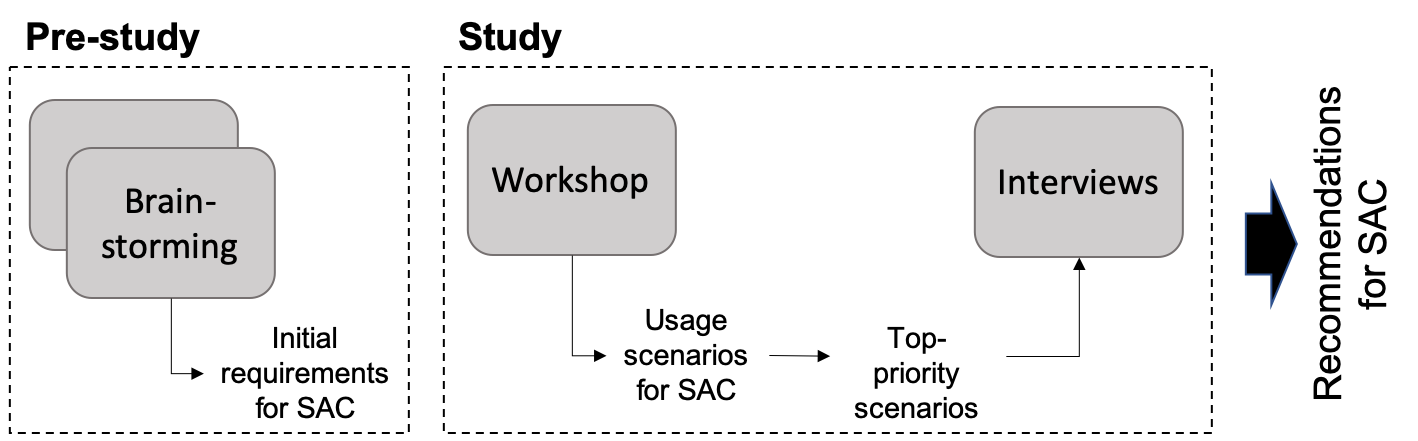}
    \caption{Method used to get the industrial needs for security assurance cases}
    \label{fig:RQ2Method}
\end{figure}

\textbf{RQ2.} To understand the internal needs for security assurance cases in the automotive industry, we used a three-steps method, as shown in Figure \ref{fig:RQ2Method}.

\paragraph{Pre-study}
The goal of the pre-study was to assess the overall industrial expectations with regards to SAC.
In particular, the pre-study reflects the point of view of the security leaders in the two companies participating to this investigation. 
In each company and independently from each other, a panel of experts performed a series of brainstorming meetings. 
The goal of the brainstorming was to form a consensual opinion of how the SAC should `look and feel', within each company and from the perspective of security people.
The results of the two panels were compared and merged into a single list of requirements and constraints, with the support of University C.

The panels consisted of R\&D personnel (i.e., technical leaders) with expertise in both security and safety.
Further, the participants were all familiar with the concept of assurance cases.
As a consequence, these groups of people were very homogeneous and might not have been aware of the full spectrum of needs and expectations in their two large companies.
Therefore, we decided to perform a larger study (comprising a workshop and a series of interviews), involving a larger and more diverse set of stakeholders\footnote{The study has been performed only in Company B, due to resource constraints.}.
In Section \ref{sec:rq2} we present the results from both the pre-study and the larger study, and we compare the observations.

\paragraph{Workshop}
The first step in our study was conducting a workshop to elicit  usage scenarios related to security assurance cases. The workshop was conducted at Company B. We invited stakeholders from different backgrounds and different parts of the organization. 
As shown in Table \ref{tbl:participants}, in total, we had 12 participants and three moderators contributing.
We started the workshop with a presentation to introduce security assurance cases to the participant which did not have previous experience with them.
We then divided the participants into three groups of 4 participants, making sure to spread similar roles and competences among the groups, e.g., we had three participant who were familiar with safety cases, so we assigned them to different groups. 
We asked the groups to brainstorm for 45 minutes on usage scenarios for security assurance cases, and to describe them as user stories, like ``As a <<role>> I would use security assurance cases for <<usage>>''~\cite{cohn2004}. We explicitly asked the participants to come up with real-life scenarios in the context of their company. 

%In the third session, we asked the participants to vote for the usage scenarios they find most interesting, and would want to work on for the last session. Each participant was given three votes, and they were given the chance to put them all on one usage scenario or spread them. Session four was then about getting a deeper understanding of the usage scenarios. The participants were asked to answer the following questions for the selected usage scenario: \emph{(i)} \textbf{what} types of security claims are associated with the scenario?;  \emph{(ii)} \textbf{who} should be responsible of the case? e.g., who is the owner, QA responsible, and maintainer?; \emph{(iii)} \textbf{how} would the practicalities be managed, e.g., the repositories of evidence and claims?; \emph{(iv)} \textbf{challenges} of applying the scenario in real life.

%Solution Train Engineer - Project management / product ownership
%Solution manager  - Project management / product ownership
%Functional safety assessor - Compliance expert
%Component owner - Purchasing expert
%Senior legal counsel - Legal expert

\paragraph{Prioritization and Interviews}

%TODO: not enough from the break-out sessions. Decision was made to switch to interviews. Due to limited resources, we had to prioritize the top 5 scenarios.

At this step, we wanted to dig deeper and get a better understanding of the most important scenarios. We also wanted to acquire the point of view of more diverse stakeholders. Hence we had to prioritize the scenario and identify stakeholders to be interviewed for the top ones.

Concerning the prioritization, we aimed at getting expert opinions on which usage scenarios are of most value to the company, from a security perspective. 
As shown in Table \ref{tbl:participants}, we sent out the scenarios collected from the workshop to 10 security experts and asked them to select the top five scenarios by assigning a rank from 1 to 5 to them, where 5 is assigned to the most valuable scenario for the company.

Afterwards, we selected, the top five usage scenarios and identified a key stakeholder at company B for each.
Finally, we conducted in-person interviews with these stakeholders. Note that at the interviews, a security expert from the company was also present as a facilitator of the discussion.
The interviewees were selected based on the relevance of their expertise to the actors of the user stories in the corresponding usage scenarios. For example, the actor of one of our top usage scenarios is a \emph{legal risk owner}, hence, we selected an interviewee who has extensive experience in law and has the role: \emph{senior legal counsel} in Company B.

We organized each interview into four parts, according to the following themes:
\begin{enumerate}[i]
\item 
\textbf{value} In the first part, we focus on the value that SAC might bring to the stakeholder in terms of, e.g., efficiency, and quality management. The objective of the discussion is to picture the ‘status quo’ (e.g., to understand how the level of security is currently appraised) and the expectations (i.e., how things should improve). 
\item \textbf{content and structure} The focus of this part is to get the interviewees' technical opinions on how the content and structure of SAC should be, e.g., in terms of level of detail and types of claims; 
\item 
\textbf{integration} This part is about understanding how SAC could be integrated with the current way of working, and whether it could fit in the current activities, or would require modifications to the process; and
\item
\textbf{challenges and opportunities} The last part of the interview is about understanding the challenges and opportunities that the stakeholders foresee in applying SAC.
\end{enumerate}

In each interview, there was an interviewer (an author), an interviewee, and a security expert who acted as a discussion enabler (also an author). We recorded the interviews, and used the recordings to extract a transcript for each interview. These were then sent to the corresponding interviewees validation, and additional comments. 
%\todo{TODO: done by whom, recorded, analized HOW, analysis done by 2+1 ppl, results sent back to interviewwes for OK}

\section{RQ1: External drivers}\label{sec:rq1}

In this section we look at external drivers that put requirements on the automotive industry. These drivers include regulations, standards, best-practises, and guidelines. The intent of the analysis is to find the relation and the motivation of a SAC from these documents. 
We look at some of the current documentation as well as some upcoming ones. We do not present a complete list, but rather look at the ones that may have a large impact on the subject and the field of interest. These are presented in~\ref{tbl:external}. The first three columns in the table indicate how SAC is referenced in the document, i.e., whether it is explicitly or implicitly mentioned, or whether it is beneficial for the purpose of the document. The rest of the table categorizes the documents in terms of type (regulation, standard, guideline, or best-practice) and market. The last four columns indicate whether the document requires compliance or conformance, and whether the document targets the process or the product. 

\afterpage{% 
\begin{landscape}
\begin{table*}
\caption{External drivers and references}
\label{tbl:external}
\begin{tabular}{llll|llllll}
\toprule
            \multicolumn{4}{l|}{SAC reference motivation}         & Categorization        &            &                &       &   &  \\ 
\midrule
         \rotatebox{45}{Explicit}  & \rotatebox{45}{Implicit}   & \rotatebox{45}{Beneficial} & \rotatebox{45}{Reference}          & \rotatebox{45}{Type}  & \rotatebox{45}{Market}         & \rotatebox{45}{Compliance} & \rotatebox{45}{Conformance} & \rotatebox{45}{Process}    & \rotatebox{45}{Product}      \\ 
\midrule
   \checkmark &            &             &  ISO/SAE 21434 DIS \cite{iso21434} & Standard       & International  &            & \checkmark  & \checkmark &  \checkmark  \\
   \checkmark &            &             &  SAE J3061 \cite{J3061}            & Guideline      & International  &            & \checkmark  & \checkmark & \checkmark   \\
              & \checkmark &             & SELF DRIVE Act \cite{SELFDRIVE}    & Regulation     & US     & \checkmark &             & \checkmark & \checkmark   \\
              & \checkmark &             & ADS 2.0 \cite{ADS2}                & Best-practise  & US     &            & \checkmark  & \checkmark & \checkmark   \\
              &            & \checkmark  & AV 3.0 \cite{AV3}                  & Best-practise  & US     &            & \checkmark  & \checkmark & \checkmark   \\
              &            & \checkmark  & GDPR \cite{gdpr}                   & Regulation     & Europe & \checkmark &             & \checkmark &  \\
              &            & \checkmark  & SPY Car Act \cite{SPYCar}          & Regulation     & US     & \checkmark &             & \checkmark & \checkmark \\
              &            & \checkmark  & CCPA \cite{CCPA}                   & Regulation     & US     & \checkmark &             & \checkmark &  \\
              &            & \checkmark  & UNECE GRVA CS \cite{UNECE-Reg}             & Regulation     &  International      & \checkmark &   & \checkmark & \checkmark   \\
              &            & \checkmark  & UNECE GRVA OTA \cite{UNECE-OTA}         & Regulation     &  International      & \checkmark &             & \checkmark & \checkmark \\ 
              &            & \checkmark  & ICV \cite{ICV}                     & Standard       & China  &           & \checkmark  &            & \checkmark   \\
              &            & \checkmark  & GB/T 35273 \cite{GBT35273}         & Standard       & China  &            & \checkmark  &        \checkmark    &    \\
              &            & \checkmark  & CSL \cite{CSL}                     & Regulation     & China  & \checkmark &             &        \checkmark    &    \\
\bottomrule
\end{tabular}
\end{table*}
\end{landscape}
}

There are both general and specific legislation and regulations regarding security and privacy that are applicable to the automotive industry. An example of the former is the European General Data Protection Regulation (GDPR) \cite{gdpr} and the Chinese equivalent GB/T 35273 \cite{GBT35273}, although a recommendation. Below we describe the relation between these documents and SAC. \\

Two of the analyzed documents explicitly mentions SAC: ISO/SAE 21434 DIS \cite{iso21434} and SAE J3061 \cite{J3061}. The former gives requirements on a cybersecurity case and that it shall provide the arguments that cybersecurity is achieved. The latter states that a cybersecurity case provides evidence and argumentation that design and implementation is sufficiently secure.

The Safely Ensuring Lives Future Deployment and Research In Vehicle Evolution (SELF DRIVE) Act~\cite{SELFDRIVE} contains requirements stating that a Cybersecurity plan that includes various security activities shall be developed. It thus implies that it shall be possible to show that the plan also has been implemented. The SAC shall therefor contain the corresponding evidence and argumentation of how the security plan has been met, including confidence in process for incident handling. 
Similarly, for the Automated Driving Systems (ADS 2.0) \cite{ADS2} it is beneficial to show evidence that cybersecurity processes are followed and show that the company adheres to industry best practice. The authors of the (ADS 2.0) document encourage documentation of how cybersecurity has been reached, including design choices, analyses, testing, etc. The SAC shall therefor cover these aspects.
The Automated Vehicles 3.0~\cite{AV3} also indicates the benefit to show evidence that cybersecurity processes are followed and show that the company adheres to industry best practice, including design principles and incident handling. The SAC shall therefor contain the corresponding evidence and argumentation of how security has been considered and handled during design and confidence in processes for incidence handling.

There is a current initiative on UN level, prepared by a subgroup of the working group on Intelligent Transport Systems / Automated Driving (IWG ITS/AD) of WP.29, referred to as ``UN Task Force on Cyber security and OTA issues'' (TF-CS/OTA), to establish regulations and type approval on Cybersecurity for vehicles that includes vehicle categories; M (standard passenger vehicles) and N (trucks). 
This include ``Regulation on uniform provisions concerning the approval of cyber security''~\cite{UNECE-Reg} and ``Regulation on uniform provisions concerning the approval of software update processes''~\cite{UNECE-OTA}.
For the former it can be seen as highly beneficial to show compliance and conformance to requirements in the regulation concerning process implementation and fulfillment as well as demonstration of performed verification activities and its outcome. The SAC shall therefor contain evidence and argumentation in regards to confidence in the implemented processes that shall cover the life-cycle of the vehicle to maintain an appropriate level of security. As well as product evidence and argumentation of an adequately secure product.
The latter one contain requirements for demonstration of evidence of a secure update process.
The SAC shall therefor contain evidence and argumentation in regards to confidence in the SW update processes in order to maintain an adequately secure product and acceptable level of risk.

Further, for the privacy related documents: General Data Protection Regulation (GDPR)~\cite{gdpr}; Security and Privacy in Your Car (SPY CAR) Act~\cite{SPYCar}; Chinese Information Security Technology – Personal Information Security Specification GB/T 35273~\cite{GBT35273}; and  California Consumer Privacy Act~\cite{CCPA}, SAC can be beneficial in regards to showing compliance to regulation, processes, methods and technologies to secure handling of data. The SAC shall therefor contain evidence and argumentation in regards to confidence in the processes as well as  implementation measures of handling private data. 
%\textbf{EU}: 
%The General Data Protection Regulation (GDPR) \cite{gdpr} came into effect in May of 2018 and concerns privacy aspects of user's private data and to withhold transparency and visibility to the user. To ensure data integrity and security of user data several aspects need to be considered. 

SAC can also be beneficial in the China Cyber Security Law (CSL)~\cite{CSL} to show that the organization has sufficient policies and processes that handle cybersecurity. Including incident handling and secure handling of private data.  The SAC shall therefor contain evidence and argumentation in regards to confidence in the post-production processes.
The CSL is a high level law and more detailed requirements will exist through the Chinese strategy and framework on Intelligent and Connected Vehicles (ICV)~\cite{ICV}. With the same reasoning it can be beneficial with the ICV in regards to showing compliance and conformance to requirements and processes, methods and design/technology. The SAC shall therefor contain evidence and argumentation of confidence in the implemented measures and processes to maintain an adequately secure product and acceptable level of risk.

\section{RQ2: Internal Needs and Oppotunities}\label{sec:rq2}

%%%%%%%%%%%%%%%%%%%%%%%%%%%%%%%%%%%%%%%%%%%%%%%%%%%%%%
\subsection{Pre-study: Expectations of Security Leaders}
\label{sec:pre-study}

The panels of security experts concluded that the internal needs of an OEM in regards to the use of assurance cases for cybersecurity can be summarized as follows. 

\noindent\textbf{Secure product argumentation.} Need to have an argumentation for adequately secure implementations of HW/SW E/E systems in their vehicle throughout their life-cycle. This can be used, among other things, to make release decisions.

\noindent\textbf{Supporting evidence.} Need to have all the necessary evidence that the implemented systems are secure enough in order to release a vehicle to the end-customers. In particular, the evidence should undergo a quality assurance process and should cover all the relevant parts included in the development, with sufficient detail for the critical parts.
    
\noindent\textbf{Legal compliance.} Comply to the legislation, regulations and type approvals on national and international levels (in order to even be allowed to sell vehicles).

\noindent\textbf{Supply chain.} Manage collaboration between OEM and suppliers in terms of requirements, structure, aggregation and level of details for SAC constituents.

\noindent\textbf{Standard conformance.} Conform to standards, guidelines and best practices and implement state of the art processes and methods (in order to be competitive w.r.t. other OEMs).

\noindent\textbf{Process harmonization.} Existing development processes and the way of working need to be harmonized with the processes and methods required by SAC. One such example may be an agile way of working and product organization.

%%%%%%%%%%%%%%%%%%%%%%%%%%%%%%%%%%%%%%%%%%%%%%%%%%%%%%
\subsection{Workshop to Identify Broad Usage Scenarios}
\label{sec:scenarios}

The workshop participants (working in three parallel group sessions) identified thirteen unique usage scenarios (US), which are listed below in no particular order. These scenarios depict, in a narrative forms, the broader set of needs and expectations of an OEM with respect to SAC. 
These scenarios cover a diverse set of stakeholders (10 different roles) within the OEM organization, which is, typically, a very large one.
%The two roles which were given more than one scenario are project manager (3 usage scenarios), and product owner (2 usage scenarios). This is not surprising as the organization in which the study was conducted is projectized, and is in a transition phase towards the Scaled Agile Freamework (SAFe).  

\begin{figure}
    \centering
    \includegraphics[width=1\linewidth]{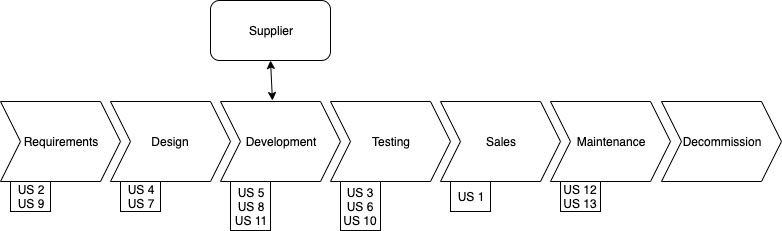}
    \caption{Usage scenarios for the life-cycle of an automotive product}
    \label{fig:lifecycle}
\end{figure}

When looking at the suggested usage scenarios, we can see that they span over multiple phases in an automotive product's life-cycle. Figure~\ref{fig:lifecycle} shows a high-level view of the different phases in the automotive product's life-cycle and the usage scenarios suggested in each phase. As the figure shows, the participant were able to identify at least one usage scenario in each phase except for the final decommission phase.

\noindent\textbf{US 1} As a salesman, I would use top-level SAC to prove to our customers that the company has considered all relevant security aspects of the final product, and has enough evidence to claim that it has fulfilled them.
 
\noindent\textbf{US 2}  As a member of the compliance team, I would use detailed SAC to prove to authorities that the company has complied to a certain standard, legislation, etc., and show them evidence of my claim of compliance.
 
\noindent\textbf{US 3}  As a project manager, I would use SAC to make sure that a project is ready from a security point of view to be closed and shipped to production.
 
\noindent\textbf{US 4}  As a project manager, I would include SAC in my project plan. I would make sure the project has the needed resources and time for creating the case (argumentation, evidence collection, etc.).
 
\noindent\textbf{US 5}  As a project would use SAC to monitor the progress of my project when it comes to fulfillment of security requirements.
 
\noindent\textbf{US 6}  As a product owner, I would use SAC to make an assessment of the quality of my product from a security perspective, and make a roadmap for future security development.
 
\noindent\textbf{US 7}  As a product owner, responsible for handling threats and vulnerabilities, I would use SAC to evaluate the effect of new threats and vulnerabilities, and evaluate whether a change is needed to the product.
 
\noindent\textbf{US 8}  As a member of the purchase team, I would include SAC as a part of the contracts made with suppliers, in order to have evidence of the fulfillment of security requirements at delivery time, and to track progress during development time.
 
\noindent\textbf{US 9}  As an action owner, I would use detailed and visual SAC to communicate with the risk owner, and decide how to update the product security in the right way (to know what to do)
 
\noindent\textbf{US 10}  As a system leader, I would use SAC to make an assessment of the quality of my system from a security perspective, and make a roadmap for future security development (same as US6, but on wider scope).
 
\noindent\textbf{US 11}  As a software developer, I would use SAC from previous similar projects as a guideline for  secure development practices.
 
\noindent\textbf{US 12}  As a legal risk owner, I would use SAC in court if a legal case is raised against the company for security related issues. I would use the SAC to prove that sufficient preventive actions were taken.
 
\noindent\textbf{US 13}  As a member of the corporate communication team, I would use SAC as a reference to answer security related questions.

%%%%%%%%%%%%%%%%%%%%%%%%%%%%%%%%%%%%%%%%%%%%%%%%%%%%%%

\subsection{Prioritisation of Scenarios and In-depth Interviews}

\begin{figure}
	\centering
\begin{tikzpicture}
\begin{axis}[ 
xbar, xmin=0,
xlabel={Score},
symbolic y coords={
{US13},{US11},{US5},{US9},{US4},{US10},{US7},{US1},{US8},{US3},{US12},{US6},{US2}
},
ytick=data,
nodes near coords, 
nodes near coords align={horizontal},
ytick=data,
]
\addplot[fill=gray] coordinates { 
(33,{US2}) 
(23,{US6})
(21,{US12})
(19,{US3})
(18,{US8})
(11,{US1})
(9,{US7})
(6,{US10})
(5,{US4})
(4,{US9})
(1,{US5}) 
(0,{US11}) 
(0,{US13}) 
 };
\end{axis}
\end{tikzpicture}
\caption{Prioritized usage scenarios} \label{fig:top5}
\end{figure}
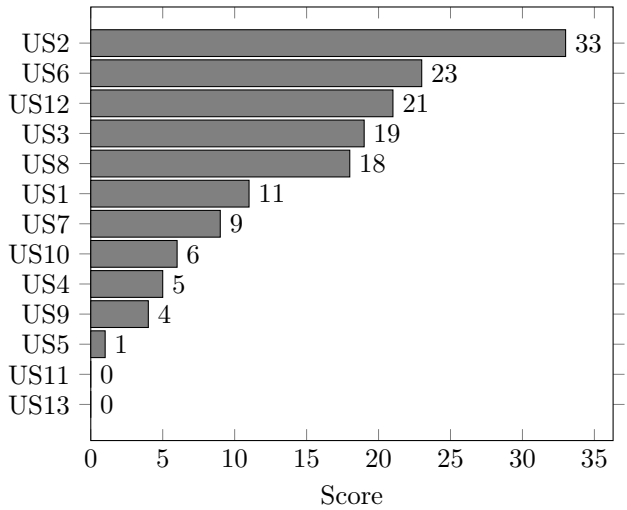

As mentioned in the methodology, after the workshop we sent the scenarios to experts and asked them to prioritize them based on the value the scenarios provide to the company. The result of the prioritization task is shown in Table \ref{fig:top5}. For the top five scenarios, we identified the key stakeholders for those scenarios and conducted in-depth interviews.

\paragraph{Interview on US3 and US6 -- Product delivery and process improvement}
The interview was conducted with two interviewees, to tackle two of the usage scenarios, as explained in the methodology~\ref{subsec:methodology}.
The interviewees were one solution train engineer and one solution manager (participants 19 and 20 in Table \ref{tbl:participants}).
\begin{aquote}{Participant 19}
[SAC provides] an opportunity for the company to build a reputation of building secure vehicles.
\end{aquote}
The interviewees said that today, the security assurance in projects is mostly based on experience, and is done by providing evidence such as test reports for some claims. These claims are derived from projects' requirements are not put together in a structured way. The interviewees also stressed that what is done today is simply not sufficient given the rapid evolution of connectivity in the vehicles. Historically there has been \emph{``a trust in the physical shell (the cab) around the items in the vehicle.''}, but today these items are connected to the outside world, and can be a target for cyber attacks.
 
As per the interviewees, having SACs as a part of scoping projects, setting the milestones and deliverables, and connecting them to the development process, would be of great value to security assurance from a project management perspective. Additionally from a product point of view, having a holistic SAC which is updated by the related projects would give the product owner an understanding of how secure the product is at different time-points, e.g., after integrating changes from different projects.

Both interviewees agreed that SACs \textbf{should be built on a product level rather than a project level}. SACs should be integrated and built within projects, but only to contribute towards the product's SAC. A product, however, may be big and complex, and may include items that are more interesting from a security perspective than others, e.g., Electrical Control Units (ECU) that are connected to the outside world in a vehicle. In this case there needs to be a severity assessment to focus on the right items.

The interviewees think that the SAC work can be integrated into the project manager's and product owner's work by capturing it in projects' requirements. They emphasized that it is important to work on quality, e.g., security as a \textbf{part of the development process, and not as a separate activity}. The workload in the beginning will be high, but with time, there will be patterns that can be reused to build SAC with less effort.

When it comes to the challenges, the interviewees consider handling the complexity of the product (vehicle in this case), and finding right competences to be the main ones. Other challenges include securing a buy-in to work with SAC all the way from upper management to development teams.

\paragraph{Interview on US2 -- Compliance} 
The interview was conducted with a functional safety assessor who has a wide range of experience in compliance (participant 21 in Table~\ref{tbl:participants}).
\begin{aquote}{Participant 21}
The security case can serve as an umbrella document for all the analysis documents.
It can be used by an assessor to find the right documents in order to assess compliance to a standard.
\end{aquote}
In the current situation, the compliance team is only concerned with safety matters, as per the interviewee. The only involvement with security issues is when there is a breach which affects the product's safety. The expectation is that SAC can \textbf{serve as an umbrella document for all the analysis documents}. Hence it can be used by an assessor to find related document to assess compliance to a certain standard.

The SAC should be \textbf{created on a whole vehicle level}, as per the interviewee. It needs, however, to be \textbf{dividable in manageable pieces}, i.e., pieces that can be managed by individuals, which would be responsible for the corresponding part of the SAC. For compliance purposes, the authorities look at system level, hence, for every project that includes changes in the system, it is important to conduct an \textbf{impact analysis} to identify affected artefacts, and update the SAC accordingly. 

How working with SAC can be integrated within the way of working of the compliance team depends much on how SACs are implemented. If they are integrated within the projects (as they should according to the interviewee), then the compliance team would have the responsibility of following them up throughout the project, as well as making sure that they are complete after the verification phase. 

The interviewee considers the main challenge to be finding resources and competences to carry out the work related to creating, maintaining, assessing and supporting SACs.

\paragraph{Interview on US8 -- Suppliers}
The interview was conducted with a component owner in Company B, who is experienced in purchasing and working with suppliers (ID 22 in Table \ref{tbl:participants}).
\begin{aquote}{Participant 22}
[Working with SAC provides] An opportunity to catch up with suppliers, which in many cases have come further in thinking about security than the company.
\end{aquote}
Today, security assurance when working with suppliers is about making sure of the fulfillment of security requirement, and running test cases on a sample of the requirements. However, \emph{``there is an uncertainty to a large extent that the received software is secure''}, as per the interviewee. On the other hand, safety critical functions, e.g., breaking, is handled differently. The suppliers are usually asked to show how requirements are broken down and implemented during regular review sessions. In some cases, suppliers are asked to provide safety cases, which are used together with the internal safety cases to make sure that the claims align. The interviewee expects that SAC can be used in the same manner as safety cases. Additionally, they can be used to \textbf{communicate} regarding security both with suppliers and internally, and as a \textbf{supporting artifact for creating security requirements}.

Regarding the granularity of the cases, the interviewee distinguishes between two types of SAC: the ones created at the supplier's end, and the ones created by the company.
On the supplier's side the cases should be on the ECU level, followed by a threat-based level. Whereas the company-owned SACs should be on the complete vehicle level, and broken down to ECU level, which is contributed by the suppliers. An important aspect mentioned by the interviewee is the weighting of the claims based on severity. This is to be able to prioritize the claims during the follow up sessions with the suppliers, and when testing the implementation.

Integrating SAC in the current way of working with suppliers would increase the workload, but there are no obvious conflicts, according to the interviewee. However, there is a \textbf{need for tools} to store, extract, and compare SACs. Additionally, a version handling tool is also required to keep track of the SACs and their changes. The interviewee also mentioned the need to \textbf{use an exchangeable format} when building the cases, on both ends (supplier and company), in order to compare and integrate them.

A challenge is to find and provide practical training on SAC in the industrial context. Another challenge mentioned by the interviewee is finding resources with the right competences to carry out the SAC related work. The interviewee emphasized that based on the experience from safety, even when there is education about the cases, it was much more complicated when actual work was done.

\paragraph{Interview on US12 -- Legal}
The interview was conducted with a senior legal counsel at Company B (participant 23 in Table~\ref{tbl:participants}). 
\begin{aquote}{Participant 23}
An evidence based structured approach to argue about security would definitely be used as evidence in court.
\end{aquote}
In the current situation, the company has not had any legal case for security related issues, but there have been functional-safety related cases claiming that feature malfunctioning have caused accidents. 
Current evidence used in court for these kinds of legal cases are of two types: \emph{(i)} usage of technology according to an acceptable standard; \emph{(ii)} implementations of the used standards and technologies are correctly done (this should be certified by a third party assessor). Hence, if the ISO 21434 \cite{iso21434} becomes an industry standards, then it can be used as an evidence in court.  However, it is very important to assure the quality of the case when it comes to completeness in the argumentation and evidence. The SAC used as an evidence will be available to the opponents, and it could be exploited to find holes and error to be used against the company.

The granularity needed for the SAC depends pretty much on the legal case according to the interviewee. However, there is a need to create \textbf{SACs for complete vehicles, and views that could be broken down}. This is to avoid cases where a legal case against the company involves a composition of systems (end-to-end function), and SACs are created for a subset of these systems. This would be a weakness if they are used as evidence.

The creation of SACs should, from a company perspective, be \textbf{during development} to assure security, according to the interviewee. However, from a legal perspective, SACs can be created once a legal case is filed, but that could lead to questions asking why the SACs were not created during development. Additionally, it would increase the probability of the risk of having insufficient evidence, as it would be harder to locate and assign them. Moreover, from a liability perspective, it is much better to create the SAC proactively.

As per the interviewee, if SACs are to be used as evidence, the relevant stakeholders in the company will be reached out and asked to provide the SACs when needed. Then the legal responsible will have meetings with the stakeholders to understand it. This means that the ownership of the SACs would not be the responsibility of the legal responsible, even if it is created specially for a legal case. The legal responsible would be a user of the SAC.

At the end, the interviewee stressed that introducing a structured way of security assurance \emph{''can lead to creating better systems which can protect the company from issues from regulators and third parties.''}. However, there has to be a buy-in on different levels in the company in order to do this in a correct way.

\section{Threats to validity}
\label{sec:threats}

In terms of \emph{external validity}, we are aware that the general validity of our results could be limited to the companies involved in the study. Also, the companies are from the same country. Therefore, the results might not directly translate to companies with a different culture.
However, the involved companies are of high profile, quite large and compete at the international level. Therefore, they are able to provide a quite broad perspective on the entire automotive industry.
In any case, the results presented in this paper are an important first important step towards a larger survey study involving more companies and professionals, internationally.

In terms of \emph{internal validity} we consider the following aspects. First, the selection of the standards and regulations investigated in RQ1 could have been incomplete. However, we are confident we are adressing the most important documents, especially for the mentioned markets (EU, China, US). 
Second, in the prioritization of the scenarios of RQ2, there is a risk that the selection of the top scenarios was biased by present market pressure towards compliance to the upcoming standards.
Third, elements of bias could be have been introduced via the selection of the participants. 
%For instance, some participants in the workshop were product development people with limited knowledge of SAC. Hence, there was a risk that their expectations were based on theoretical knowledge more than in-the field experience.
%This situation is natural, as we were interested in gathering a cross-sectional perspective on SAC, i.e., we wanted to go beyond the views expressed in the pre-study by a relatively small group of security experts.
%We mitigated this limitation by having a session held by a safety expert at the beginning of the workshop to explain (safety) assurance cases from a practical point of view. Additionally, each group in the workshop had at least one member with hands-on experience working with (safety) assurance cases. 
The same limitation applies to the participants of the interviews, as the selection of participants was based on expertise and availability (convenience sampling). 
%However, in this part of the study, we considered the knowledge of the business around the scenarios to be the most important factor. 
All in all, we have a balance mix of participants with different types of expertise: security, product development, business, and legal. This provides us with enough confidence that the results are representative of the expectations and needs across the studied companies. 
%Forth, and finally, for participants that took part to multiple phases of the study, especially the workshop-prioritization ones, there could have been a maturation effect risking a biased view.

\section{Discussion}
\label{sec:discussion}

As this study contains different parts, we think it might be useful to the reader to if we illustrated how the different results relate to each other. 
Further, we translate the results into a concrete set of recommendations for the OEMs.

\subsection{Mapping of Results}

\begin{table*}
\caption{Traceability of the results in this study}
\label{tbl:coverage}
\small
%\begin{tabular}{lll}
\begin{tabular}{@{}P{3.2cm}P{2.9cm}P{4.9cm}@{}}
\toprule
         Internal needs: pre-study & Internal needs: Usage scenario & External needs \\ 
\toprule
  Secure product argumentation & US3, US5, US6, US7, US9, US10, US11 
                               & \cite{iso21434},
                                 \cite{J3061}    \\
                                 \midrule
  Supporting evidence          & US2, US8, US11, US12 
                               & \cite{iso21434},
                                 \cite{J3061}  \\\midrule
  Legal compliance             & US2, US8 
                               & \cite{gdpr},
                                 \cite{SELFDRIVE}, 
                                 \cite{SPYCar}, 
                                 \cite{CSL},
                                 \cite{CCPA}, 
                                 \cite{UNECE-Reg},  
                                 \cite{UNECE-OTA}    \\\midrule
  Supply chain                 & US2, US3, US5, US8 
                               & \cite{gdpr},
                                 \cite{iso21434},
                                 \cite{ADS2},
                                 \cite{J3061}, 
                                 \cite{GBT35273}, 
                                 \cite{UNECE-Reg},
                                 \cite{UNECE-OTA},
                                 \cite{AV3}   \\\midrule
  Standard conformance         & US6, US8, US10, US11 
                               & \cite{ICV},
                                 \cite{iso21434},
                                 \cite{ADS2},
                                 \cite{J3061},
                                 \cite{GBT35273},
                                 \cite{AV3} \\\midrule
  Process harmonization        & US4, US5  &     \\\midrule
  ---						   & US1, US13 & \\
\bottomrule
\end{tabular}
\end{table*}

Table~\ref{tbl:coverage} shows a mapping of the results we got from the different parts of this study: the internal needs from the pre-study, the usage scenarios, and the external needs. 
As an example of the mapping, using SAC as \emph{Supporting evidence} is identified as an internal need in the pre-study. It relates to multiple usage scenarios, e.g., \emph{US12}, where SAC can be used as a piece of evidence in case of a legal suit. It also relates to the ISO/SAE 21434 DIS~\cite{iso21434} standard and the SAE J3061~\cite{J3061} best practice which explicitly require SAC with evidence on secure design and implementation.

As shown in the table, the results are not heterogeneous, but rather align to a large extent. 
Every internal needs identified by the experts in the pre-study is linked to at least one of the usage scenarios, as well as one external driver, with the exception of the \emph{Process harmonization} need.
The table shows how the scenarios proved to be useful tool to obtain a deeper view into the internal needs and in detailing the high-level needs identified in the pre-study.
On the other hand, the table also show that the scenarios go beyond the high-level needs identified by the security leaders, by broadening the scope of the analysis.
In particular, two additional scenarios (\emph{US1} and \emph{US13}) suggest using an abstracted level of SAC to communicate and answer security related questions both from inside the company and from potential customers.

\subsection{Recommendations}

As a summary of the findings of this study, we provide our recommendations for companies (particularly, OEMs) that are starting to work with security assurance cases. 
These recommendations are not complete solutions, but rather steps towards establishing a ground to integrate SAC with the companies' way of working, and to help different stakeholders make use of SAC.

\emph{Standards and Regulations -- Cover both process and product.}
Several of the security-related standards/regulations contain both requirements on processes and the product. The processes include how to develop the product in a secure manner as well as keeping the product secure after its release. In some cases the product requirements even suggests what kind of measure that should be considered. Since both standards and regulations put requirements on audits and assessments of processes and products, including certification of security processes, the SAC can function as the tool for showing both compliance and conformance to those requirements.
Privacy related standards/regulations foremost imply that processes shall be in place in order to handle private data in a secure manner. This also means having the appropriate measure in place in order to accomplish this. The conclusion is thus that the SAC shall contain arguments and evidence both that the processes are sufficient and that the implementations of mechanisms that handle private data are secure. Some of the regulations mentioned in \ref{tbl:external} that foremost requires processes for handling of data, such as \cite{gdpr},  \cite{CCPA} and \cite{GBT35273} can implicitly be considered as product requirements since there will also be a need for implementation of security mechanisms that for example encrypt data during transfer. 

\emph{Granularity of SAC -- Whole product over sub-projects.} 
In industries producing complex products, e.g., automotive, it is common that the products are organized in multiple projects. Additionally, the changes to these products are also done using projects (commonly called delta projects). In this case, SACs should be created on a product level rather than a project level to fulfill the usage scenarios identified in our study. When a SAC is created within a project, it has to be integrated later with the product's SAC. They should, however, be built in a way which allows different stakeholders to have different views corresponding to different abstraction levels.

\emph{Creation of SAC -- No retro-fitting and no silos.} 
It is possible to build SAC for existing products, but going forward, it is important to embed the work on SAC into the development process at the organization.
Security cases can provide real value to an organization if it is not just considered as a ``check-in-the-box'' activity, or, worse, an overhead.
Security cases can be used as a methodology for guiding the work of cybersecurity and communicating its importance across the teams.
The work on Security cases should include a large part of the organization, with different teams working on different parts of the security case, e.g., teams working on the system, component, and functional levels. 
This requires to have clear collaboration interfaces to avoid issues such as inconsistencies, and conflicts. 
%These interfaces could be defined for example in the cybersecurity planning. 
Moreover, each case should have an owner (even SACs that consist of multiple sub-cases), which should be made explicit to all teams working on the cases. This owner would drive the work on the SAC and be responsible for its maintenance and quality control.
However, in order for this to work, there must be a buy-in on different level of the organization, starting from top management, and continuing all the way through the organization to the development teams. Additionally, the importance of tool support and automation should not be underestimated.

%The interviews confirmed the usage scenarios to a large extent. It was obvious that the cases could be used in multiple different ways by different stakeholders to achieve different goals. The cases can be used to prove that security has been considered and covered to external parties (US2 for compliance, and US12 for legal). They can also be used to show internal parties that security has been considered and provide evidence that threats and risks and handled (US6 and US3). It can also be used as a document to follow up work with external parties (US8). 

\emph{Quality of SAC -- Actively assess completeness and confidence.} 
Security assurance cases are going to serve multiple purposes within the organization (see the diversity in the usage scenarios) and they can be used in contexts that have different levels of criticality (e.g., legal case vs process improvement). Therefore, it must be clear what the quality level of each SAC is, so that they are not used in the wrong context.  
We recommend to introduce measures to assess the quality of the SACs. These may include, for example, the \emph{completeness} of the argumentation, and the level of \emph{confidence} in the evidence. This is emphasized by the security experts in the pre-study (second bullet in Section \ref{sec:pre-study}), as well as all the interviewees. However, in some of our identified usage scenarios, this becomes even more important. For example when SAC are used as evidence in court, they would be available to the oppositions as well, meaning that any flaw could back-fire against the company. Hence, it is very important to be able to assess the quality of a SAC before using it in court.

\emph{SAC and suppliers -- A common language is key to smooth collaboration.} When it comes to working with suppliers, the SAC should be built using an exchangeable format. This is to enable the SAC created by the suppliers to be integrated with the SAC of the corresponding product. Another important aspect is to add an assessment of the severity level of the claims. This is to be able to followup on the most severe items, when it is not possible to do a complete followup. 

\emph{SAC and suppliers -- Plan for shared ownership.}
The suppliers might require to keep parts of the SAC private (e.g., some evidence). In this case, it is important to have a mechanism to keep ensuring the overall quality of the SAC, e.g., by introducing a black-box with meta-information such as the validity of the items behind the box. Additionally, the ownership of the whole case has to be considered, as the complete SAC would not be in the hands of a single stakeholder. For instance, using a SAC in a legal case would require a disclosure process in order to compile a SAC from multiple sources and multiple owners.

\emph{Miscellanea -- With opportunities come challenges.}
Working with SAC is not trivial and comes with many challenges. Traceability and change analysis were considered main challenges by the majority of the participants. Additionally, finding the right competences to carry out the SAC-related work, role identification and description, and acquiring the right tools and integrating them in the organizations tool chain were also considered major challenges.

%\todo{There is a lot of same conclusions in RQ2 from both the pre-study and the workshop -> shown in the table. Maybe this could be somewhat more highlighted? Otherwise I (Alex) think the discussion section covers the RQs quite good. There are however, some text that is currently commented out on the RQ1 topic in regards to CS reg./Standard not sure if I did that or someone else..?}

\section{Conclusion}\label{sec:end}

% Safe and Secure Automotive Over-The-Air
%Updates
%Thomas Chowdhury1, Eric Lesiuta1, Kerianne Rikley1,
%Chung-Wei Lin2, Eunsuk Kang2, BaekGyu Kim2, Shinichi Shiraishi2,
%Mark Lawford1, and Alan Wassyng1

In this study, we have analyzed the requirements around the use of security assurance cases in the automotive domain. In particular, we have listed the constraints coming from standards \& regulations and have identified the internal needs of OEMs.
We have concluded this work by translating the results in a number of pragmatic recommendations for OEMs.
These are valuable and necessary contributions for the overall furthering of the automotive industry and for a more effective and secure development of future road vehicles.

In future work, we plan on extending this work with a survey including a larger group of automotive companies and automotive security experts.
Additionally, we are addressing the requirements identified in this paper by means of a systematic methodology to create security assurance cases for the automotive industry.

\chapter{Paper B}
\label{chap_paper_b}
\thispagestyle{empty}
\subsection*{Security Assurance Cases -- State of the Art of an Emerging Approach}
\subsubsection*{M. Mohamad, J.P. Steghöfer, R. Scandariato}
\subsubsection*{{\em Empirical Software Engineering Journal} 26, 70 (2021).} \cite{paperB}

\newpage
\thispagestyle{empty}
\mbox{}
\newpage
\addtocounter{page}{-2}
\newpage
\section*{Abstract}
%Context
Security Assurance Cases (SAC) are a form of structured argumentation used to reason about the security properties of a system. After the successful adoption of assurance cases for safety, SAC are getting significant traction in recent years, especially in safety-critical industries (e.g., automotive), where there is an increasing pressure to be compliant with several security standards and regulations.
Accordingly, research in the field of SAC has flourished in the past decade, with different approaches being investigated.
%what we did
In an effort to systematize this active field of research, we conducted a systematic literature review (SLR) of the existing academic studies on SAC. 
%how we did it
Our review resulted in an in-depth analysis and comparison of 51 papers.
%results
Our results indicate that, while there are numerous papers discussing the importance of SAC and their usage scenarios, the literature is still immature with respect to concrete support for practitioners on how to build and maintain a SAC. More importantly, even though some methodologies are available, their validation and tool support is still lacking.\\
\linebreak
\textbf{Keywords: } security; assurance cases; systematic literature review

\newpage

%As an alternative to including the tex files, you can just paste the entire paper in here. Up to you!

\section{Introduction}
\label{sec:intro}

A security assurance case (a.k.a.\ security case, or SAC) is a structured set of arguments that are supported by material evidence and can be used to reason about the security posture of a software system.
SACs represent an emerging trend in the secure development of critical systems, especially in domains like automotive and healthcare.
The adoption of security cases in these industries is compelled by the recent introduction of standards and legislation. 
For instance, the upcoming standard ISO/SAE~21434 on ``Road Vehicles -- Cybersecurity Engineering'' includes the explicit requirement to create `cybersecurity cases' to show that a vehicle's computing infrastructure is secure.

The creation of a security case, however, is far from trivial, especially for large organizations with complex product development structures.
For instance, some technical choices about the security case might require a change of the development process.
The security case shown in Figure~\ref{fig:sacExample} (and discussed in Section~\ref{sec:relWork}), e.g., requires that a thorough threat analysis is conducted throughout the product structure and at different stages of the development. 
If this analysis is not yet created during development, either a thorough re-organization of the way of working is necessary or the security case should have been structured in a different way. 
Also, the construction of a security case often requires the collaboration of several stakeholders in the organization, e.g., to ensure that all the necessary evidence is collected from the software and process artifacts.

Companies are thus facing the conundrum of making both urgent and challenging decisions concerning the adoption of SACs.
In order to facilitate such an endeavor, this paper presents a systematic literature review (SLR) of research papers on security cases.
It summarises academic research which has published a relatively large number of papers on the topic in recent years and therefore provides practitioners an overview of the state of the art.
To the best of our knowledge, this is the first study of this kind in this field.
This SLR collects most relevant resources (51 papers) and presents their analysis according to a rich set of attributes like, the types of argumentation structures that are proposed in the literature (threat identification -- used in Figure~\ref{fig:sacExample} -- being one option), the maturity of the existing approaches, the ease of adoption, the availability of tool support, and so on.

Ultimately, this paper presents a reading guide geared towards practitioners.
To this aim, we have created a workflow describing the suggested activities that are involved in the adoption of security cases.
Each stage of the workflow is annotated with a suggested reading list, which refers to the papers included in this SLR. 
We remark that the SLR also represents a useful tool for academics to identify research gaps and opportunities, which are discussed in this paper as well.

The rest of the paper is structured as follows. 
In Section~\ref{sec:relWork} we provide some background on assurance cases and discuss the related work.
In Section~\ref{sec:resmethod} we describe the research questions and the methodology of this study.
In Section~\ref{sec:res} we list the papers included in this study and present the results of the analysis.
In Section~\ref{sec:wf} we present a workflow for SAC creation and a reading guide for practitioners who want to adopt them.
In Section~\ref{sec:disc} we further discuss the results and the lessons learnt from them.
In Section~\ref{sec:valTh} we discuss the threats to validity of this study.
Finally, Section \ref{sec:con} presents the concluding remarks.
\section{Background and Related Work}
\label{sec:relWork}
In this section, we first present background information about SACs, their main elements and their application areas as well as a simple example of a SAC. Afterwards, we discuss the related work.
\subsection{Assurance cases}

\begin{figure}[tb]
\begin{center}
\includegraphics[width=\columnwidth]{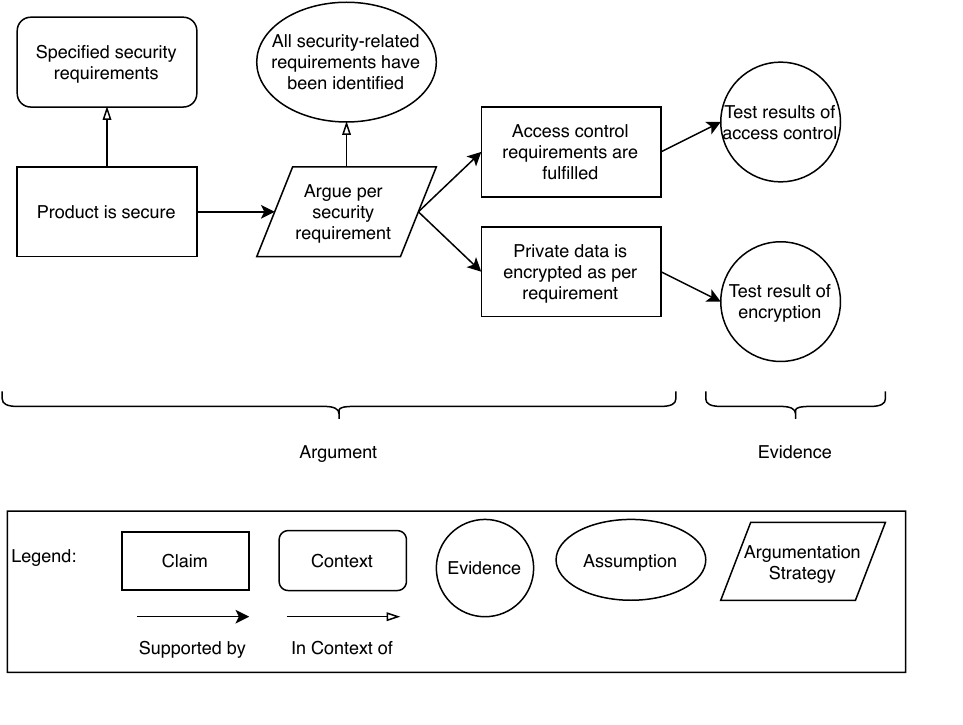}
\end{center}
\caption{An example of a SAC}
\label{fig:sacExample}
\end{figure}

Assurance cases are defined by the GSN standard~\cite{GSN_standard} as \emph{``A reasoned and compelling argument, supported by a body of evidence, that a system, service or organisation will operate as intended for a defined application in a defined environment.''} 
Assurance cases can be documented in either textual or graphical forms.
Figure~\ref{fig:sacExample} depicts a very simple example of an assurance case and its two main parts, i.e., the argument and the evidence.
The case in the figure follows the GSN notation~\cite{gsn}, and consists of the following nodes: claim (also called goal), context, strategy, assumption (also called justification), and evidence (also called solution). 
At the top of the case, there is usually a high-level \emph{claim}, which is broken down to sub-claims based on certain strategies.
The claims specify the goals we want to assure in the case, e.g., that a certain system is secure.
An example of a \emph{strategy} is to break down a claim based on different security attributes.
Claims are broken down iteratively until we reach a point where \emph{evidence} can be assigned to justify the claims/sub-claims. 
Examples of evidence are test results, monitoring reports, and code review reports. 
The \emph{assumptions} made while applying the strategies, e.g., that all relevant threats have been identified, are made explicit using the assumption nodes. 
Finally, the \emph{context} of the claims is also explicitly defined in the context nodes. An example of a context is the definition of an acceptably secure system. 
%All of the claims, sub-claims, context, strategies, and assumptions build the argument part of the assurance case, whereas evidence (solutions) to the claims makes up the evidence part.

Assurance cases have been widely used for safety-critical systems in multiple domains \cite{bloomfield2010}. An example is the automotive industry, where safety cases have been used for demonstrating compliance with the functional safety standard ISO 26262~\cite{palin2011,birch2013,iso26262}. 
However, there is an increasing interest in using these cases for security as well. 
For instance, the upcoming automotive standard ISO 21434~\cite{iso21434} explicitly requires the creation of cyber-security arguments. 
SACs are a special type of assurance cases where the claims are about the security of the system in question, and the body of evidence justifies the security claims. 

\subsection{Related work}
To the best of our knowledge this study is the first systematic literature review on SACs. However, there have been studies covering the literature on safety assurance cases.

Nair et al. \cite{nair2013} conducted a systematic literature review to classify artefacts which can be considered as safety evidence. The researchers contributed with a taxonomy of the evidence, and listed the most frequent evidence types referred to in literature. The results of the study show that the structure of safety evidence is mostly induced by the argumentation and that the assessment of the evidence is done in a qualitative manner in the majority of cases in contrast to quantitative assessment. Finally, the researchers list eight challenges related to safety evidence. The creation of safety cases was the second most mentioned one in literature according to the study. 
In our study, we focus on security rather than safety cases. We also review approaches for creating complete assurance cases, meaning that we look into both the argumentation and the evidence parts, in contrast to the study of Nair et al.~\cite{nair2013} which focuses on the evidence part only.\\

Maksimov et al.~cite{maksimov2018} contributed with a systematic literature review of assurance case tools, and an extended study which focuses on assurance case assessment techniques~\cite{maksimov2}. The researchers list 37 tools that have been developed in the past two decades and an analysis of their functionalities. The study also includes an evaluation of the reported tools on multiple aspects, such as creation support, maintenance, assessment, and reporting. In our study, we also review supporting tools for the creation of assurance cases, but we focus on the reported tools specifically for SAC.

Gade et al.~\cite{gade2015} conducted a literature review of assurance-driven software design. The researchers provide a review of 15 research papers
with an explanation of the techniques and methodologies each of these papers provide with regards to assurance-driven software design. This work intersects with our work in that assurance-driven software design can be used as a methodology or approach for creating assurance cases. However, unlike Gade et al.~our study focuses on SAC, and is done in a systematic way.

Ankrum et al.~\cite{9_ankum2005} created a non-deterministic workflow for developing a structured assurance case. However, the proposed flow does not include anything related to tools or patterns usage. It does not consider the preliminary stage of considering a SAC either.
Cyra and Gorski~\cite{1_cyra2007} present the life-cycle, derivation procedure, and application process for a trust case template. All these artifacts, however, build on the argumentation strategy being derived from a standard, which is not always the case.

\section{Research Method}
\label{sec:resmethod}

We conducted a systematic literature review following the guidelines introduced by Kitchenham et al.~\cite{kh2007guidelines}.

%%%%%%%%%%%%%%%%%%%%%%%%%%%%%%%%%%%%%%%%%%%%%%%%%%%%%%%%%%%%
%%%%%%%%%%%%%%%%%%%%%%%%%%%%%%%%%%%%%%%%%%%%%%%%%%%%%%%%%%%%
%%%%%%%%%%%%%%%%%%%%%%%%%%%%%%%%%%%%%%%%%%%%%%%%%%%%%%%%%%%%

\subsection{Research questions and assessment criteria}
\label{sec:resmethod:rq}

This study aims at answering the following four research questions.

[\textbf{RQ1}] \textbf{RATIONALE --- In the literature, what rationale is provided  to support the adoption of SAC?}

In particular, we are interested in whether there are statements that go beyond the intuitive rationale of using SAC ``for security assurance''. 
For instance, our initial research \cite{industrialNeeds} indicated that compliance with security standards and regulations is also an important driver.
As shown in Table~\ref{method_charrq1}, to answer this research question we analyze the surveyed papers and extract two characteristics: 
\begin{itemize}
\item 
\emph{Motivation}, i.e., the reason for using SACs as stated by the researchers. We used two criteria for determining whether a certain study provides a motivation for using SAC. That is, the wording has to be explicit (i.e., there must be a reference to usage or advantage) and specific (i.e., providing some details). 
\item 
\emph{Usage scenario}, i.e., scenarios in which SAC could be used to achieve additional goals, next to security assurance. We used the same criteria (explicit and specific mention) used for the motivation.
\end{itemize}

\begin{table}[tb]
\centering
\caption{Assessment criteria for RQ1 (rationale)}
\label{method_charrq1}
\begin{tabular}{ll}
\toprule
\textbf{RQ1 criteria} &  \textbf{Value}  \\
\toprule
Motivation     & E.g., compliance to standards, \\
               & ensuring the fulfillment of security requirements, \\ 
               & documenting security claims, \dots\\
              % & \dots \\ 
\midrule
Usage scenario & E.g., support for court case, \\ 
               & assess security level of product or service,\\
               & obtain certification, \dots\\
              % & \dots \\ 
\bottomrule
\end{tabular}
\end{table}

[\textbf{RQ2}] \textbf{CONSTRUCTION --- In the literature, which approaches are reported for the construction of SACs and which aspects do the approaches cover?} 

This question aims at inventorying the existing approaches for creating SAC, which is a challenging task for adopters. As shown in Table~\ref{method_charrq2}, we also assess the \emph{coverage} of the approach, i.e., whether it can be used for creating the argumentation, for collecting the evidence, or both. 
Finally, for each covered part of the SAC, we summarise the approach with respect to the suggested \emph{argumentation} strategy and the types of \emph{evidence} to be used in creating SACs. 

\begin{table}[tb]
\centering
\caption{Assessment criteria for RQ2 (construction)}
\label{method_charrq2}
\begin{tabular}{ll}
\toprule
\textbf{RQ2 criteria } & Values \\ 
\bottomrule
Coverage & Argumentation, Evidence,  Generic (i.e., both) \\
       %  & Evidence     \\ 
      %   & Generic (i.e., both) \\
\midrule
Argumentation & E.g., based on threat avoidance, \dots \\
(if covered)  &  \\
\midrule
Evidence     & E.g., collect test results, \dots \\
(if covered) &  \\
\bottomrule
\end{tabular}
\end{table}

[\textbf{RQ3}] \textbf{SUPPORT --- In the literature, what practical support is offered to facilitate the adoption of SAC?}

The purpose of this question is to understand the practicalities of creating and working with SAC.
With reference to Table~\ref{method_charrq3}, first we study the approaches and identify the conditions (i.e., \emph{prerequisites}) that have to be met in order for the outcome of the paper to be applicable.
Second, we check whether the papers propose libraries of \emph{patterns} or templatized SAC, as these are extremely useful for non-expert adopters.
Third, we analyze the \emph{tool support}. We check whether the paper suggests the usage of a tool for any of the activities related to SAC. In case it does, we extract the description of that tool, and whether it was created by the researchers or if it is a third party tool used in the paper.
The last characteristic in this research question is the \emph{notation} used to represent the SAC. The most common ones are GSN~\cite{gsn}, and CAE~\cite{CAE}, but there are other notations such as plain text.

\begin{table}[tb]
\centering
\caption{Assessment criteria for RQ3 (support)}
\label{method_charrq3}
\begin{tabular}{lll}
\toprule
\textbf{RQ3 criteria } &  & Values \\

\bottomrule

Prerequisites &                 & E.g., threat modeling is performed, \dots \\
            %  &                 & \dots \\
\midrule
Patterns      &                 & E.g., a catalog or argumentation patterns \\
              &                 & \dots \\
\midrule
Tool support  & Tool mentioned  & Yes / No        \\
              & Type of tool    & Created  / Used \\
\midrule
Notation      &                 & Graphical (GSN, CAE), Textual, \dots \\
            %  &                 & Graphical (CAE)  \\
            %  &                 & Textual          \\
            %  &                 & \dots            \\
\bottomrule
\end{tabular}
\end{table}

[\textbf{RQ4}] \textbf{VALIDATION --- In the literature, what evidence is provided concerning the validity of the reported approaches?} 

Our interest is to understand how the approaches and usage scenarios of SAC are validated (or supported by evidence). 
We aim at identifying: 
\begin{itemize}
\item
The \emph{type} of validation conducted in the study, e.g., case study, or experiment. Note that `case study' is a widely used term to refer to worked examples~\cite{caseStudies,runeson2009}. In this work, we consider a validation conducted in an industrial context to be a case study~\cite{yin2003}, and those done within a research context to be illustrations.  Experiments are studies in which independent variables are manipulated to test their effect on dependent variables~\cite{caseStudies}.
\item
The \emph{domain} (i.e., application area) in which the validation is conducted.
\item
The \emph{source} of the data used for the validation, e.g., a research project or a commercial product.
\item
Whether or not a SAC is \emph{created} as part of the validation process. 
\item
In case a SAC is created, we look for its \emph{creators}. This characteristic has three possible values: academic authors, authors with industrial background, or third-party experts.
\item
The \emph{validators}, i.e., the parties that conducted the validation with values the same as for creators.
\end{itemize}

\begin{table}[tb]
\centering
\caption{Assessment criteria for RQ4 (validation)}
\label{method_charrq4}
\begin{tabular}{ll}
\toprule
%\multicolumn{2}{c}% \\
\textbf{RQ4 criteria} & Values \\
\bottomrule
Type & Illustration, Case study, Experiment, Other \\
    % & Case study \\
   %  & Experiment \\
  %   & Other \\
\midrule
Domain & Medical, Automotive, Software engineering, \dots \\
      % & Automotive \\
      % & Software engineering \\
      % & \dots \\
\midrule
Data source & Research project, Commercial product, \dots \\
           % & Commercial product \\
           % & \dots \\
\midrule
SAC created & Yes / No   \\
\midrule
Creators   & Academic authors, Industrial authors, third-party experts \\
           %& Industrial authors \\
           %& 3rd party experts \\
\midrule
Validators & Academic authors, Industrial authors, third-party experts \\
          % & Industrial authors \\
           %& 3rd party experts \\
\bottomrule
\end{tabular}
\end{table}

%%%%%%%%%%%%%%%%%%%%%%%%%%%%%%%%%%%%%%%%%%%%%%%%%%%%%%%%%%%%
%%%%%%%%%%%%%%%%%%%%%%%%%%%%%%%%%%%%%%%%%%%%%%%%%%%%%%%%%%%%
%%%%%%%%%%%%%%%%%%%%%%%%%%%%%%%%%%%%%%%%%%%%%%%%%%%%%%%%%%%%

\subsection{Performing the systematic review}

We performed a search for papers related to SAC by means of 3 scientific search engines: IEEE Xplore, ACM Digital Library, and Elsevier Scopus. 
Weselected these libraries, and did not include Google Scholar as, in our own prior experience which was confirmed by a preliminary search, the results from this search engine overlap with the results of the engines we mentioned. 

\subsubsection{Constructing the search string}

To maximize the chance of obtaining all relevant papers in the field, the search string used in the search engines must contain keywords that are commonly used in said papers.
Therefore, prior to constructing the search string, we familiarized ourselves with the specific terminology used by researchers in the field of SAC.
To do so, we conducted a manual search for papers related to SAC that were published in the past five years in the following venues: SAFECOMP, CCS, SecDev, ESSOS, ISSRE, ARES, S\&P, Asia CCS, and ESORICS.
The selection of the venues was based on their high visibility in the security domain.

Next, we created the search string for the selected libraries to identify papers that are potentially relevant for this study. 
In particular, we used two groups of keywords. 
The first group (line 1 below) is meant to scope the area of the study, while the second group (lines 2--4) included the terms referring to the parts of an assurance case. As a result, we formed the search string as follows:

\begin{lstlisting}[language=Python]
( security OR privacy OR trust ) AND 
( claim OR argument OR evidence 
  OR justification OR 'assurance case' 
  OR assurance )
\end{lstlisting}

As a quality check for our search string, we ensured that we would find three relevant, known studies \cite{8_finnegan2014,10_othmane2014,20_xu2017} with the search string. This was to make sure that our search string would return all three relevant studies, hence confirming its validity.
We ran the query in IEEE Xplore and confirmed that the papers were returned.

\subsubsection{Inclusion and exclusion criteria}

\begin{table}[tb]
\centering
\caption{Inclusion and exclusion criteria}
\label{method_inex}
\begin{tabular}{l}
\toprule
%\multicolumn{2}{c}% \\
\textbf{Inclusion criteria} \\
\bottomrule
1. Studies addressing the creation, management, or application of SAC. \\
2. Studies related to security/privacy/trust assurance. \\
3. Studies related to security/privacy/trust argumentation. \\
\toprule
\textbf{Exclusion criteria}\\ 
\toprule
1. Studies written in any language other than English. \\
2. Studies published before 2004. \\
3. Short papers (less than 3 pages). \\
4. Studies focusing on risk/threat/hazard detection. \\
5. Studies addressing risk/threat/hazard analysis. \\
6. Studies addressing cryptography. \\
7. Studies focusing on security assessment/evaluation. \\
8. Studies about (only) safety assurance. \\
\bottomrule
\end{tabular}
\end{table}

The inclusion and exclusion criteria are shown in Table~\ref{method_inex}. This list has been created and fine-tuned by means of a calibration exercise involving two authors. We have invested significant time in performing an initial search of papers (prior to the systematic search) and discussing what papers should be included / excluded and why. This calibration made the application of this criteria straightforward later on, when filtering the results of the systematic search (as discussed below).
The inclusion criteria are rather straightforward, considering the nature of this SLR.
Concerning the exclusion criteria, we have decided to only consider studies written in English language, as this is the common language among the authors of this SLR. 
Further, SAC have been the focus of research only in recent times (although assurance cases, in general, have been around for much longer) and the field is rapidly evolving. Hence, we restricted our SLR to the past 15 years to avoid outdated results.
We also excluded short papers, as answering our research questions requires studies with results rather than only ideas.
Finally, exclusion criteria 4--8 exclude studies that focus on topics that are marginally related to SAC but would not help us answer our research questions.

\subsubsection{Searching and filtering the results}

We executed the query on three libraries (IEEE Xplore, ACM Digital Library, and Scopus) in January 2019, and got the results shown in Table~\ref{method_res}.
In the case of Scopus, we limited the search to the domains of computer science and engineering. Also, because of the high number of returned results from Scopus, we decided to limit the included studies to the first 2000 after ordering the results based on relevance. We believe that the considered studies were sufficient, as the last 200 papers of the retained set from Scopus (i.e., papers 1801-2000) were all excluded when we applied the first filtering round (see below).

\begin{longtable}{lllll}
\label{method_res}\\
\caption{Number of included studies after each round of filtration}\\
\toprule
& & \multicolumn{3}{l}{\textbf{After filtering round}} \\
	        \textbf{Library} & \textbf{Papers} & \textbf{1st}  & \textbf{2nd} & \textbf{3rd} \\
\toprule
\endfirsthead
\multicolumn{5}{c}%
{\tablename\ \thetable\ -- \textit{Continued from previous page}} \\
\toprule
& & \multicolumn{3}{l}{\textbf{After filtering round}} \\
	        \textbf{Library} & \textbf{Papers} & \textbf{1st}  & \textbf{2nd} & \textbf{3rd} \\
\toprule
\endhead
\midrule \multicolumn{5}{r}{\textit{Continued on next page}} \\
\endfoot
\bottomrule
\endlastfoot
IEEE Xplore & 4513 & 118 & 23 & 22 \\
			ACM DL      & 1927 &  35 &  3 &  3 \\
			Scopus      & 2000 &  68 & 23 & 19 \\

			            &      &     &    &\textbf{+7} (snowballing) \\
		\bottomrule
			\textbf{Total}& \textbf{8440} &   &   &\textbf{51} \\
\end{longtable}

After the systematic search had been applied, one author, who has been working in industrial projects about SAC with multiple partners in multiple domains (automotive and medical), performed an initial filtering (based on the inclusion and exclusion criteria) and tracked their confidence (high, medium and low) with each included / excluded paper. For the cases of medium to low confidence we held a series of meetings after each filtration round, where the three authors jointly discussed whether such papers should be included / excluded.

In the first filtering round, we applied the inclusion and exclusion criteria to the titles and keywords of all results (8440 papers). As shown in Table~\ref{method_res} this round reduced the number of studies to 211 papers. 
In the second filtering round, we applied the inclusion and exclusion criteria\footnote{Except for exclusion criteria 1,2, and 8, which only needed to be applied once.} to the abstracts and conclusions of the 211 remaining studies. After this step, the number of studies was reduced to 49.
In the last filtering round, we fully read the remaining 49 papers, applied the inclusion and exclusion criteria on the whole text, and ended up with 43 included studies.

We also looked at the references mentioned by the included papers and performed  \emph{backward snowballing}~\cite{snowballing}. In this step, we did not restrict the search to only peer-reviewed studies in order to allow for potential gray literature to be included. This resulted in additional 7 papers (including 2 technical reports) being included in our review. We looked into the references of these 7 papers, but this did not result in the inclusion of additional papers and we terminated the snowballing.

Finally, the authors kept monitoring the literature on the topic of SAC after the search was performed. This led to the inclusion of one additional paper, which is accessible through Scopus.
In total, we thus included \textbf{51 studies}.

\subsection{Analysis of the included papers}
%\mazsays{Enough?}
Once the final list of included studies was ready, we started the analysis phase. This was done in an iterative manner, where one author would use the infrastructure provided in Tables~\ref{method_charrq1}, \ref{method_charrq2}, and and~\ref{method_charrq4} to prepare the analysis of a batch of papers (approximately 10 at a time). The outcome is then discussed in a group of the three authors as a means of quality control and calibration for the next batch.

\section{Results}
\label{sec:res}

In this section, we provide a descriptive analysis of the included papers in this SLR, and then present the results and answers to our four research questions.
%Venues.
%Domains of venues.

\subsection{Descriptive statistics}
\begin{figure}
	\centering
	\begin{tikzpicture}
		\begin{axis}[
		        width=\columnwidth,
                height=4cm,
				title={},
				xlabel={Year},
				ylabel={Number of publications},
				xmin=2003, xmax=2020,
				ymin=-2, ymax=12,
				xtick={2004,2006,2008,2010,2012,2014,2016,2018},
				ytick={0,3,6,9,12},
				legend pos=north west,
				ymajorgrids=true,
				grid style=dashed,
			/pgf/number format/.cd,
            use comma,
            1000 sep={}]
			\addplot[
				color=blue,
				mark=square,
			]
			coordinates {
				(2005,2)(2006,0)((2007,3)(2008,1)(2009,1)(2010,0)(2011,3)(2012,3)(2013,5)(2014,7)(2015,10)(2016,4)(2017,6)(2018,5)
			};
		\end{axis}
	\end{tikzpicture}
	\caption{Publication year of the included studies} \label{fig:M1}
\end{figure}
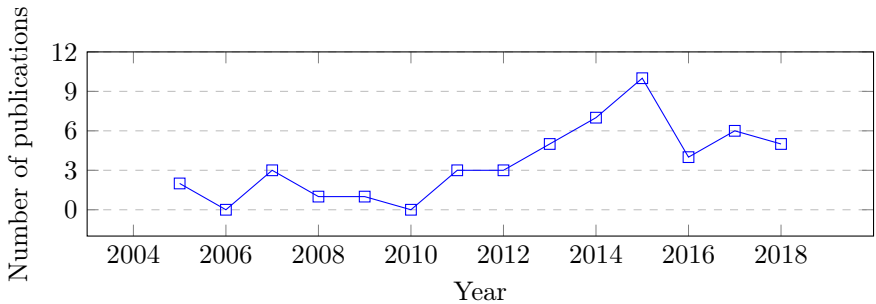

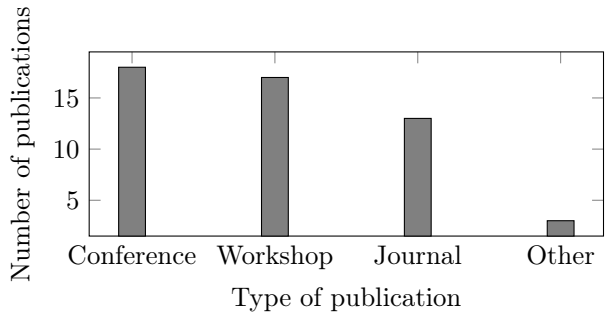
\begin{figure}
	\centering
\begin{tikzpicture}
        \begin{axis}[
            width=\columnwidth*0.7,
            height=4cm,
            symbolic x coords={Conference, Workshop, Journal, Other},
            xtick=data,
            xlabel={Type of publication},
			ylabel={Number of publications},
            grid style=dashed
          ]
            \addplot[ybar,fill=gray] coordinates {
                (Conference,   18)
                (Workshop,  17)
                (Journal,   13)
                (Other, 3)
            };
        \end{axis}
\end{tikzpicture}
\caption{Types of publication of the included studies } 
\label{fig:M2}
\end{figure}
Figure~\ref{fig:M1} shows the years when our 51 included studies were published. The graph shows a peak of 10 publications in 2015, which indicates an increase in interest in the research field compared to previous years, especially the time between 2005 and 2012 where the number of publications was three or less each year. We decided to exclude the studies from 2019 in Figure \ref{fig:M1}, as our search was conducted in that year and thus, results would necessarily be only partial. Including the results from that year would thys give a false indication of the trend compared to previous years.

Figure~\ref{fig:M2} shows the venues where the included studies were published. The graph shows that most of the publications were in conferences and workshops (18 and 17 respectively). 13 of the papers were published in journals, and three were technical reports.
%	Approach statistics. Cited? Where? How old?
We also looked into the authors of the selected papers to find the portion of the papers with at least one author from industry. We found that less than 25\% (12 papers)~\cite{2_cockram2007,3_goodger2012,14_netkachova2015,19_netkachova2016,20_xu2017,24_gacek2014,25_rodes2014,29_bloomfield2017,30_netkachova2014,37_gallo2015,39_cheah2018,47_ionita2017} included at least one author from industry.

\begin{figure}
\centering     %%% not \center
\subfigure[Ranking of conferences]{\begin{tikzpicture}
        \begin{axis}[
            width=0.5\columnwidth,
            height=5cm,
            symbolic x coords={A*,A,B,C,NA},
            xtick=data,
            xlabel={},
			ylabel={Number of publications},
            grid style=dashed
          ]
            \addplot[ybar,fill=gray] coordinates {
                (A*,   0)
                (A,  2)
                (B,   5)
                (C, 2)
                (NA, 9)
            };
        \end{axis}
\end{tikzpicture}}
\subfigure[Ranking of journals]{\begin{tikzpicture}
        \begin{axis}[
            width=0.5\columnwidth,
            height=5cm,
            symbolic x coords={A*,A,B,C},
            xtick=data,
            xlabel={},
			ylabel={Number of publications},
            grid style=dashed
          ]
            \addplot[ybar,fill=gray] coordinates {
                (A*,   2)
                (A,  1)
                (B,   7)
                (C, 3)
            };
        \end{axis}
\end{tikzpicture}}
\caption{Ranking of the venues of included journal papers according to the Core ranking portal}
 \label{fig:ranking}
\end{figure}
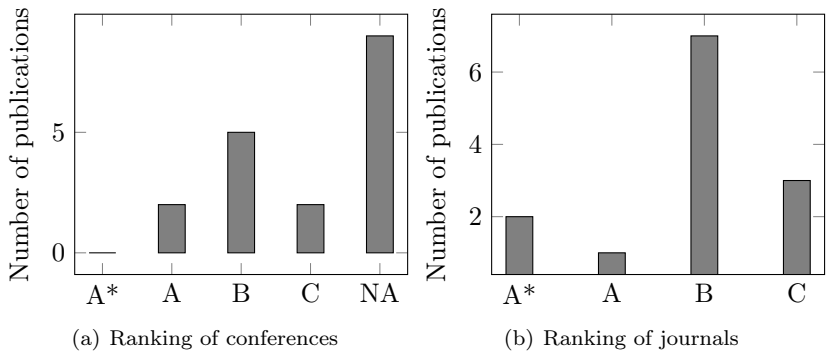

To get an overview of the quality of the papers, we looked at the ranking of the venues for both conference and journal publications. We used CORE~\cite{core}, which has search portals for conferences and journals. The site gives the following ranking categories: A* -- flagship venue in the discipline, A -- Excellent venue, B -- Good venue, and C -- Other ranked venue. The ranking is based on the ERA ranking process~\cite{arc_2018}. For journals that were not ranked in Core (8 studies), we compared their impact factors to similar journals listed in CORE and assigned a ranking accordingly. Figure~\ref{fig:ranking} shows the rankings of the venues that could be found in the portal's database. The column NA refers to conferences that were not found in the database.

%\mazsays{Riccardo: is this ok?}

%\begin{figure}
%  \begin{subfigure}[b]{0.5\textwidth}
   
%\begin{tikzpicture}
%        \begin{axis}[
%            width=\columnwidth,
%            height=4cm,
%            symbolic x coords={A*,A,B,C,NA},
%            xtick=data,
%            xlabel={Ranking of conferences},
%			ylabel={Number of publications},
%            grid style=dashed
%          ]
%            \addplot[ybar,fill=gray] coordinates {
%                (A*,   0)
%                (A,  2)
%                (B,   5)
%                (C, 2)
%                (NA, 9)
%            };
%        \end{axis}
%\end{tikzpicture}
%  \end{subfigure}
  %
%  \begin{subfigure}[b]{0.5\textwidth}
%    
%	\centering
%\begin{tikzpicture}
%        \begin{axis}[
%            width=\columnwidth,
%            height=4cm,
%            symbolic x coords={A*,A,B,C},
%            xtick=data,
%            xlabel={Ranking of journals},
			%ylabel={Ranking},
         %   grid style=dashed
       %   ]
       %     \addplot[ybar,fill=gray] coordinates {
      %          (A*,   2)
      %          (A,  1)
      %          (B,   7)
      %          (C, 3)
      %      };
    %    \end{axis}
%\end{tikzpicture}
%  \end{subfigure}
 % \caption{Ranking of the venues of included journal papers according to the Core ranking portal}
 % \label{fig:ranking}
%\end{figure}

\subsection{RQ1: Motivation}
In order to find the rational reported in literature for the adoption of SAC, we looked into motivations and usage scenarios, as explained in Section~\ref{sec:resmethod:rq}. 
Some of the identified motivations in RQ1 could also be seen as usage scenarios. For example, \emph{compliance with standards and regulation} could be seen as a motivation for using SAC, but also as a purpose for which SAC could be used.  

\subsubsection{Motivation}
\label{subSec:res:motivation}
%Generally speaking, the main reason that motivates usage of SAC is to perform security assurance on a system. 
In the literature, papers often refer to the use of SAC as a means to build security assurance, which is a generic (and rather obvious) motivation.
Instead, we looked for more specific motivations. 
%We extracted this piece of information by looking for motivations to use SAC rather than the suggested approach or method for creating them. 
In some of the papers, the motivation was made explicit in a separate section, or as the focus of the whole study (e.g. \cite{50_knight2015,53_alexander2011}). However, in most papers, this was briefly discussed either in the introduction and background sections, or as a part of motivating the used or suggested approach for creating SAC. 
If a study discusses only the generic SAC benefits, or is not being specific about the motivation (e.g., states that SAC provide security assurance in general), then we have categorised this paper as one that does not discuss any motivations for using SAC.

Table~\ref{res_rq1_tab_motive} shows all motivations found in our 51 sources.
The results show that about 73\% of the studies included at least one motivation for using SAC.

Categorizing the motivations resulted in the following categories:
\begin{itemize}
    \item External forces: Compliance with standards and regulation (9 mentions), and compliance with requirements in case of suppliers (4 mentions).
    \item Process improvement: SAC helps in integrating security assurance with the development process (6 mentions). Moreover, they help factoring work per work items, and analyzing complex systems (2 mentions). 
    \item Structure and documentation: The structure of SAC implies a way of work that reduces technical risks, and enhances security communication among stakeholders (7 mentions). 
    \item Security assessment: SAC help in assessing security and spotting weaknesses in security for the systems in question (6 mentions). Hence, they help building confidence in the those systems (3 mentions).
    \item Knowledge transfer: It is a proven approach in safety which has been used effectively for a long time, and could be similarly in security (5 mentions).
\end{itemize}
%\todo[inline]{Mazen: add the number of mentions for the other motivations}

%External forces 
%Process improvement
%Structure and documentation
%Security assessment
%Knowledge transfer

\begin{normalsize}
\begin{longtable}{@{}P{4cm}P{7.6cm}@{}}
\label{res_rq1_tab_motive}\\
\caption{RQ1 --- Papers stating the motivations for using SAC}\\

\toprule
\textbf{Study}  & \textbf{Motivation}  \\
\toprule
\endfirsthead
\multicolumn{2}{c}%
{\tablename\ \thetable\ -- \textit{Continued from previous page}} \\
\toprule
\textbf{Study}  & \textbf{Motivation}  \\
\toprule
\endhead
\midrule \multicolumn{2}{r}{\textit{Continued on next page}} \\
\endfoot
\bottomrule
\endlastfoot
                                    \multicolumn{2}{l}{\textit{External forces}}
                                    \\\midrule
                                    \textbf{Ankrum et al.} \cite{9_ankum2005} &  Comply with standards and regulation
                                
			                      \\ \textbf{Calinescu et al.} \cite{55_calinescu2017}    & Comply with security requirements of safety-critical systems
			                      \\ \textbf{Cyra et al.} \cite{1_cyra2007}   & Comply with standards and regulation
			                      \\ \textbf{Finnegan et al.} \cite{8_finnegan2014}   & Comply with regulation and maintain confidence in the product in question
			                      \\ \textbf{Finnegan et al. (2)} \cite{45_finnegan2014}   & Comply with regulation 
			                      \\ \textbf{He et al.} \cite{4_he2012}  & Reason about cybersecurity policies and procedures
			                      \\ \textbf{Mohammadi et al.} \cite{51_mohammadi2018}  &  Learn from the safety domain where it is a proven approach
			                      \\ \textbf{Ray et al.} \cite{16_ray2015}  & Comply with regulation and internal needs from cyber-physical systems' manufacturers
			                      \\ \textbf{Sklyar et al. (2)} \cite{21_sklyar2017,28_sklyar2017,31_sklyar2019}  & Comply with standards
			                      \\ \textbf{Sljivo et al.} \cite{23_sljivo2016}  & Comply with standards and regulation
			                      \\ \textbf{Strielkina et al.} \cite{22_strielkina2018}& Comply with security regulation
			                      \\\midrule
			                      \multicolumn{2}{l}{\textit{Knowledge transfer}}
                                    \\\midrule
			                      %knwoledge transfer
			                       \textbf{Goodger et al.} \cite{3_goodger2012}  & Learn from the safety domain to integrate oversight for safety and security
			                      \\ \textbf{Ionita et al.} \cite{47_ionita2017}  & Learn from the safety domain where it is a proven approach
			                      \\ \textbf{Netkachova et al. (2)} \cite{30_netkachova2014}  & Learn from the safety domain where it is a proven approach
			                      \\ \textbf{Poreddy et al.} \cite{26_poreddy2011} & Learn from the safety domain, where it is a proven approach
			                      \\ \textbf{Sklyar et al.} \cite{17_sklyar2016}  & Learn from the safety domain, where it is a proven in-use approach
			                      \\\midrule
			                      \multicolumn{2}{l}{\textit{Process improvement}}
			                      \\\midrule
			                      %process improvement
			                       \textbf{Ben Othmane et al.} \cite{18_othmane2016}  & Trace security requirements and assure security during iterative development.
			                      \\ \textbf{Ben Othmane et al.} \cite{10_othmane2014}  & Assure security during iterative development
			                      \\ \textbf{Cheah et al.} \cite{39_cheah2018}  & Cope with the increasing connectivity of systems 
			                      \\ \textbf{Cockram et al.} \cite{2_cockram2007}  & Reduces both technical and program risks through process improvement
			                      \\ \textbf{Gallo et al.} \cite{37_gallo2015}   & Factor analytical and implementation work per component, requisite, technology, or life-cycle
			                      \\ \textbf{Lipson et al.} \cite{56_lipson2008} & Help analyzing complex systems
			                      \\ \textbf{Netkachova et al.} \cite{19_netkachova2016}  & Tackle security issues which have intensified challenges of engineering safety-critical systems.
			                      \\ \textbf{Weinstock et al.} \cite{52_weinstock2007}  & Include people and processes in security assurance in addition to technology
			                      \\\midrule
			                      \multicolumn{2}{l}{\textit{Security assessment}}
			                      \\\midrule
			                      %security assessment
			                       \textbf{Alexander et al.} \cite{53_alexander2011}  & Help security evaluators to focus their attention on critical parts of the system
			                       \\ \textbf{Bloomfield et al.} \cite{29_bloomfield2017}  & Ensure the fulfillment of security requirements
			                       \\ \textbf{Finnegan et al.} \cite{45_finnegan2014}  & Improve overall security practices and demonstrate confidence in security
			                      \\ \textbf{Hawkins et al.} \cite{15_hawkins2015}  & Justify and assess confidence in critical properties
			                      \\ \textbf{Knight} \cite{50_knight2015}& Spot security related weaknesses in the system
			                      \\ \textbf{Poreddy et al.} \cite{26_poreddy2011} &  Assist in identifying security loopholes while changing the system
			                      \\ \textbf{Rodes et al.} \cite{25_rodes2014}  & Measure software security
			                      \\ \textbf{Strielkina et al.} \cite{22_strielkina2018}  & Acquire an input for decision making of requirement conformity 
			                      \\ \textbf{Vivas et al.} \cite{34_vivas2011}  & Acquire confidence that the security of the system meets the requirements
			                      \\\midrule
			                      \multicolumn{2}{l}{\textit{Structure and documentation}}
			                      \\\midrule
			                      %structure and documentation
			                      \textbf{Agudo et al.} \cite{38_agudo2009}   & Incorporate certifications and evaluation methods in an evidence-based structure
			                      \\ \textbf{Alexander et al.} \cite{53_alexander2011}  &   Summarize security thinking when vendors are involved
			                      \\ \textbf{Finnegan et al.} \cite{43_finnegan2013}   & Communicate and report achieved security level
			                      \\ \textbf{Knight} \cite{50_knight2015} & Document rational for security claims
			                      \\ \textbf{Netkachova et al.} \cite{14_netkachova2015} & Aid in communication as it provides a summary of issues and their interrelationship
			                      \\ \textbf{Patu et al.} \cite{5_patu2013} & Aid in the survival of modern system, with respect to security challenges
			                      \\ \textbf{Ray et al.} \cite{16_ray2015} & Comply with internal needs from cyber-physical systems' manufacturers

\end{longtable}
\end{normalsize}

\subsubsection{Usage Scenarios}
\label{subSec:res:usage}
While SACs are usually used to establish evidence-based security assurance for a given system, researchers have reported cases where SAC could be used to achieve different goals. We looked into studies that focus on using SAC for a purpose other than security assurance, or for a purpose that is specific to a certain domain (e.g., security assurance for medical devices) or context (e.g., security assurance within the agile framework).
Table \ref{res_rq1_tab_us} shows the usage scenarios of SAC found in literature. We were able to extract usage scenarios from 14 different papers (28\% of the total number of papers).  The usage scenarios we found show a wide range of applications of SAC. 
Seven of the papers suggest using SAC for \emph{evaluating different parts of the system or its surroundings}. For five papers, the use of SAC can be categorised as \emph{providing process and life-cycle support}. One paper suggests to use SAC to communicate between organisations involved in developing and using medical devices and one paper uses SAC to teach students about information security.

\begin{longtable}{@{}lP{7cm}@{}}
\label{res_rq1_tab_us}\\
\caption{RQ1 --- Papers relevant to understanding the usage scenarios}\\
\toprule
\textbf{Study} & \textbf{Usage Scenario} \\
\toprule
\endfirsthead
\multicolumn{2}{c}%
{\tablename\ \thetable\ -- \textit{Continued from previous page}} \\
\toprule
\textbf{Study} & \textbf{Usage Scenario} \\
\toprule
\endhead
\midrule \multicolumn{2}{r}{\textit{Continued on next page}} \\
\endfoot
\bottomrule
\endlastfoot
     \multicolumn{2}{@{}l}{\emph{Evaluating different parts of the system or its surroundings}}
  \\ \midrule
     \textbf{Graydon et al.} \cite{46_graydon2013}  & Evaluation of security standards, i.e., to make sure that conformance with a security standard is sufficient to achieve an acceptable and adequate level of security.
  \\ \textbf{Haley et al.} \cite{54_haley2005} & Proving the achievement of security requirements satisfaction. The SAC arguments are constructed as formal claims about the system behaviour called outer arguments, and informal inner arguments, which include but are not limited to: sub-claims, supporting facts, reliability, and trustworthiness claims. 
  \\ \textbf{Masumoto et al.} \cite{6_masumoto2013}  & Validation of service service grade achievement. The grade index value is used to quantify security claims, e.g., availability operation level, and the SAC is used to assure that the service meets that value.
  \\ \textbf{Mohammadi et al.} \cite{51_mohammadi2018}  & Ensuring trustworthiness in cyber-physical systems using trustworthiness cases. These cases are incorporated with the development process to actively evaluate trustworthiness while developing the system
  \\ \textbf{Netkachova et al. (2)} \cite{30_netkachova2014}  & Evaluation of security of critical infrastructures. The scenario suggests using an inter-dependency analysis method to assign a quantitative evaluation of the reliability of the evidence of the SAC. This would be used to support decisions related to the CI system's security
  \\ \textbf{Rodes et al.} \cite{25_rodes2014}  & Measuring software security based on confidence in security argument. The claims are annotated with a confidence level, and the SAC is supplemented with confidence arguments providing an assurance of the quality of the SAC's evidence and context items to make sure that these items properly support their associated claims
  \\ \textbf{Yamamoto} \cite{49_yamamoto2015}  &  Evaluation of system architecture based on security claims. It suggest assigning values to evidence ranging from -2 to 2 based on how satisfied the evidence is. 
  \\ \midrule
     \multicolumn{2}{@{}l}{\emph{Providing process and life-cycle support}}
  \\ \midrule
     \textbf{Agudo et al.} \cite{38_agudo2009}    & Integrating security engineering and assurance based development using SAC to help addressing security in a systematic and comprehensive way throughout the life-cycle.  
			                       
  \\ \textbf{Ben Othmane et al.} \cite{18_othmane2016,10_othmane2014} &  Controlling the impact of incremental development on security assurance using SAC. The SACs are developed along-side the security features and are used to actively ensure the completeness of security tests of these features.

	\\ \textbf{Bloomfield et al.} \cite{29_bloomfield2017}   & Using SAC in the development of security strategies and policies. This is done by using the arguments of the SAC as a structuring mechanisms for the objectives of the intended policy / strategy.
	\\ \textbf{Goodger et al.} \cite{3_goodger2012}  & supporting the protection of critical information infrastructure assets by providing a method to manage the life-cycles of the assets using SAC. Each asset is represented by one SAC, and a larger SAC is used to group multiple individual ones. The suggested SACs are living documents which take a defined evidence body as input, and have defined monitoring points.
  \\ \midrule
     \multicolumn{2}{@{}l}{\emph{Other uses}}
  \\ \midrule
     \textbf{Finnegan et al.} \cite{43_finnegan2013}  & Reporting the achieved security level. The proposed SACs help to show the achievement of security capability and communicating them between Health Delivery Organizations (HDO) and Medical Device Manufacturers (MDM)
  \\ \textbf{Gallo et al.} \cite{37_gallo2015}   & Using SAC to teach information security. The students are provided with a SAC and asked to learn from it and extract security requirements and goals in order to use them in their own projects.

\end{longtable}

\subsection{RQ2: Approaches}
\label{res:approaches}
We were able to find 26 different approaches in the literature. These studies focus on creating either complete SACs, or parts of them (argumentation or evidence).
Table~\ref{res_rq2_tab_approaches} shows these approaches, which part/s of SAC they cover, which argumentation strategies they use to divide the claims and create the arguments, and the evidence used to justify the claims in the approaches.
We categorize the approaches as follows:

\begin{itemize}
    \item \emph{Integrating SAC in the development life-cycle}: These approaches suggest mapping the SAC creation activities to the development activities to integrate SACs in the development and security processes~\cite{38_agudo2009,10_othmane2014,16_ray2015,34_vivas2011}, as well as assurance case driven design~\cite{17_sklyar2016,21_sklyar2017,28_sklyar2017,31_sklyar2019}.
    In general, these approaches suggest that the different stages of software development (requirements, design, implementation, and deployment) correspond to different abstraction levels of the security claims that can be made on the system. The hierarchical structure of SAC makes it possible to document these claims at every development stage as well as the dependencies to claims in the later or earlier stages~\cite{34_vivas2011}. This also applies to incremental development, e.g., using the SCRUM method~\cite{10_othmane2014}. Updating SACs during the development life-cycle is, however, essential for these approaches to work. Hence, conducting these updates has to be included as a mandatory activity in the security life-cycle of the system under development~\cite{17_sklyar2016}.
    \item \emph{Using different types of AC for security}: These approaches suggest using different types of assurance cases other than SAC for security assurance. These types are: \emph{(i)} trust cases, which are based on assurance cases templates derived from the requirements of security standards~\cite{1_cyra2007}; \emph{(ii)} trustworthiness cases, which focus mainly on addressing users' trust requirements~\cite{42_gorski2012,51_mohammadi2018}; and \emph{(iii)} combined safety and security cases~\cite{2_cockram2007}. This approach combines safety and security principles to create assurance cases with the main goal of achieving acceptable safety. The resulting cases have separate top claims for safety and security followed by separate argumentation; \emph{(iv)} dynamic assurance cases~\cite{55_calinescu2017}, an approach for generating arguments and evidence based on run-time patterns for the assurance cases of self-adaptive systems; \emph{(v)} multiple viewpoint assurance cases where security is treated as an assurance viewpoint \cite{23_sljivo2016}. The approach suggests to reuse AC artefacts by building multiple-viewpoint AC using contracts, and introduces an algorithm for a model transformation from a contract meta model into an argumentation meta model; and \emph{(vi)} dependability cases with focus on security \cite{41_patu2013}. 
    
    \item Documenting and visualizing SAC: These studies give guidelines of how to document a SAC, and visualize it \cite{26_poreddy2011,40_coffey2014,52_weinstock2007}. In this category there are papers that focus on a specific part of SAC. These are:
    \subitem Argumentation-centric: These approaches focus on the argumentation part of the SACs.  Different strategies are suggested in literature: security standards-based argument \cite{8_finnegan2014,45_finnegan2014,9_ankum2005}, and satisfaction argument \cite{54_haley2005}. Structures of argumentation found in literature are: model-based \cite{15_hawkins2015}, and layered structure \cite{14_netkachova2015,20_xu2017}. Moreover, we have one study which suggests an automatic creation of argument graphs \cite{7_tippenhauer2014}. 
    As we can see, there is a variety of argumentation strategies used in these approaches, which shows that SAC arguments can be flexible and fit for most security artefacts present at organizations. However, this is not necessarily a positive characteristic when applied in industry, as it might result in heterogeneous SACs created in different parts of an organization. In consequence, it would be hard to apply quality metrics to the SACs and to combine SACs created for sub-systems. Hence, companies need to find a way to choose a suitable approach, but there is a lack of comparison of SAC creation approaches in literature, especially for different industries and in different contexts. This is further discussed in \ref{disc:approaches}.
    \subitem Evidence-centric: These approaches focus mainly on different aspects of SACs' evidence. These aspects are: searching for evidence \cite{11_chindamaikul2014}, collecting and generating evidence \cite{36_shortt2015,56_lipson2008}, and rating of potential artifacts to be used as evidence \cite{39_cheah2018}. We conclude that even though the approaches cover main evidence-related activities, i.e., searching, locating, and rating, there are still essential parts missing, which are for example: assigning the evidence to claims, storing the evidence, and updating it over time. Similar to the argumentation-centric studies, the evidence-centric ones need to be more focused on the contexts in which they are applicable. Apart from the work of Cheah et. al \cite{39_cheah2018} which is done in the automotive domain, there is no focus towards domain specific SAC evidence work. We discuss this further in \ref{disc:coverage_slr}
  
\end{itemize}

\subsubsection{Coverage}
\label{res:coverage}
As shown in Table~\ref{res_rq2_tab_approaches}, 16 of the found approaches cover the creation of complete SACs, six focus on argumentation, and the remaining four on evidence.
Five out of the 16 studies to create complete security cases did not include any examples of evidence to justify the claims.
In general, the level of detail in the studies varies significantly. For example in the studies which cover the creation of complete SAC, we found papers covering the main elements of SAC and providing a very high-level description of both how to create them and what to use them for~\cite{16_ray2015,26_poreddy2011}, while other papers had very detailed descriptions of how to extract the claims and divide them to create the arguments. However, the latter is often related to a specific context, e.g., self-adaptive systems~\cite{55_calinescu2017}. We also observed that these studies focus significantly more on the argument part than the evidence part. This is further discussed in Section~\ref{disc:coverage_slr}.

\subsubsection{Argumentation}
Argumentation is a very important part of SAC. The argumentation starts with a security claim, and continues as the claim is being broken down into sub-claims. The strategy is used to provide a means by which claims are broken down. Each level of the argumentation could be done with a specific strategy. Hence, one SAC might have one or more argumentation strategies as is the case in some of the included studies in this SLR, e.g., \cite{38_agudo2009,51_mohammadi2018}.%\todo{Provide examples!} 

We looked for an explicit mention of the used strategy. If none was provided, we analysed the example cases to find the used argumentation strategy. Table \ref{res_rq2_tab_approaches} shows the approaches we found in literature with the respective argumentation strategies used in each of them.

When regarding argumentation strategies in the context of the different approaches, we could not find any correlation between the two. For instance, different approaches which integrate SAC within the development life-cycle use different argumentation strategies (e.g., requirements~\cite{38_agudo2009} and development phases~\cite{16_ray2015}).
The most common strategy depends on the output of a threat, vulnerability, asset or risk analysis (8 papers)~ \cite{2_cockram2007,40_coffey2014,1_cyra2007,51_mohammadi2018,41_patu2013,34_vivas2011,20_xu2017,52_weinstock2007}. 
Other popular strategies are breaking down the claims based on the requirements or more specifically quality requirements and even more specifically security requirements (5 papers)~ \cite{38_agudo2009,55_calinescu2017,54_haley2005,14_netkachova2015,28_sklyar2017}, 
 and arguing based on security properties, e.g., confidentiality, integrity and availability (5 papers)~\cite{11_chindamaikul2014,8_finnegan2014,51_mohammadi2018,26_poreddy2011,28_sklyar2017}. 
Additionally, researchers also used system and security goals (4 papers) ~\cite{38_agudo2009,10_othmane2014,51_mohammadi2018,7_tippenhauer2014}, 
software components or features (3 papers)~\cite{38_agudo2009,15_hawkins2015,28_sklyar2017}, 
security standards and principles (2 papers)~\cite{9_ankum2005,23_sljivo2016}, 
pre-defined argumentation model (1 paper)~\cite{42_gorski2012}, 
and development life-cycle phases (1 paper)~\cite{16_ray2015}.    

\subsubsection{Evidence}
Even though evidence is a very important and complex part of SAC, only four of 26 included approaches focused on it. Even in the complete approaches, there was a much deeper focus on the argumentation than the evidence. This explains why five out of the 16 complete approaches did not even include an example of what evidence would look like.
We found evidence either by looking for explicit mentions in the articles or by extracting the evidence part from the reported SACs.
Table \ref{res_rq2_tab_approaches} shows the approaches we found in literature with the respective evidence types used in each of them.

The most common types of evidence reported in literature are \emph{test results (TR)} (12 papers)~\cite{18_othmane2016,55_calinescu2017,39_cheah2018,11_chindamaikul2014,56_lipson2008,26_poreddy2011,36_shortt2015,17_sklyar2016,21_sklyar2017,28_sklyar2017,31_sklyar2019,23_sljivo2016} and different types of analysis. These analysis include threat and vulnerability (TVA) \cite{2_cockram2007,8_finnegan2014,45_finnegan2014,41_patu2013}, code (CA) and bug (BA) \cite{11_chindamaikul2014,18_othmane2016,17_sklyar2016,21_sklyar2017,28_sklyar2017,31_sklyar2019}, security standards and policies (PA) \cite{38_agudo2009,14_netkachova2015}, risk (RA) \cite{51_mohammadi2018}, and log analysis (LA) \cite{51_mohammadi2018,41_patu2013}.
Cheah et al.~\cite{39_cheah2018} present a classification of security test results using security severity ratings. This classification can be included in the security evaluation, which may be used to improve the selection of evidence when creating SACs. Chindamaikul et al.~\cite{11_chindamaikul2014} investigate how information retrieval techniques, and formal concept analysis can be used to find security evidence in a document corpus. 
Shrott and Weber \cite{36_shortt2015} present a method to apply fuzz testing to support the creation of evidence for SACs.

Other types of evidence reported in literature include \emph{process documents (PD)} \cite{56_lipson2008}, \emph{design techniques (DT)}~\cite{51_mohammadi2018}, and \emph{security awareness and training (SA)} \cite{41_patu2013,56_lipson2008,52_weinstock2007}.
Lipson and Weinstock \cite{56_lipson2008} describe how to understand, gather, and generate multiple kinds of evidence that can contribute to building SAC.

\begin{center}
%\begin{longtable}{@{}P{7.5cm}P{1.6cm}P{6.4cm}P{3cm}@{}}
\begin{longtable}{@{}P{3.4cm}P{1.6cm}P{4.3cm}P{1.6cm}@{}}
\caption{RQ2 --- Papers presenting approaches to construct SAC (TR: Test Results, TVA: Threat and Vulnerability Analysis, CA: Code Analysis, BA: Bug Analysis, PA: Security Standards and Policies, RA: Risk Analysis, LA: Log Analysis, PD: Process Document, SA: Security Awareness and Training)}
\label{res_rq2_tab_approaches}\\
\toprule
 \textbf{Approach} & \textbf{Coverage} & \textbf{Argumentation} & \textbf{Evidence} \\
\toprule
\endfirsthead
\multicolumn{4}{c}%
{\tablename\ \thetable\ -- \textit{Continued from previous page}} \\
\toprule
 \textbf{Approach} & \textbf{Coverage} & \textbf{Argumentation} & \textbf{Evidence} \\
\toprule
\endhead
\midrule \multicolumn{4}{r}{\textit{Continued on next page}} \\
\endfoot
\bottomrule
\endlastfoot

			                      \multicolumn{4}{@{}l}{\textit{Integrating SAC in the development life-cycle}}
			                      \\\midrule
			                      Assurance-based development \cite{38_agudo2009} & Complete    & Requirements, system goals, system views and models & PA 
			                      \\   Security assurance for incremental SD \cite{10_othmane2014} & Complete & Security goals & TR, CA
			                      \\  Integrating security engineering and AC development \cite{16_ray2015} & Complete & Development life-cycle phases & - 
			                      \\  Assurance Case Driven Design \cite{17_sklyar2016,28_sklyar2017,21_sklyar2017,31_sklyar2019} & Complete & Quality requirements, security properties, features, components, software layers, green IT principles & CA, TR
			                      \\  Security assurance driven SD \cite{34_vivas2011} & Complete & Threats, vulnerabilities & -
			                      \\\midrule
			                      \multicolumn{4}{@{}l}{\textit{Using different types of AC for security}}
			                      \\\midrule
			                           Dynamic assurance cases \cite{55_calinescu2017}   & Complete &  Requirements   & TR 
			                      \\   TRUST-IT - trustworthiness arguments \cite{42_gorski2012}  & Complete & Toulmin's argument~\cite{toulmin2003}& - 
			                      \\   Trustworthiness cases\cite{51_mohammadi2018} & Complete & Availability, threat analysis, goals satisfaction & RA, LA, DT, TVA
			                      \\   Evidence-based dependability case \cite{41_patu2013} & Complete & Vulnerabilities & TVA, SA, LA
			                      \\   Multiple-viewpoint AC \cite{23_sljivo2016} & Complete & Contracts (pair of assumptions and guarantees) & TR 
			                      \\   Trust-cases for security standards compliance \cite{1_cyra2007}  & Argument & Risks & -
			                      \\Dependability by contract \cite{2_cockram2007} & Argument & Vulnerabilities, threats, and mitigation & TVA
			                      \\\midrule
			                      \multicolumn{4}{@{}l}{\textit{Documenting and Visualizing SAC}}
			                      \\\midrule
			                          Mapping SAC to standards \cite{9_ankum2005}   &  Complete & Security standard description & - 
			                      \\   Concept map-based \cite{40_coffey2014}   & Complete & Vulnerabilities & -
			                      \\  Risk based approach \cite{8_finnegan2014,45_finnegan2014}  & Complete & Security capabilities, mitigation controls & TVA, LA 
			                      \\  Layered Approach \cite{14_netkachova2015} & Complete & Source of security requirements, changes during life-cycle & TVA, PA
                                  \\   Documenting AC for Security \cite{26_poreddy2011} & Complete & Security properties & TR 
			                      \\   Arguing security \cite{52_weinstock2007} & Complete & Prevention and detection & TR, TVA, SA 
			                       \\   Satisfaction arguments \cite{54_haley2005} & Argument & Security requirements & - 
			                       \\   Model-based assurance \cite{15_hawkins2015} & Argument & Software components & - 
		                        \\   Automatic generation of argument graphs \cite{7_tippenhauer2014} & Argument & Security goals  & -
			                    \\   Layered Argument strategy \cite{20_xu2017} & Argument & Assets, threats & -   
			                     \\ Systematic Security Evaluation \cite{39_cheah2018}   & Evidence & - & TR   
			                      \\  Document retrieval and concept analysis \cite{11_chindamaikul2014}  & Evidence &  Security properties   & TR, BA 
			                      \\   Evidence-based security properties' assurance \cite{56_lipson2008}& Evidence & - & PD, TR, SA
			                      \\  Hermes Targeted fuzz testing \cite{36_shortt2015} & Evidence & - & TR 
			                     
\end{longtable}
\end{center}
%\end{landscape}

\subsection{RQ3: Support}
\label{res:support}
In this section, we list our results from reviewing the practical support to facilitate the adoption of SAC reported in literature. Specifically, we report on the tools used to assist in any of the SAC activities, e.g., creation and maintenance, the prerequisites of the approaches, and patterns for creating SAC. 

\subsubsection{Tools:}
We found 16 software tools which have been used one way or another in the creation of SAC in literature. 
Seven of the found tools were created by researchers. Four of these seven target assurance cases in general~\cite{35_fung2018,24_gacek2014,15_hawkins2015,7_tippenhauer2014}, while the remaining three are created to be used in the creation of SAC specifically~\cite{18_othmane2016,39_cheah2018,36_shortt2015}.
Table~\ref{res_rq3_tab_tools} shows the tools and the respective studies in which they are used. A brief description of the main functionalities of the tools, as well as whether the tools are created or used by the authors are also presented.
There are four main types of reported tools. In the following, we list the tools of each type, and we discuss the main features of each tool as reported in the studies:

\begin{itemize}
    \item Creation tools: used to create and document assurance cases in general. 
    \item Argumentation tools: focus mainly on the creation of the argumentation part of SAC. 
    \item Evidence tools: focus on the creation of SAC evidence.
    \item Support tools: several studies reported supporting tools to assist the creators of SAC in the analysis needed for creating them, e.g., by helping users determine the relevance of a given document to be used as evidence~\cite{11_chindamaikul2014}.
\end{itemize}

\begin{longtable}{@{}P{2.2cm}P{2.5cm}P{5.9cm}P{.3cm}@{}}
\label{res_rq3_tab_tools}\\
\caption{RQ3 --- Tools supporting the creation, documentation, and visualization of SAC (U: Used, C: Created)}\\
\toprule
\textbf{Study}
			& \textbf{Tool support}
			& \textbf{Description}
			& \textbf{}
			\\
\toprule
\endfirsthead
\multicolumn{4}{c}%
{\tablename\ \thetable\ -- \textit{Continued from previous page}} \\
\toprule
\textbf{Study}
			& \textbf{Tool support}
			& \textbf{Description}
			& \textbf{}
			\\
\toprule
\endhead
\midrule \multicolumn{4}{r}{\textit{Continued on next page}} \\
\endfoot
\bottomrule
\endlastfoot
\multicolumn{4}{@{}l}{\emph{Creation Tools}}\\ \midrule
    \textbf{Poreddy et al.}~\cite{26_poreddy2011}&Adelard Safety Case Editor (ASCE)~\cite{tool_adrelard} & Supports the creation and visualisation of AC. It supports multiple notations, and enables assigning multiple formats of data in the bodies of AC nodes, e.g., text, tables, and images. Additionally, allows validation of the AC structure based on the rules of a notation or based on user-defined rules. & U
    \\ \textbf{Ankrum et al.}~\cite{9_ankum2005}&Adelard Safety Case Editor (ASCE)~\cite{tool_adrelard} &  & U
    \\ \textbf{Finnegan et al.}~\cite{43_finnegan2013} & TurboAC~\cite{tool_turboac} & Enables converting artefacts of different formats used to create assurance cases, e.g., tabular or XML into HTML files which can be viewed and navigated with web browsers. Also allows electronically submitting assurance cases to external authorities. & U 
    \\ \textbf{Gacek et al.}~\cite{24_gacek2014}&Resolute & An open source software which enables to automatically construct assurance cases based on models which use the Architecture Analysis and Design Language (AADL). & C
    \\ \textbf{Patu et al.}~\cite{41_patu2013}& D-Case Editor~\cite{d-case} & Used to document and visualize assurance cases. Includes a library of patterns which can assist the users in creating the cases. & U
 \\ \midrule \multicolumn{2}{@{}l}{\emph{Argumentation Tools}}\\ \midrule
    \textbf{Hawkins et al.}~\cite{15_hawkins2015}&Instantiation program (no specific name) & A model-based tool which takes GSN argument patterns and different information models as input. The information models hold relevant information for AC arguments, such as design models. The tool identifies the elements required to instantiate the GSN model, and outputs an instantiated model. & C
    \\ \textbf{Ionita et al.}~\cite{47_ionita2017}&OpenArgue~\cite{tool_openArgue}  & Provides  editors and the ability to derive graphical arguments from textual requirements specifications. Users can specify inter-argument relationships, e.g., where one argument can mitigate another. & U 
    \\ & ArgueSecure~\cite{tool_arg_sec} & Allows users to collaboratively work on argumentation spreadsheets, designed to decompose the arguments into claims, assumptions and facts. & U
    \\ \textbf{Tippenhauer et al.}~\cite{7_tippenhauer2014}&CyberSAGE~\cite{tool_CyberSAGE}& Allows users to integrate information from different sources, e.g., network topology and attacker models, to automatically generate security arguments. & C
    \\ \textbf{Calinescu et al.}~\cite{55_calinescu2017} &UPPAAL~\cite{tool_uppaal}& A verification tool suite used to generate evidence to show the achievement of a claimed goal. Also verifies that a model satisfies pre-defined correctness properties. & U
    \\ \textbf{Shortt et al.}~\cite{36_shortt2015} &Hermes& Provides dynamic code coverage analysis which can be used as SAC evidence. & C
    \\ \textbf{Cheah et al.}~\cite{39_cheah2018} &Software tool (no specific name)& Semi-automated tool for penetration testing with features such as: identifying open ports, spoofing a device and scanning of log files. The output of the tool is used to create the body of evidence to be used in a SAC. & C
\\ \midrule \multicolumn{2}{@{}l}{\emph{Support Tools}} \\ \midrule
    \textbf{Ben Othmane et al.}~\cite{18_othmane2016}&Meld\cite{tool_meld}&  Visualizes differences between different files and helps merging these files. & U
    \\  &SECUREAGILE & Traces the impact of code changes on security to support the iterative development of security features with help of SAC. & C
    \\ \textbf{Chindamaikul et al.}  \cite{11_chindamaikul2014}&Concept lattice & Helps users to determine the relevance of a given document. & U
    \\ \textbf{Fung et al.}~\cite{35_fung2018}&MMINT-A & Automated change impact assessment for SAC. & C
    \\ \textbf{Go\`rski et al.}~\cite{42_gorski2012}&NOR-STA~\cite{tool_norsta}& A set of services used for editing and assessing argumentation of assurance cases. Also acsts as a repository to store evidence used in SAC.& U
    
\end{longtable}

\subsubsection{Prerequisites:}

Prerequisites are the conditions that need to be met before an approach presented in a study can be applied. We found prerequisites in the included studies by checking the inputs of the proposed outcomes (approaches, usage scenarios, tools, and patterns). If an input is not a part of the outcome itself, we considered it to be a prerequisite to that outcome. 
Table~\ref{res_rq3_tab_prereq} shows the prerequisites we found along with the respective type of study for each. There are 17 reported prerequisites. The majority belong to approaches (11) \cite{11_chindamaikul2014,2_cockram2007,1_cyra2007,15_hawkins2015,41_patu2013,9_ankum2005,39_cheah2018,23_sljivo2016,7_tippenhauer2014,34_vivas2011,20_xu2017} while the remaining ones belong to usage scenarios (2) \cite{29_bloomfield2017,3_goodger2012}, patterns (2) \cite{5_patu2013,4_he2012}, and tools (1) \cite{24_gacek2014}.
We categorize prerequisites as follows:
\begin{itemize}
    \item Usage of specific format~\cite{24_gacek2014,15_hawkins2015,23_sljivo2016}: In this category, studies require the use of artefacts which have specific formats to achieve the purpose of the study. 
    \item Usage of specific documents and repositories~\cite{11_chindamaikul2014,2_cockram2007,4_he2012,41_patu2013,7_tippenhauer2014,34_vivas2011}: The studies in this category use specific repositories and documents for retrieving required data for building or using SAC. 
    \item Usage of security standards~\cite{9_ankum2005,1_cyra2007}: The studies in this category require the use of security standards to create SAC or make use of them.
    \item Existence of analysis and modelling~\cite{39_cheah2018,3_goodger2012,5_patu2013,20_xu2017}: The studies in this category require the existence or performing certain analysis and models to achieve their purpose.  
    \item Existence of special expertise~\cite{29_bloomfield2017}: The one study in this category relies on expertise provided by an external safety regulator.
\end{itemize}

\begin{longtable}{@{}P{2.6cm}P{1.6cm}P{7.1cm}@{}}
\label{res_rq3_tab_prereq}\\
\caption{RQ3 --- Papers discussing the prerequisites of SAC approaches, usage scenarios, and tools}\\
\toprule
\textbf{Study} & \textbf{Type} & \textbf{Prerequisites}	\\
\toprule
\endfirsthead
\multicolumn{3}{c}%
{\tablename\ \thetable\ -- \textit{Continued from previous page}} \\
\toprule
\textbf{Study} & \textbf{Type} & \textbf{Prerequisites}	\\
\toprule
\endhead
\midrule \multicolumn{3}{r}{\textit{Continued on next page}} \\
\endfoot
\bottomrule
\endlastfoot
    \multicolumn{3}{@{}l}{\emph{Usage of specific format}} \\ \midrule
    \textbf{Sljivo et al.}  \cite{23_sljivo2016}& Approach & Re-usability of assurance cases by using \emph{safety contracts created using the Safety Element out-of-context Meta-model (SEooCMM)}.
 \\ \textbf{Hawkins et al.}  \cite{15_hawkins2015}& Approach & Model-based approach for creating assurance cases based on GSN and an extended model of the structured assurance case meta-model by the OMG~\cite{omg}. Requires {reference information models} (e.g., design and analysis models) and a {weaving model} (which connects the reference models to the GSN pattern) as inputs.
 \\ \textbf{Gacek et al.}   \cite{24_gacek2014}&Tool& Proposes a tool for automatic generation of SAC which requires a {system model specified in the Architecture Analysis and Design Language (AADL).~\cite{aadl}}.
 \\ \midrule \multicolumn{3}{@{}l}{\emph{Usage of specific documents and repositories}} \\ \midrule
    \textbf{Chindamaikul et al.}   \cite{11_chindamaikul2014}&Approach & Creation of SAC using information retrieval techniques based on an \emph{existing repository of evidence}.
 \\ \textbf{Patu et al.}  \cite{41_patu2013}& Approach & Creation of SAC based on a {pre-defined list of common risks, vulnerabilities and solutions in the domain}.
 \\ \textbf{Cockram et al.}  \cite{2_cockram2007}&Approach&Creation of SAC argument uses \emph{Module boundary contracts} identified from the specification of the system in question. 
 \\ \textbf{Tippenhauer et al.}  \cite{7_tippenhauer2014}&Approach&Automatic SAC argument generation requires the use of \emph{extension templates}. These templates are formalization of sub-argument patterns.
 \\ \textbf{Vivas et al.}  \cite{34_vivas2011}&Approach&The approach suggests integrating SAC within SDLC ((Software Development Life-Cycle)), which requires the existence of a \emph{Well defined SDLC process}.
 \\ \textbf{He et al.}  \cite{4_he2012}&Pattern& Patterns are created based on a \emph{repository of lessons learned and recommendations} from previous security incidents.
 \\ \midrule \multicolumn{3}{@{}l}{\emph{Usage of security standards}} \\ \midrule
    \textbf{Cyra et al.}    \cite{1_cyra2007}&Approach& Cyra et al. Trust case templates are based on \emph{security standards} and restructure the standards' information, e.g., their requirements.
 \\ \textbf{Ankrum et al.}  \cite{9_ankum2005}&Approach& Applied assurance cases to the requirements of three standards in order to study the applicability and problems of that approach. The authors used the learning outcome to create a practical SAC, but also used {artefacts from one of the standards}.
 \\ \midrule \multicolumn{3}{@{}l}{\emph{Existence of analysis and modelling}} \\ \midrule
    \textbf{Cheah et al.} \cite{39_cheah2018} &Approach& Construction and severity classification of SAC evidence based on a scripted attack tree and manual threat modelling. \emph{Asset analysis and models} are required.
 \\ \midrule \multicolumn{3}{@{}l}{\emph{Existence of special expertise~}} \\ \midrule
    \textbf{Xu et al.}  \cite{20_xu2017}&Approach& Requires an \emph{asset analysis and model} as a basis for a layered approach for creating SAC.
\\ \textbf{Bloomfield et al.}  \cite{29_bloomfield2017}&Usage scenario& Suggests using SAC for developing security strategy and policies. Most of the arguments and evidence are derived from the \emph{expertise of a safety regulator}.
\\ \textbf{Goodger et al.}  \cite{3_goodger2012}&Usage Scenario& Requires an \emph{asset analysis} to identify the assets for which the SAC will be created in order to protect critical infrastructure.
\\ \textbf{Patu et al. (2)} \cite{5_patu2013}&Pattern& Describe an \emph{asset analysis} as the basis for identifying security patterns at the requirements phase of the development life-cycle.

\end{longtable}

\subsubsection{Patterns} Reoccurring claims and arguments in SAC can be subsumed in patters. They can save the creators of SACs a lot of time and effort. We found ten studies which deal with patterns. Six of these create their own argumentation patterns~\cite{8_finnegan2014,45_finnegan2014,4_he2012,5_patu2013,26_poreddy2011,20_xu2017}.
The remaining four include usage of patterns~\cite{15_hawkins2015,7_tippenhauer2014}, a guideline for creating and documenting security case patterns \cite{52_weinstock2007}, and a catalogue of security and safety case patterns \cite{48_taguchi2014}.
Since we we only considered patterns created and used for SAC, we excluded those studies in which patterns are borrowed from the safety domain, e.g., \cite{55_calinescu2017}.

Table \ref{res_rq3_tab_patterns} shows the studies that deal with SAC patterns. While the created patterns cover an important aspect, namely abstraction, it is not clear how re-usable or generalize-able they are. Some patterns are derived from various security standards, e.g., \cite{8_finnegan2014,48_taguchi2014} (these are usually from the medical domain where security standardization is more mature compared to other security-critical domains), and one from lessons learned from security incidents~\cite{4_he2012}, but none is derived from previous applications of SAC in industry.
Another observation we made is that the patterns focus heavily on the argumentation part of SAC in contrast to the evidence part. Only few studies provided examples of evidence that can be used in a given pattern \cite{26_poreddy2011,48_taguchi2014,52_weinstock2007}. However, these examples are specific to the context of the studies, and leaves the abstraction to the reader, with the notable exception of the examples provided by Weinstock et al.~\cite{52_weinstock2007}.

\begin{longtable}{@{}lP{7.7cm}@{}}
\label{res_rq3_tab_patterns}\\
\caption{RQ3 --- Papers presenting patterns.}\\
	
\toprule
\textbf{Study}	& \textbf{Description of the pattern-based approach}\\
\toprule
\endfirsthead
\multicolumn{2}{c}%
{\tablename\ \thetable\ -- \textit{Continued from previous page}} \\
\toprule
\textbf{Study}	& \textbf{Description of the pattern-based approach}\\
\toprule
\endhead
\midrule \multicolumn{2}{r}{\textit{Continued on next page}} \\
\endfoot
\bottomrule
\endlastfoot
 \multicolumn{2}{@{}l}{\emph{Creation of patterns}} \\ \midrule
 \textbf{Finnegan et al.}  \cite{8_finnegan2014,45_finnegan2014} &Creation of security capability argument pattern using a risk-based approach. Argues for each security capability defined in a technical report for risk management in medical devices.
                                \\ \textbf{He et al.} \cite{4_he2012}&Creation of generic cases which use security arguments that are informed by security incidents in healthcare organizations. 
                                %\\ \textbf{Lipson and Weinstock} & \cite{56_lipson2008}&A template for evidence collection and documentation is presented.
                                \\ \textbf{Patu et al.}  \cite{5_patu2013}&Creation of security patterns during the requirement phase of system development. One suggested pattern argues over security attributes.
                                %Link which includes the set of cases is broken!
                                \\ \textbf{Poreddy et al.}  \cite{26_poreddy2011}&Creation of assurance case patterns. Suggested argumentation strategies are: integrity, availability, reliability, confidentiality and maintainability.
                                \\ \textbf{Xu et al.}  \cite{20_xu2017}& Creation of different argument patterns to be used in different layers to form a layered argument structure.
\\ \midrule \multicolumn{2}{@{}l}{\emph{Usage of patterns}} \\ \midrule
   \textbf{Hawkins et al.}  \cite{15_hawkins2015}&Usage of argument patterns as input to the model-based approach for building assurance case arguments. A suggested pattern argues over individual software components.
\\ \textbf{Tippenhauer et al.}  \cite{7_tippenhauer2014}&Usage of argument patterns to automatically generate argument graphs. The paper includes five different patterns categorized into the categories inter-type and intra-type.
\\ \midrule \multicolumn{2}{@{}l}{\emph{Other}} \\ \midrule
   \textbf{Taguchi et al.}  \cite{48_taguchi2014}&A catalogue of safety and security case patterns. The patterns are derived from process patterns through a literature survey.
\\ \textbf{Weinstock et al.}  \cite{52_weinstock2007}&A guideline of how to create and use SAC patterns. An example pattern is also presented.
\end{longtable}

\subsubsection{Notations}

Out of 51 studies, 41 specify at least one notation to be used for expressing and documenting a SAC. 
Table \ref{res_notationRef} shows the number of studies that use each notation, and lists them.
The most common notation is the Goal Structure Notation (GSN)~\cite{gsn} which is suggested by 27 studies. Another popular notation is the Claim Argument Evidence (CAE)~\cite{CAE} notation which is suggested by nine studies. Other notations are: text (6 studies), concept maps~\cite{40_coffey2014} (1), and Claim-Argument-Evidence Criteria (CAEC)~\cite{19_netkachova2016,14_netkachova2015,30_netkachova2014} notation which is extension of the CAE notation (3 studies of the same authors).

\begin{longtable}{P{2.2cm}P{1.5cm}l}
\label{res_notationRef}\\
\caption{Studies which use each notation}\\
\toprule
\textbf{Notation} &  \textbf{Number} & \textbf{Study}  \\
\toprule
\endfirsthead
\multicolumn{3}{c}%
{\tablename\ \thetable\ -- \textit{Continued from previous page}} \\
\toprule
\textbf{Notation} & \textbf{Number} & \textbf{Study}  \\
\toprule
\endhead
\midrule \multicolumn{3}{r}{\textit{Continued on next page}} \\
\endfoot
\bottomrule
\endlastfoot
GSN     & 27 & \cite{53_alexander2011,9_ankum2005,18_othmane2016,10_othmane2014,55_calinescu2017,11_chindamaikul2014,2_cockram2007,8_finnegan2014,45_finnegan2014,35_fung2018,3_goodger2012,46_graydon2013,15_hawkins2015,4_he2012} \\ 
 & & \cite{47_ionita2017,6_masumoto2013,51_mohammadi2018,41_patu2013,5_patu2013,26_poreddy2011,16_ray2015,25_rodes2014,23_sljivo2016,48_taguchi2014,52_weinstock2007,20_xu2017,49_yamamoto2015} \\
CAE & 9 & \cite{53_alexander2011,9_ankum2005,29_bloomfield2017,43_finnegan2013,3_goodger2012,47_ionita2017,30_netkachova2014,19_netkachova2016,14_netkachova2015}\\ 
Text & 6 & \cite{39_cheah2018,1_cyra2007,24_gacek2014,37_gallo2015,42_gorski2012,47_ionita2017} \\ 
CAEC & 3 & \cite{17_sklyar2016,31_sklyar2019,28_sklyar2017}\\ 
Concept maps & 1 & \cite{40_coffey2014} \\ 
\end{longtable}

\subsection{RQ4: Validation}
\label{res:validation}

\afterpage{
\begin{landscape}
\begin{longtable}{@{}P{4.6cm}P{2.5cm}P{5.4cm}lll@{}}
\label{res_rq4_tab_validation}\\
\caption{RQ4 --- Papers presenting a form of validation}\\
\toprule
\textbf{Study}
		%	&\textbf{Validation}
			& \textbf{Domain}
			& \textbf{Data source}
			& \textbf{SAC}
			& \textbf{Creators}
			& \textbf{Validators}\\
\toprule
\endfirsthead
\multicolumn{6}{c}%
{\tablename\ \thetable\ -- \textit{Continued from previous page}} \\
\toprule
\textbf{Study}
		%	&\textbf{Validation}
			& \textbf{Domain}
			& \textbf{Data source}
			& \textbf{SAC}
			& \textbf{Creators}
			& \textbf{Validators}\\
\toprule
\endhead
\midrule \multicolumn{6}{r}{\textit{Continued on next page}} \\
\endfoot
\bottomrule
\endlastfoot
			                      \multicolumn{6}{@{}l}{\emph{Case Study}}
			                      \\\midrule
			                      \textbf{Ben Othmane et al.} \cite{18_othmane2016}&Software Engineering&E-Commerce product&\ON&Authors&3rd party
			                      \\ \textbf{Bloomfield et al.}  \cite{29_bloomfield2017}&Safety&Regulatory organization&\ON&Authors*&3rd party
                                \\ \textbf{Calinescu et al.} \cite{55_calinescu2017} &Marine, Trading&Underwater Vehicle System, Trading System&\ON&Authors&Authors
                                \\ \textbf{Cheah et al.} \cite{39_cheah2018} &Automotive&Vehicle infotainment system, diagnostics tool&\ON&Authers*&3rd party
                                 \\ \textbf{Fung et al.} \cite{35_fung2018}&Automotive&Power sliding door -- Case from ISO26262 standard&\ON&Authors&Authors
                                 \\ \textbf{Goodger et al.} \cite{3_goodger2012}&Critical infrastructure&Critical information Infrastructure&&NA&Authors*
                                \\ \textbf{G\`orski et al.}    \cite{42_gorski2012}&Medical&A software for patient monitoring&\ON&Authors&Authors
                                \\ \textbf{Graydon et al.}  \cite{46_graydon2013}&Security&Security standards&\ON&Authors&Authors
                                 \\ \textbf{He et al.} \cite{4_he2012}&Medical&Lessons learned from security incidents, security standards, policies, and procedures&\ON&Authors&Authors
                                 \\ \textbf{Xu et al.} \cite{20_xu2017}&Software Engineering&IM server&\ON&Authors*&Authors*
                                 
			                      \\\midrule
			                      \multicolumn{6}{@{}l}{\emph{Illustrative Case}}
			                      \\\midrule
			                      \textbf{Ankrum et al.} \cite{9_ankum2005}&Security&Research security project&&Authors&Authors
			                      \\ \textbf{Ben Othmane et al. (2)} \cite{10_othmane2014}&Telecom&Commercial project&\ON&Authors&Authors
			                      \\ \textbf{Cockram et al.}  \cite{2_cockram2007}&SafSec&Command and control system for locating persons&\ON&Authors*&Authors*
                                \\ \textbf{Coffey et al.}  \cite{40_coffey2014}&Software Engineering&SOA composite application&\ON&Expert group&3rd party
                                \\ \textbf{Cyra et al.}    \cite{1_cyra2007}&Security&Security standard BS 7799-2 &\ON&Authors&Authors
                                \\ \textbf{Gacek et al.}  \cite{24_gacek2014}&Embedded Systems&Research project for unmanned air vehicles&\ON&Authors*&Authors*
			                    \\ \textbf{Haley et al.}  \cite{54_haley2005}&Software Engineering&Example HR system&&NA&Authors
                                \\ \textbf{Hawkins et al.}  \cite{15_hawkins2015}&Model-Based Engineering&Cryptographic controller system&\ON&Authors&Authors \\ \textbf{Mohammadi et al.} \cite{51_mohammadi2018}&Medical&OPerational Trustworthiness Enabling Technologies (OPTET) research project &\ON&Authors&Authors
                                \\ \textbf{Netkachova et al.}  \cite{14_netkachova2015}&Aviation&A security gateway data-flow controller&\ON&Authors*&Authors*
			                     \\ \textbf{Patu et al. (2)}  \cite{41_patu2013}&Networking&Research e-learning project&\ON&Authors&Authors
			                     \\ \textbf{Poreddy et al.}  \cite{26_poreddy2011}&Aviation&Avionic mission control computer system&\ON&Authors&Authors
                                \\ \textbf{Ray et al.}  \cite{16_ray2015}n&Medical&A medical cyber-physical system for pumping insulin&\ON&Authors&Authors
                                \\ \textbf{Rodes et al.}  \cite{25_rodes2014}&Security&Example scenario with confidence properties measurement&\ON&Authors*&Authors*
                                \\ \textbf{Shortt et al.} \cite{36_shortt2015} &Software Engineering&Java-based open source library (Crawler4J)&&NA&Authors
                                \\ \textbf{Sklyar et al. (3)}  \cite{28_sklyar2017}&SafSec&Requirements derived from safety and security standard&\ON&Authors&Authors
                                \\ \textbf{Sklyar et al. (4)}  \cite{31_sklyar2019}&Medical&Example medical system&\ON&Authors&Authors
                                \\ \textbf{Sljivo et al.}  \cite{23_sljivo2016}&Aviation&Wheel breaking system&\ON&Authors&Authors
                                \\ \textbf{Strielkina et al.}  \cite{22_strielkina2018}&Medical&Healthcare IoT system&&NA&Authors
                                \\ \textbf{Tippenhauer et al.}  \cite{7_tippenhauer2014}&Electrical&An electrical power grid use case&\ON&Authors&Authors
                                \\ \textbf{Vivas et al.}  \cite{34_vivas2011}&Software Engineering&The research project PICOS (Privacy and Identity Management for Community Services)&\ON&Authors&Authors
                                
			                      \\\midrule
			                      \multicolumn{6}{@{}l}{\emph{Experiment}}
			                      \\\midrule
			                      \textbf{Chindamaikul et al.}   \cite{11_chindamaikul2014}&Information Retrieval&Open source software development project&\ON&Expert group&Authors
			                      \\ \textbf{Gallo et al.}   \cite{37_gallo2015}&Education&Course in information security&\ON&NA&Authors*
			                      \\ \textbf{Masumoto et al.}  \cite{6_masumoto2013}&Software Engineering&Commercial web application&\ON&Authors&Authors
                                
			                      \\\midrule
			                      \multicolumn{6}{@{}l}{\emph{Observation}}
			                      \\\midrule
			                      \textbf{Finnegan et al. (2)}   \cite{45_finnegan2014}&Medical&Technical report&\ON&Authors&3rd party
\end{longtable}
\end{landscape}
}

We consider validation to be the process to show that an approach or tool for creating SAC works in practice or that an SAC can actually be used for a suggested usage scenario. In case validation is performed in a selected study, we looked for the type of validation, the domain of application, the source of data, whether a SAC is created during the validation, the creators of the SACs, and who performed the validation. 

Table~\ref{res_rq4_tab_validation} shows these different aspects for the 36 studies which include a validation of the outcome. The majority of the outcomes were validated using illustrative cases (21), 11 were validated using case studies, and the remaining four used experiments (3) and observation as a part of an Action Design Research (ADR)~\cite{ADR} study. 

The data sources vary among the validations, as can be seen in Table \ref{res_rq4_tab_validation}. We categorize these sources into three main categories:
\begin{itemize}
    \item Research, open source, and in-house projects (20)~ \cite{9_ankum2005,11_chindamaikul2014,2_cockram2007,40_coffey2014,24_gacek2014,54_haley2005,15_hawkins2015,51_mohammadi2018,14_netkachova2015,41_patu2013,26_poreddy2011,16_ray2015,25_rodes2014,36_shortt2015,31_sklyar2019,23_sljivo2016,22_strielkina2018,7_tippenhauer2014,34_vivas2011,37_gallo2015}
    \item Commercial products / systems (9)~ \cite{18_othmane2016,10_othmane2014,55_calinescu2017,39_cheah2018,3_goodger2012,42_gorski2012,6_masumoto2013,20_xu2017,30_netkachova2014}
    \item Standards, regulation, and technical reports (7)~ \cite{29_bloomfield2017,1_cyra2007,45_finnegan2014,35_fung2018,46_graydon2013,4_he2012,28_sklyar2017}
\end{itemize}

SACs were presented in 31 out of the 36 validations. Representing a complete SAC is mostly not possible even in small illustrative cases due to the  amount of information required to build one. However, how much of an SAC is represented in the included validations varies to a large extent. Some validations present an example of a full branch of SAC, i.e., a claim all the way from top to evidence (e.g., He and Johnson~\cite{4_he2012}), while others present very brief examples of SACs (e.g., Gallo and Dahab~\cite{37_gallo2015}). \\ \linebreak
Table~\ref{res_rq4_tab_validation} also shows who created the SACs in each study. In only two cases, experts were used to create the SACs. In the majority of the studies (28), the authors created the SACs. However, eight of the studies included authors from industry. These are shown in Table \ref{res_rq4_tab_validation} as ``Authors*'' in the Creators column. \\ \linebreak
Table~\ref{res_rq4_tab_validation} also shows the domains in which the validation was conducted. The most common domains are Software Engineering (7) and Medical (7). \\ \linebreak
The last column in Table~\ref{res_rq4_tab_validation} shows the persons which performed the validation in each study. Out of the 36 included validations, only five used third parties to validate the outcomes. These were industrial partners in two cases~\cite{18_othmane2016,39_cheah2018}, an external regulator~\cite{29_bloomfield2017}, one security expert~\cite{40_coffey2014}, and a group of security experts \cite{45_finnegan2014}.  In the remaining 31 validations, the authors performed the validation. However, eight of the studies included authors from industry. These are shown in Table \ref{res_rq4_tab_validation} as ``Authors*'' in the Validator column.
\pagebreak
\section{SAC creation workflow}\label{sec:wf}
\begin{figure}[H]
\begin{center}
  \includegraphics[width=\textwidth]{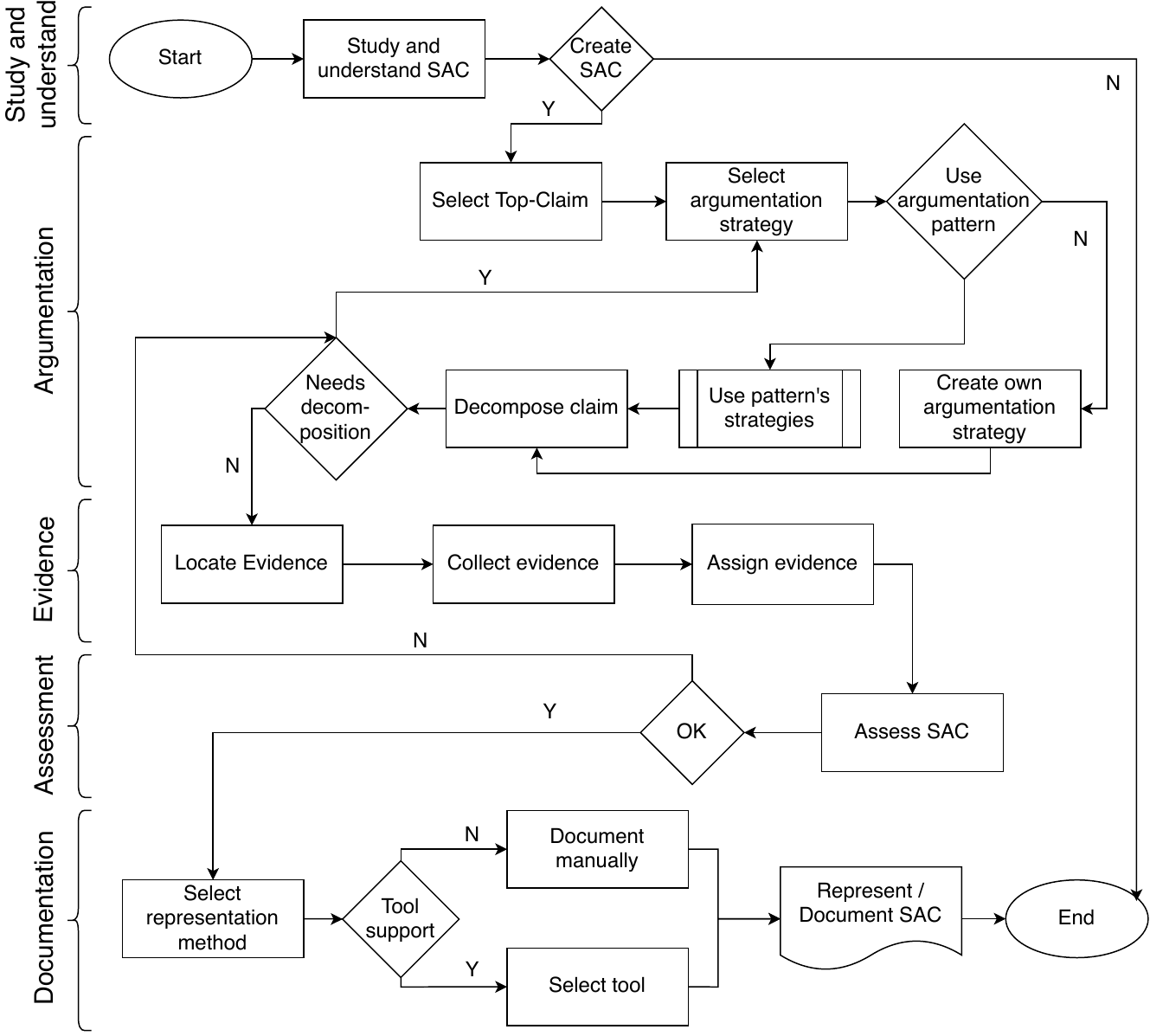}
  \caption{Flowchart of SAC creation}
  \label{fig:flowchart}
\end{center}
\end{figure}

Based on the results of this systematic literature review, we have found that the outcomes described in the literature fall into one or more parts of the workflow depicted in Figure~\ref{fig:flowchart}.

We also realised that there is agreement in the literature that SAC are to be created in a top-down manner. This means that one starts from a top-claim which represents a high-level security goal and work their way through strategies and sub-claims all the way to the evidence. We have not seen approaches that, e.g., start from the existing evidence of a certain system and constructs claims out of them in a bottom-up fashion. However, this agreement is not expressed in sufficient level of detail in any one paper yet.

Hence, we have synthesised the existing knowledge into a generic workflow for the construction of SAC. 
Even though the literature might have some gaps and fallacies, this workflow is useful as a contextual learning guide for the readers to familiarize themselves with the different aspects of SAC creation.

There are five main blocks in the workflow. We will list and describe them in the remainder of this subsection. Additionally, Table~\ref{wf:reading} lists recommended papers from our systematic literature review which focus on aspects related to the individual blocks and together provide a thorough investigation of each.

\paragraph{Study and understand SAC:} Building SACs is not trivial and requires significant effort. Hence, before going ahead and creating them, it is important to understand what they are and what they can be used for. This step includes studying the structure of SACs, their benefits, what needs to be in place to create them, and their potential usage scenarios, e.g., standards and regulation compliance. Block number 1 in Figure~\ref{fig:flowchart} shows the corresponding entity in the workflow.

\paragraph{Argumentation:} This block includes selecting the top claim to achieve, and the strategy to decompose this claim into sub-claims. This is a very important step, as selecting an argumentation strategy decides to a big extent which activities are needed to complete the SAC. For example, if a strategy using decomposition based on vulnerabilities is adopted, a vulnerability analysis of the system in question has to be conducted. 
A sub-block of the argumentation is the usage of patterns. Patterns help the creators of SACs to save time and effort by using pre-defined and proven structures. The creators could, however, decide not to use a pattern, and create their own unique structure if the situation requires that.
A pattern is created based on the knowledge gathered while creating SACs. It is outside the scope of this workflow. However, this is discussed in the recommended papers in Table~\ref{wf:reading}.

\paragraph{Evidence:} This block includes locating, collecting, and assigning evidence to the claims of the SAC. 
In some cases, the evidence is not present when the SAC is being built; hence, they need to be created. In our workflow, this would be a part of the \texttt{collect evidence} activity. Moreover, these activities might be done in an iterative manner, including the assessment.

\paragraph{Assessment:} This block focuses on assessing SACs. This is done to check the quality of the created SAC, and, e.g., to determine whether a claim needs extra evidence to reach a certain confidence level. Assessment starts after the claims have been identified and the evidence is assigned to the corresponding claims. The result of this step might require the creators of the SAC to go back to the point where they assess a claim and make a decision whether or not to further decompose it or assign evidence to it.
Since there is a lack of studies that focus on quality assurance of SAC, we have recommended studies which include some metrics to help assessing SACs in Table~\ref{wf:reading}.

\paragraph{Documentation:} This block includes making a decision of whether or not to use a tool for modelling the argument and documenting the SAC. If a tool is used, then the notation to be used is limited to the one/s supported by the tool. If the documentation will be created manually, the creators will have the freedom to use an existing notation, extend one, or even create their own.

\begin{longtable}{ll}
\label{wf:reading}\\
\caption{Recommended reading material for each block of the SAC creation workflow}\\
\toprule
\textbf{}\textbf{Block} & \textbf{Recommended reading}  \\
\toprule
\endfirsthead
\multicolumn{2}{c}%
{\tablename\ \thetable\ -- \textit{Continued from previous page}} \\
\toprule
\textbf{} \textbf{Block} & \textbf{Recommended reading}  \\
\toprule
\endhead
\midrule \multicolumn{2}{r}{\textit{Continued on next page}} \\
\endfoot
\bottomrule
\endlastfoot
Study and understand SAC  & \cite{38_agudo2009,53_alexander2011,10_othmane2014,37_gallo2015,50_knight2015,19_netkachova2016,52_weinstock2007} \\ 

Argumentation & \cite{38_agudo2009,18_othmane2016,40_coffey2014,15_hawkins2015,51_mohammadi2018,7_tippenhauer2014,34_vivas2011,20_xu2017} \\

Evidence & \cite{39_cheah2018,11_chindamaikul2014,56_lipson2008,36_shortt2015} \\

Assessment & \cite{11_chindamaikul2014,25_rodes2014} \\
Documentation & \cite{18_othmane2016,47_ionita2017,26_poreddy2011,7_tippenhauer2014} \\
\end{longtable}

\section{Discussion}\label{sec:disc}
In this section we discuss the main findings and insights we gathered while reading the papers included in this study.
In summary the main observations are the following:
\begin{itemize}
    \item There are potential benefits of SAC adoption, but further investigation is need.
    \item There is a rich variety of approaches, with room for improvement.
    \item Knowledge transfer from the safety domain should take into consideration differences between safety and security.
    \item There is a lack of quality assurance of the outcomes, which should be avoided in future studies.
    \item There is imbalanced coverage in literature, which requires more academic research.
    \item There is room for improvement when it comes to support, which requires companies or the open-source community to step up.
    \item There is a lack of a mature guidelines for SAC adoption, which might require a standardization activity.
\end{itemize}

\subsection{Potential for a wide range of benefits}
\label{disc:potential}
The literature is full of motivations for using SAC, as well as suggestions for where to use them, as our results of RQ1 show in Sections~\ref{subSec:res:motivation} and \ref{subSec:res:usage}. 
We consider this to be a positive factor. 
However, our impression is that these motivations are on a high level and lack detailed studies to show how realistic and applicable they are.
For example, many papers motivate the adoption of SAC as a way to establish compliance with regulation and standards without pinpointing the regulations' and standards' specific constraints for using SAC. Without having an in-depth knowledge of the specific regulation or standard, it is hard to determine whether SACs are explicitly required, or rather just recommended as a way to create a structured argument for security.
Incidentally, in our own previous work we have tried to demystify this issue in the context of automotive systems \cite{industrialNeeds}. 

Furthermore, some studies suggest that SACs can be used for evaluating the level of security of a system by assigning measurements to the elements of the cases, e.g., the evidence. However, these studies lack detailed guidelines of how to create these quantitative attributes. An example is the usage scenario that suggests to use attribute values for evaluating an architecture~\cite{49_yamamoto2015}. The approach suggests to assign values to evidence, ranging from $-2$ to $2$, based on how satisfying the evidence are to the claims, i.e., to what degree the provided evidence justify the claims. However, there is no specific criteria for determining the attribute value, making the exercise subjective and nontransparent.

Another example is when SACs are suggested as tools to aid in information security education~\cite{37_gallo2015}. This very interesting concept is not supported by a discussion on the required level of detail in the SACs presented to students.

We believe that there is a substantial gap between the potential of SAC reported in literature and their application in industry. An obvious question to ask is: why are SACs not more widely adopted in industry even though there are so many motivations and usage scenarios for them in literature? It has already been shown that adopting SACs is non-trivial~\cite{industrialNeeds}. It requires a substantial amount of effort and time, which grows as the systems become more complex. It also comes with many challenges, such as finding the right expertise to create them. Furthermore, the challenges do not stop at the creation of SACs, but are extended to updating, maintaining, and making them accessible at the right level of abstraction to the right users. We believe that these matters need to be addressed in studies that suggest the usage of SACs in different domains.

\subsection{Wide variety of approaches}
\label{disc:approaches}
The literature includes a rich variety of studies which explore approaches for creating SACs, especially when it comes to the argumentation part, as shown in the results of RQ2 in Section \ref{res:approaches}.
This gives organizations the possibility to choose those approaches that fit their way of working and the security artefacts they produce.
For example, a company that works according to an agile methodology could choose to adopt an SAC approach for iterative development~\cite{18_othmane2016}.
However, this choice has to consider constraints of the applicability of the approach, including benefits and challenges of its adoption. 
These aspects are not discussed in the literature and the burden is left to the adopter.

Another example is the question of conformance with different standards. While this has been discussed in literature, there is a lack of studies which systematically assess different approaches based on their ability to help achieving conformance with a certain standard. To generalize this, we observed that there is a lack of studies which compare different approaches in different contexts. In consequence, from an industrial perspective, organizations need to select suitable approaches in an exploratory way, which can be confusing. 

The studies presenting new approaches also lack the discussion of the granularity level that is possible, or required to achieve using each approach. We believe that future studies should take into consideration the possible usages for SACs created using different approaches, and discuss the required granularity level based on that. For example, would a SAC created through the security assurance-driven software development approach \cite{34_vivas2011} be useful to companies which outsource parts of their development work to providers? In that case, on which level should these cases be created, e.g., on the feature level or on the level of the complete product?

Lastly, we believe that there is room for exploration of hybrid approaches which combine two or more of the approaches reported in literature. This becomes especially important when different approaches target individual parts of SAC, e.g., argumentation and evidence.

\subsection{Security might differ from safety}
We have seen in many cases that the approaches presented in literature treat security and safety cases as the same, e.g., \cite{11_chindamaikul2014,46_graydon2013,15_hawkins2015,23_sljivo2016,3_goodger2012,35_fung2018,9_ankum2005,24_gacek2014,31_sklyar2019,28_sklyar2017}. 
We believe that since assurance cases in general are mature in the safety domain and have been used for a long time, it is natural to consider the gained knowledge and transfer it into other domains, such as security. 
However, this knowledge transfer has to take into consideration the differences between safety and security, e.g., in terms of field maturity and nature. 
For example in safety, there is usually a wide access to information in contrast to security, where threat and risk analysis are considered sensitive information \cite{safetyvssecurity}. 
%Although we decided to include these studies in this SLR, we still think that their applicability in security needs further investigation, taking into consideration the differences between the domains. 
Alexander et al.~\cite{53_alexander2011} provide a discussion on the differences between safety and security both from theoretical, and practical aspects. Other studies combine security and safety assurance by creating combined arguments or security-informed safety arguments~\cite{48_taguchi2014,19_netkachova2016,2_cockram2007,14_netkachova2015}. We have also seen that some studies use different types of assurance cases to argue for security in Section \ref{res:approaches}. The results do not show any noticeable differences to SAC. This means that we were not able to find any special characteristics in the different types of ACs that distinguish them from SAC, when they are applied on security. However, the approaches for creating the argumentation part differ among the types according to their focus, e.g., trustworthiness and depend-ability.

\subsection{Lack of quality assurance}
Quality assurance is the weaker part of the literature reviewed in this study. We talk here about three main things. First is the quality of the outcomes when it comes to their applicability in practice. We have seen in the results of RQ4 in Section~\ref{res:validation} that \emph{illustrative cases} are often preferred over types of more empirically grounded validation. This indicates scarcity of industrial involvement. The reason might be a lack of interest, which contradicts with the reported motivations and usage scenarios, or simply because it is hard to get relevant data from industrial companies to validate the outcomes, as security-related data is considered to be sensitive (as we mentioned earlier). Furthermore, with the exception of a few cases, the creation and validation of SAC in literature is done by the authors of the studies. We believe that this contributes heavily to the lack of information addressing challenges and drawbacks of applying SACs in a practical context.

The second issue is the generalize-ability of the approaches with regards to their used argumentation strategies. The approaches we reviewed use a wide variety of argumentation strategies, e.g., based on threat analysis, requirements, or risk analysis. However, they lack validations and critical discussions as to whether the approaches work only with the used strategies or can use other strategies as well. We suggest to validate these approaches based on different types of strategies in future research.

The last point is the lack of mechanisms for building-in quality assurance within the SACs. We believe that it is essential for the argumentation provided in SACs to be complete in order for them to be useful. For that there needs to be a mechanism to actively assess the quality of the arguments to gain confidence in them. This is not addressed in literature apart from a few studies, e.g., \cite{11_chindamaikul2014,25_rodes2014}.  
Similarly, the evidence part also needs to be assessed. e.g., by introducing metrics to assess the extension to which a certain evidence justifies the claim it is assigned to. The inter-relation between claims and evidence need to be addressed. For example each claim can have a certain saturation level to be achieved, and each evidence provides a degree of saturation. Hence, it would be possible to assess whether the claim is fully satisfied or not by the assigned evidence.

\subsection{Imbalance in coverage}
\label{disc:coverage_slr}
The coverage of matters related to SAC in literature is imbalanced to a large extent. When it comes to the approaches, our results in \ref{res:coverage} indicate a tendency towards covering the argumentation part more than the evidence part. This indicates a weakness in the approaches, as elements of SAC cannot be evaluated in silos. For example, if we take an approach to create security arguments, how would we know which evidence to associate with these. Moreover, we will not be able to assess whether we actually reach an acceptable level of granularity for the claims to be justified by evidence. Same thing applies for the evidence part. If we only look at the evidence we will not be able to know which claims the suggested evidence can help justify. To be able to evaluate the evidence, they have to be put in context with the rest of the SAC. When reviewing the studies that focus on one element of SAC, we were not able to find any links to related studies focusing on the remaining elements, which indicates incompleteness of the approaches especially for putting them into practice.

When it comes to other areas, the assessment and quality assurance of SAC is rarely covered, as we discussed in the previous sub-section. Furthermore, there is a lack of studies covering what comes after the creation of SAC. In particular, for SAC to be useful, they have to be updated and maintained throughout the life-cycles of the products and systems they target, otherwise, they become obsolete \cite{industrialNeeds}. Particularly, there need to be traceable links between the created SACs and the artefacts of these products and systems. Many SAC approaches use GSN, which allows to reference external artefacts using the context and assumption nodes. However, these nodes are rarely exploited in the examples provided in the studies we reviewed. 

\subsection{Room for support improvement}
The tools reported in literature cover activities related to AC, such as creation, documentation and visualization, as shown in the results of RQ3 in Section \ref{res:support}. Some of these tools have features such as the validation of AC based on consistency rules related to the used notation, or even user-specified rules \cite{tool_adrelard}. Other tools assist in the maintenance of AC through change impact analysis \cite{35_fung2018}, and assessment of AC \cite{tool_norsta}. When it comes to automatic creation of SAC, there were only coverage for the argumentation part \cite{7_tippenhauer2014,15_hawkins2015}, which reflects the imbalance in coverage we discussed earlier.

What we observed is that most of the tools are originally created for supporting safety cases, and not in particular SAC. As a consequence, they lack specific features which can be very helpful while building SAC. In particular, we note the fact that security assurance cases need to be treated as living documents (more so than their safety counterparts) due to a continually shifting threat landscape. 
For example, there is no tool that integrates with other security tools, e.g., an intrusion detection system, to actively update evidence. In general, we note that the tools lack integration with other systems, which agrees with what Maksimov et. al.~reported in their study\cite{maksimov2018}.

Moreover, even though some studies have reported the demonstration of created tools using a case study, e.g., \cite{7_tippenhauer2014}, it is not clear how flexible they are to be tailored for specific needs of a certain organization, and to be integrated with their tool-chain. We believe that in order for practitioners to use these tools, there needs to be a certain amount of confidence, which is absent due to little reported usage or replications in industry. The same thing applies for the reported patterns. For a specific artefact to be qualified as a pattern, it needs to be used in several studies and in several contexts, which is not the case.   
Additionally, as we discussed earlier, some important aspects, e.g., traceability is not covered in literature, and this is also the case when it comes to the reported supporting tools.

We also believe that there is room for creativity in the development of the tools. For instance, there are no supporting tools which use machine learning techniques to predict whether a requirement or test case qualifies to be a part of a SAC. This opens up opportunities for companies and the open-source community to step up and close the gap between the potential and the current support.

\subsection{Need for a guideline}
Finally, we believe that there is a need for an explicit guideline for on-boarding a SAC-based approach in an industrial context. 
We believe that with the current level of maturity in related literature, companies which want to adopt SAC approaches have to account for a high cost, as they have to learn, experiment and develop a lot internally. This is due to the lack of reported validation and lessons learned from industry, but another sign is the lack of tool support specific for SAC (as mentioned above).

Standardization bodies are aware of the importance of SAC, as they are being mentioned as requirements in some security standards and best practice documents, e.g., the upcoming standard for cyber-security in automotive ISO21434 \cite{iso21434}. However, these standards do not provide any specific guideline or constraints for how SAC should be created and used. It is important that key players in selected domains (e.g., automotive and healthcare) put together efforts to standardize the scope and requirements related to SAC. We believe that this would elevate the maturity in the field.

\section{Validity Threats}\label{sec:valTh}

In this study, we consider the internal and external categories of validity threats as defined in \cite{ttv1}, and described in \cite{ttv3,kh2007guidelines}. 
The work of conducting the review was done by one researcher. This means that applying the inclusion / exclusion criteria in each of the four filtering rounds was done by one person. This imposes a risk of subjectivity, as well as a risk of missing results, which might have affected the internal validity of this study. To mitigate this, a preliminary list of known good papers was manually created and used for a sanity check of the selected and included papers. Additionally, a quality control was performed periodically by the other authors to check the included and excluded studies.

Restricting our search to three digital libraries could have increased the probability of the risk of missing relevant studies. This was mitigated by performing the snowballing search to search for papers that are not necessarily included in the databases of the three considered libraries.

Another threat to validity is publication bias \cite{kh2007guidelines}. This is due to the fact that studies with positive results are more likely to get published than those with negative results. This could compromise the conclusion validity of this SLR, as in our case we did not find any study that is, e.g., against using SAC, or which reported a failed validation of its outcome. 
In our study, we have partially mitigated this threat by also including a few technical reports (i.e., non peer-reviewed material). These papers have been identified as part of the snowballing, as we did not restrict to peer-reviewed papers.

External validity depends on the internal validity of the SLR \cite{kh2007guidelines}, as well as the external validity of the selected studies. We did scan gray literature to mitigate publication bias, but we excluded studies that are under 3 pages, and old studies to mitigate the risk of including studies with high external validity threats.

When it comes to the reliability of the study, we believe that any researcher with access to the used libraries will be able to reproduce the study, and get similar results plus additional results for the studies which get published after the work of this SLR is done.

\section{Conclusion and future work}\label{sec:con}

In this study, we conducted a systematic review of the literature on security assurance cases. We used three digital libraries as well as snowballing to find relevant studies. We included 51 studies as primary data points, and extracted the necessary data for the analysis. 

The main findings of our study show that many usage scenarios for SAC are mentioned, and that several approaches for creating them are discussed. However, there is a clear gap between the usage scenarios and approaches, on one side, and their applicability in real world, on the other side, as the provided validations and tool support are far from being sufficient to match the level of ambition. 
Based on the results of this systematic literature review, we created a workflow for working with SAC, which is a useful tool for practitioners and also provides a guideline on how to approach the study of the literature, i.e., which paper is relevant in each stage of the workflow.

Based on our results and findings, in the future we will be working to close the gap between research and industry when it comes to applying security assurance cases. We will be looking into exact needs and challenges for these cases in specific domains, e.g., automotive. We believe that introducing SAC in large organizations needs appropriate planning to, e.g., find suitable roles for different tasks related to SAC, and integrating with current activities and way of working. Hence, we see a potential direction of future work in that area. 

When it comes to the technical work, we believe that there is room for improvement in the approaches for SAC creation, especially when it comes to the evidence part. For instance, a possible future work direction is to look into ways to automatically locate, collect, and assign evidence to different claims.

Finally, we believe that quality assurance of SAC has not been addressed sufficiently in literature. As a future work, we will look into ways to ensure the completeness of a security case when it comes to the argumentation, as well as the confidence in how well the provided evidence justify these claims.

\chapter{Paper C}
\label{chap_paper_c}
\thispagestyle{empty}
\subsection*{Asset-driven Security Assurance Cases with Built-in Quality Assurance}
\subsubsection*{M. Mohamad, Ö. Askerdal, R. Jolak, J.P. Steghöfer, R. Scandariato}
\subsubsection*{{\em In International Workshop on Engineering and Cybersecurity of Critical Systems (ENCYCRIS), 2021}.} \cite{paperC}
\newpage
\thispagestyle{empty}
\mbox{}
\newpage
\addtocounter{page}{-2}
\newpage
\section*{Abstract}
Security Assurance Cases (SAC) are structured arguments and evidence bodies used to reason about security of a certain system.
SACs are gaining focus in the automotive domain as the needs for security assurance are growing. % internally and externally.
%objective
In this study, we present an approach for creating SAC. The approach is inspired by the upcoming security standards ISO/SAE-21434 as well as the internal needs of automotive Original Equipment Manufacturers (OEMs).
%method
We created the approach by extracting relevant requirements from ISO/SAE-21434 and illustrated it using an example case of the head lamp items provided in the standard.
%results
We found that the approach is applicable and helps to satisfy the requirements for security assurance in the standard as well as the internal compliance needs in an automotive OEM.\\
\linebreak
\textbf{Keywords: } security; assurance cases; automotive systems

\newpage

\section{Introduction}
%SAC
Assurance cases are structured bodies of arguments and evidence used to reason about a certain property of a system. Security Assurance Cases (SAC) are a type of assurance case for the field of cyber-security.
%Importance to industry
%Why security assurance is important for organizations producing safety critical systems.
%In the past years, a substantial amount of work has been done to study and apply safety cases for safety critical systems. 
In this paper, we turn our attention to the creation of a SAC, with particular focus on the domain of automotive applications. As vehicles become more advanced and connected, security scrutiny has increased in this domain. Furthermore new standards and regulations push towards assuring security for vehicular systems by using SAC. 
Similarly to safety cases, which are required in safety standards, e.g., ISO-26262~\cite{iso26262_2ed}, SACs are explicitly \emph{required} in ISO/SAE-21434~\cite{iso21434}. 
Additionally, SACs are required for all systems in production.

%Conformance with standards is not only an external need for automotive OEMs, but also an internal need. In our previous work \cite{industrialNeeds}, we have identified internal drivers for SAC in the automotive domain, and compliance was one of the top five needs. 

%Lack of approaches to achieve compliance for automotive companies
In literature, there are some studies that suggest the creation of SAC based on requirements derived from security standards~\cite{9_ankum2005,1_cyra2007}Cover . However, there is no approach which helps achieving conformance with the upcoming ISO/SAE-21434 standard. Additionally, since the requirements for SAC are new, there is no evidence in the literature that the knowledge base in industry is mature enough to achieve conformity to these requirements.
%Lack of approaches which build quality in the assurance cases themselves.
%Additionally,
Moreover, quality assurance of the SACs is missing in the reported approaches in literature, even though it is a very important aspect. In order for different stakeholders to use an SAC, it is essential to trust that the SAC's argument is built with a sufficient level of completeness, and that the evidence provides a sufficient level of confidence to actually justify the targeted claims.
%Important to close the gap to industry. Hence, we work with a large OEM in Sweden.
Finally, we identified that the lack of industry involvement is a significant issue in current approaches. This results in gaps between research and industry. 

%What we did
To bridge these gaps, we have worked together with Volvo Trucks, an international automotive OEM, to develop CASCADE, the asset-driven approach for SAC creation presented in this paper. CASCADE is based on the requirements and work products of ISO/SAE-21434. It is asset-driven, i.e., the resulted SACs have assets as drivers of the structure of the security arguments. Therefore, it allows creating security assurance based on what is valuable in the system. 
%CASCADE was built in collaboration with a major automotive OEM, and takes the internal needs of the OEM into consideration. 
Additionally, we integrated quality assurance in SACs created with CASCADE by distinguishing between product-related claims and quality claims, as well as building arguments for both.

From a methodological standpoint, we created and validated our approach as follows. First, we created a high-level structure of an asset-driven SAC, which included the identification of the assets, the tracing of such assets to system elements (e.g., processing, communication, and storage operations), and the identification of the relevant security assets for each asset. Second, we analyzed the ISO/SAE-21434 standard and extracted the requirements and work products that are relevant to SAC. Accordingly, we mapped the extracted items to the elements of our asset-driven SAC. We then illustrated the approach using the exemplary case study mentioned in ISO/SAE-21434. Finally, we presented the resulting approach to the security experts from an industrial automotive OEM and gathered their feedback. The results are presented in Sections~\ref{sec:app}--\ref{sec:validation}, after discussing the related work in Section \ref{sec:rw_cascade}.

\section{Background and Related work}\label{sec:rw_cascade}
%\todo[inline]{Add a background about assets in automotive}%What are assets? reference 
In this section, we provide background information about SAC, security assets in automotive, and %related
their corresponding security threats and attacks.
% on these assets.
We also review related work of asset-based approaches in literature.

\subsection{Security Assurance Cases} \label{subsec:sac_bg}
%A description of what they are and how they look like (from the generic papers).
Assurance cases are bodies of evidence organized in structured arguments, used to justify that certain claims about a systems property hold \cite{goodenough2007}. 
The argumentation in a Security Assurance Case (SAC) consists of claims about security for the system in question, and the evidence justifies these security-related claims.
SAC consist of the following primary components: (i) security claims, (ii) the context in which the claims should hold, (iii) an argument about the security claim, (iv) the strategy used to build the argument, and (v) a body of evidence to prove the claims \cite{knight2015,alexander2011}.
SAC can be expressed in a textual or graphical format \cite{alexander2011}. The most common graphical formats are the Goal Structure Notation (GSN, \cite{gsn}), and the Claims, Arguments, and Evidence notation (CAE, \cite{CAE}). 
%(focusing on security) Which aspects we want to highlight (Main decomposition strategy: security goals vs security threat) or other paper that argue over processes.

%Also add papers that take arguments from standards.
\subsection{Automotive Assets and Related Security Threats %and Attacks
}
According to~\cite{rosenstatter2020remind}, there are four categories of assets in automotive systems that are targeted by security threats and attacks. 
These assets are hardware, software, network and communication, and data storage. 
\begin{itemize}
    % 1
    \item \emph{Hardware}: This asset category includes sensors, actuators, and the hardware part of the Electronic Control Units (ECUs).
    These assets are often threatened by disruption or direct interventions that influence their availability and integrity.
    Examples of attacks on these assets include fault injection and information leakage.
    
    % 2
    \item \emph{Software}: This category includes external libraries, Operating Systems (OS), applications, virtualization, and the software part of the ECUs.
    Security threats and attacks on software assets include the manipulation of software, such as tampering attacks which often target software availability and integrity.
    
    % 3
    \item \emph{Network/Communication}: Refers to internal or external communication.
    Internal communication assets are busses such as CAN, FlexRay, LIN, MOST, and automotive Ethernet.
    External communication assets are WiFi, Bluetooth, and Vehicle to Everything (V2X) communication. 
    Examples of attacks on these assets includes fabrication or jamming attacks, spoofing, message collision, eavesdropping, hijacking, and denial of service (DoS).
    These attacks target the confidentiality, integrity, availability, and privacy of the these assets.
    
    % 4
    \item \emph{Data storage}: Sensitive data including user data, backups, cryptographic keys, forensics logs, and system information and reports. 
    These assets are targeted by unauthorized access and malicious manipulation that often influence the confidentiality, integrity, availability, and privacy of the data.

\end{itemize}

\subsection{Asset based approaches}
Researchers have been exploring several asset-based approaches for creating the argument part of SAC. Biao et al.~\cite{20_xu2017} suggest dividing the argument into different layers, and using different patterns (one per layer) to create the part of the argument that corresponds to each layer. 
Assets are considered as one of these layers, and the pattern used to create it includes claims that the assets are ``under protection'', and strategies to break down critical assets. Biao et al.~\cite{20_xu2017}, however, do not consider the quality of the cases and only focus on creating arguments without touching upon the evidence part.
%This paper does not regard QA nor evidence.

%Another approach was used by Poreddy et al.~\cite{poreddy2011}, which uses different security properties as argumentation strategy. Other used approaches argue by development life-cycle phases \cite{ray2015}, standard recommendations \cite{he2012}, and product components \cite{hawkins2015}.
%2 papers from Mazen's SLR

Luburic et al.~\cite{luburic2018} also present an asset-centric approach for security assurance.
The info used in their approach is taken from: \emph{(i)} asset inventories; \emph{(ii)} Data Flow Diagrams (DFD) of particular assets and the components that manipulate them; and \emph{(iii)} the security policy that defines protective mechanisms for the components from the previous point.
They propose a domain model where assets are the center pieces. The assets are linked to security goals. The argument considers the protection of the assets throughout their life-cycles by arguing about protecting the components that store, process, and transmit those assets.
The SAC they provide is very high level and includes two strategies: ``reasonable protection for all sensitive assets'' and arguing over the data-flow of each related component.
The authors illustrate the approach with a conference management system example. They state that the main limitations of their are asset and data flow granularity.
In our study, we consider the the assets to be the driver of our approach, but we extend the argument to reach the level of concrete security requirements. We also derive our strategies from an industrial standard and validate our approach in collaboration with an OEM. Furthermore, we extend our approach to include case quality aspects.
\section{CASCADE}\label{sec:app}

\afterpage{
\begin{figure}
\begin{center}
  \includegraphics[width=0.9\linewidth]{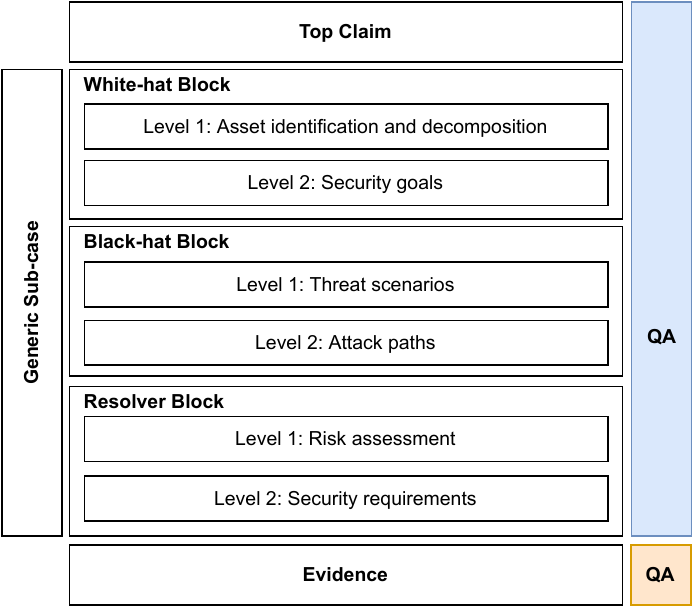}
  \caption{The CASCADE approach for creating security assurance cases} 
  \label{fig:assetApproach}
\end{center}
\end{figure}
}
%\todo[inline]{Talk about reusability and generic arguments.}

In general terms, assets are artifacts of interest to a certain entity. In computer security, these artifacts can be hardware, software, network and communication, or data~\cite{rosenstatter2020remind}. The importance of assets makes them the target of attackers. 

The CASCADE approach for creating security assurance cases takes the importance of assets to organizations into consideration. Hence, it builds the argumentation by putting assets in focus, with the goal to show that these assets are secure from cyber security attacks. Our aim is to prove that a given artefact is secure by arguing that its assets are secure.

An important design principle in the CASCADE approach is the integration of quality assurance of the cases in terms of argumentation completeness and evidence confidence. Each level of argumentation (i.e., strategy) is associated with at least one claim about completeness, and each level of evidence is associated with at least one claim about confidence. A similar concept is used by Hawkins et al.~\cite{hawkins2011} to argue about the confidence of safety cases.

\subsection{Elements of an SAC in CASCADE}
We use GSN \cite{gsn} to create SAC using the CASCADE approach. The elements of the notation are:
%described in Section~\ref{subsec:sac_bg}.\todo[inline]{No, they are not!}
\emph{(i)} Claim \footnote{In GSN, the terms goal and subgoal are used to refer to high and low abstraction levels of argumentation claims respectively. To avoid confusion, we refer to these as claims.}: a security claim about the artefact in question;
\emph{(ii)} Strategy: a method used to decompose a claim into sub-claims;
\emph{(iii)} Evidence / Solution: a justification of a Claim / set of claims;
\emph{(iv)} Context: used to set a scope of a given claim;
and \emph{(v)} Assumption: used to document the assumptions made for a certain claim.

In addition to these, we have created additional types of elements to be used in our approach:
\emph{(vi)} Case Quality-claims (CQ-claims): represent claims about the quality of the created case itself;
\emph{(vii)} Case Quality-evidence (CQ-evidence): represent evidence used to justify CQ-claims; and
\emph{(viii)} Generic sub-case: consist of generic claims, strategies, contexts, assumptions, and evidence that are not bound to a specific artifact, but instead are applicable to a wider range of artifacts in the context of a product, program, or organization. 

%What are assets
\subsection{Building blocks of the CASCADE approach}
The asset-based approach consists of building blocks, as shown in Figure~\ref{fig:assetApproach}. Each block contains a sub-set of the case. In the following sub-sections, we explain the blocks and their contents.

\subsubsection{Top claim}
This block consists of the top security claim of the artefact in question. It also includes the context of the claim and assumptions made to set the scope of the claim. If we are considering a software system, e.g., we might make an assumption that the hardware is secure.
The top claim differs between different organizations and drives the granularity of the SAC. For example a service provider might consider the security of a service to be the top claim, but an automotive OEM might need to consider the whole vehicle's security, which requires the incorporation of different services or user functions. Similarly, depending on the intended usage of the SAC, the top claim might include back-end systems, or only on-board systems. For example to assure the security of a complete vehicle, it is important to make sure that not only the vehicle's components are secure, but also the back-end systems which communicate with the vehicle. In contrast, to ensure that a certain end-user function in the vehicle is secure, it might be enough to only consider the corresponding sub-systems in the vehicle itself.

\subsubsection{Generic sub-case}
This block contains a sub-case that is applicable not only to the artefact for which the SAC is being created, but instead to a larger context. For example, if a company defines a cybersecurity policy, enforced by cybersecurity rules and processes, then the policy can be used in security claims for all its products. These claims can be re-used when creating SAC for individual artefacts. Another example is when certain claims can be made on a product level. Then these claims can be reused for all SAC of individual components of that product. Our aim with this block is to make the approach scalable in larger organizations with complex products and multiple teams. Each team can work on a part of the SAC which corresponds to their artefact. On a higher level, these SAC can be combined together, and generic arguments that are applicable to the sub-SAC can be provided.

\subsubsection{White-hat block}
This block starts with the identification of assets, which is the driver of our approach.
Asset identification is done by conducting an analysis to find the artefacts of the system that are likely to be subject to an attack.

When the assets are identified, they can be further decomposed during the different phases of the development life-cycle. For example in an OEM, a high-level asset analysis is done at the concept phase, and later a low-level analysis is conducted during implementation, where more information about the assets and their usage is known.

\paragraph{Linking assets to higher-level claims} To link the assets to the main claim, we identify which assets exist and which components use or have access to these assets. For example, in a vehicle, the driver's information can be considered an asset which is accessible by the infotainment system of the vehicle. Hence, we link this asset to the claims of the security of the infotainment system. To make this more concrete, we look at the traceability of the asset. For example, we consider the assets \emph{(i)} ``at rest'', which refers to where the assets are stored; \emph{(ii)} ``on the move'', when the asset is in transition between two entities, e.g., when sensor data is being transferred from the sensor to an ECU; and \emph{(iii)} ``in use'', which is when the asset is being used, e.g., when some diagnostics data is being processed by a back-end system. 

\paragraph{Decomposition of assets} To decompose assets, we look into the types of the identified assets. This gives an indicator whether the asset would have implications on the local part of the vehicle (one electronic control unit/ECU), or on a bigger part of the vehicle (multiple ECUs). We also look into the relations among assets, e.g., dependability.

\paragraph{Linking assets to the lower level} To link the asset to the lower level in the approach, i.e., the security goals, we identify the relevant security properties for the assets. Specifically, we look into the Confidentiality, Integrity, and Availability (CIA) triad. For example, the vehicle engine's start functionality is an asset which has relevant integrity and availability properties.  

\paragraph{Identification of security goals} When we have identified the relevant security properties for each asset, we create claims representing the security goals\footnote{A security goal is preserving a security concern (CIA) for an asset \cite{secGoals}}. Following our example of the engine start request, a claim about the achievement of a security goal would be that the availability of the request is preserved. One combination of asset/security property might lead to several goals, for example that the engine start is available using a connected mobile app and a web portal. 
To make sure the relevant properties are covered when identifying security goals, we consider damage scenarios that lead to compromising the security goals, e.g., that the engine start request is unavailable, or an unintended start of the engine occurs, which would damage the integrity of the asset.
%\todo[inline]{The difference between claim and security goal is not clear. Is there one? Why is the ability to start a vehicle with a key a security goal?}

\subsubsection{Black-hat block}
In this block, we aim to identify the scenarios that might lead to not fulfilling the identified security goals and hence cause harm to our identified assets. 

%\paragraph{Identification of damage scenarios:} Damage scenarios are scenarios that lead to compromising of a security goal. For example that the engine start request is unavailable, or an unintended start of the engine occurs. In our SAC, the scenario is translated to a claim by negating the possibility of the damage scenario, e.g., ``Unintended request for engine start is not possible''. 

\paragraph{Identification of threat scenarios} When we have identified the claims about the achievement of security goals, we proceed by identifying the threat scenarios and creating claims for negating the possibility of these scenarios. We connect these claims to the corresponding claims about achieving security goals. 
For example, a claim handling a threat scenario connected to the claim ``Unintended request for engine start is not possible'', might be identified by considering a threat model, e.g., STRIDE~\cite{howard2006security}. %\todo{Citation needed}. 
Hence a claim might look like: ``Spoofing a request for engine start is not possible''. 

\paragraph{Identification of possible attack paths} In this step, we identify possible attack paths which can lead to the realization of a threat scenario. Each threat scenario might be associated with multiple attack paths. We then claim the opposite of these attack paths. An example of an attack path is ``An attacker compromises the cellular interface and sends a request to start the engine'', and the claim would be to negate the possibility for that.

\subsubsection{Resolver block}
This block is the last one in the argumentation part of the CASCADE approach. It links the claims derived from the attack paths to the evidence.

\paragraph{Risk assessment} In this level, we assess the risk of the identified attack paths. Based on the risk level, the creators of the SAC create claims to treat the risk by, e.g., accepting, mitigating, or transferring it.

\paragraph{Requirements} At this point, requirements of risk treatments identified in the previous level are to be expressed as claims. This level may contain multiple decomposition of claims, based on the level of detail the creators of the SAC wish to achieve, which is driven by the potential usage of the SAC. For instance, if the SAC is to be used by a development team to assess the security level, this might require a fine grained requirement decomposition which might go all the way to the code level. In contrast, if the SAC is to be used to communicate security issues with outside parties, a higher level of granularity might be chosen. In either case, it is important to reach an \emph{``actionable''} level, meaning that the claims should reach a point where evidence can be assigned to justify them.

%An asset analysis might consist of one or more of the following items:
%\begin{itemize}
%    \item Asset id: what assets exist and are identified.
%    \item Asset relations: the relationships among the assets, e.g. dependability.
%    \item Asset usage: which components use or has access to the asset.
%%    \item Asset traceability (where the assets are). For example:
%%        \begin{itemize}
%            \item While asset is “at rest”. This is when the asset is stored, e.g. in the storage unit of an ECU.
%            \item While asset is “on the move”. This is when the asset is in transition between two entities. For example when a sensor data is being %transferred from the sensor to an ECU.
%            \item While asset is “in use”. This is when the asset is being used. For example when some diagnostics data is being processed at a backend %system.
%        \end{itemize}
%    \item Asset type (signals, user data… etc). This also indicates whether the asset would have implication on the local part of the vehicle (one ECU), %or on a bigger part of the vehicle (multiple ECUs).
%    \item Asset's relevant security properties, e.g., CIA.
%\end{itemize}

%When the assets are identified, we identify the security goals for each of the identified assets. 
%This is done by considering the 

\subsubsection{Evidence}
The evidence is a crucial part of an SAC. The quality of the argument does not matter if it cannot be justified by evidence. In our approach, evidence can be provided at any block of the argumentation.%\todo{Revise sentence}. 
For example, if it can be proven in the black-hat block that a certain asset is not subject to any threat scenario, then evidence can be provided, and the corresponding claims can be considered as justified.
%\todo[inline]{I don't think the notion of ``branches'' has been introduced yet.} 
If the creators of the SAC cannot assign evidence to claims, this is an indicator that either the argument did not reach an actionable point or that there is a need to go back and make development changes to satisfy the claims. For example if we reach a claim which is not covered by any test report, then there might be a need to create test cases to cover that claim. 

\subsubsection{Case Quality Assurance}
We consider two main aspects of quality assurance for SAC in CASCADE.
The first aspect is \emph{completeness} which refers to the level of coverage of the claims in each argumentation level of the SAC.
%We determine a level in the security assurance case by the strategy used to decompose the claims. 
Each level in CASCADE includes at least one strategy. 
%If a claim afterwards is decomposed using a different strategy, then we consider this to be a different level.
For each strategy, we add at least one completeness claim that refines it.
%In each level of argumentation, we add at least one completeness claim that refines the strategy used in that level. 
The role of this claim is to make sure that the strategy covers all and only the relevant claims on the argumentation level.
The completeness also relates to the context of the argumentation strategy. The context provides the information needed to determine if the completeness claim is fulfilled or not.

The second aspect is \emph{confidence} which indicates the level of certainty that a claim is fulfilled based on the provided evidence.
This is used in each level of a security assurance case where at least one claim is justified by evidence. The confidence aspect is expressed as a claim, which takes the form: ``The evidence provided for claim X achieves an acceptable level of confidence''. What makes an acceptable level of confidence is defined in the context of the strategy. The confidence claim itself must be justified by evidence.

\section{Example Case}\label{sec:illustration}
To validate our approach, we apply CASCADE on the headlamp item use case from ISO/SAE-21434 which includes the headlamp system, navigation ECU, and gateway ECU.
%top claim

\subsection{Top Claim}
We start by constructing the Top Claim block %(See Figure~\ref{fig:ill_topClaim}) 
consisting of:
%, which consists of a top security claim, its context, and the identified assumptions associated with the claim.
%\begin{figure}[t]
%\begin{center}
%  \includegraphics[width=1\linewidth]{Images/topClaim.pdf}
%\end{center}
%        \caption{Top Claim block of the headlamp use case} 
%        \label{fig:ill_topClaim}
%\end{figure}
%Figure \ref{fig:ill_topClaim} shows the top claim block 
\begin{itemize}
    \item \emph{C:1} the top security claim for the headlamp item.
    \item \emph{Cnxt:1.1} a context node setting the scope of the claim.
    \item \emph{Assmp:1.1} an assumption node, stating that the item is physically protected.
\end{itemize}
The context node refers to an external document, which is the item boundary and preliminary architecture of the headlamp item, as identified in ISO/SAE-21434.

%White hat block
\begin{figure*}
\begin{center}
  \includegraphics[width=0.85\linewidth]{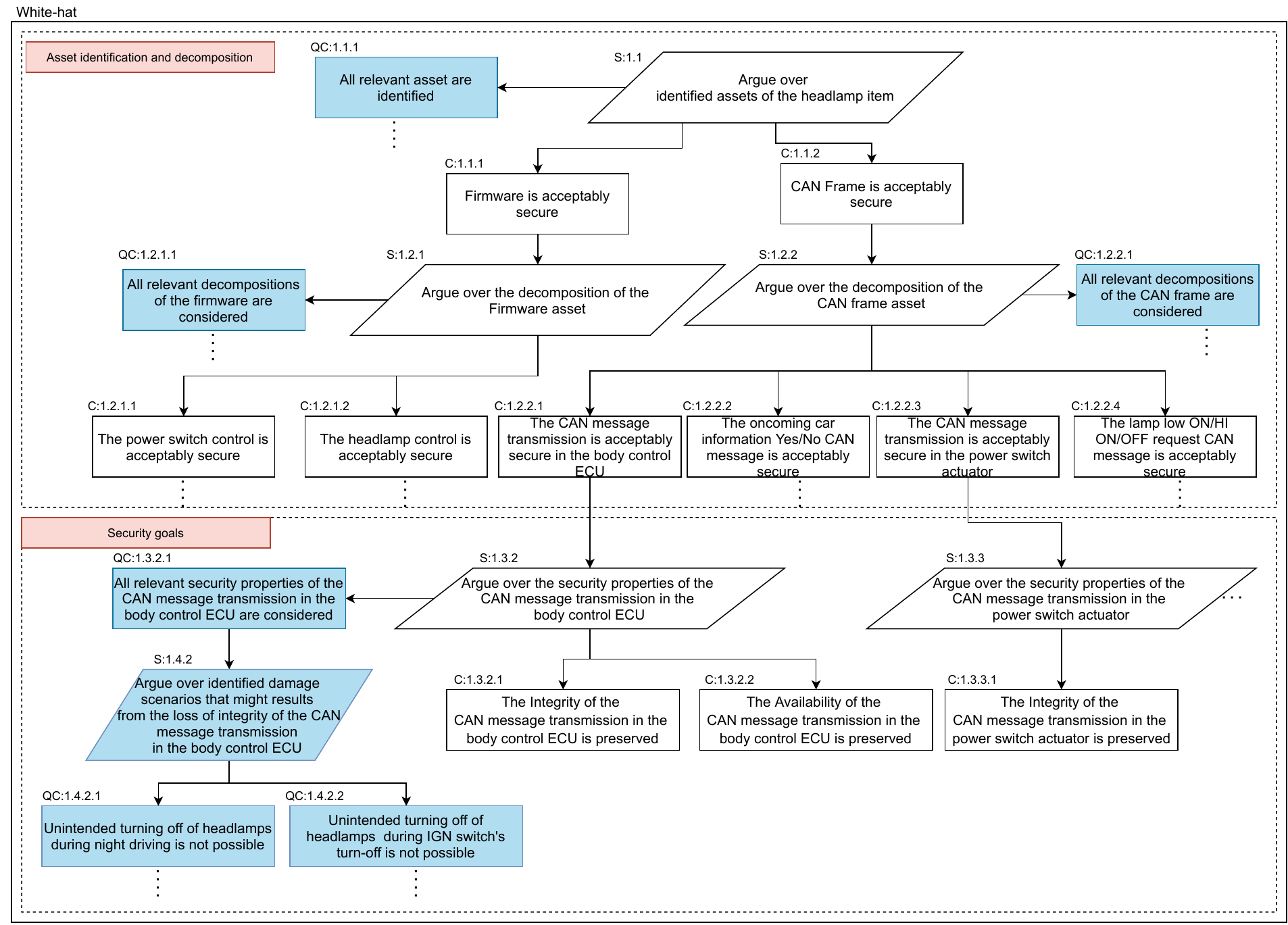}
\end{center}
        \caption{White-hat block of the headlamp use case} 
        \label{fig:ill_white}
\end{figure*}

\subsection{White-hat Block}
%Moving to 
The White-hat block is presented in Figure \ref{fig:ill_white}.
We first apply a strategy \emph{S:1.1} to decompose our main claim based on the identified assets of the headlamp item. 
In our example, the main assets are the \texttt{CAN Frame}, which holds transmitted messages, and the \texttt{Firmware} which includes control functions of the artifacts inside the headlamp system, e.g., the power switch.
We create two claims \emph{C:1.1.1} and \emph{C:1.1.2} indicating that the two assets are acceptably secure. 
The strategy \emph{S:1.1} is associated with a quality claim \emph{QC:1.1.1}, to ensure the completeness of the decomposition associated with it, and hence the completeness of the case in general. 

The two identified assets are further decomposed into sub-assets. 
This decomposition is based on the components and functions the asset belongs to. 
For example, based on claim \emph{C:1.1.2} we apply strategy \emph{S:1.2.2} and decompose the \texttt{CAN Frame} asset into a number of sub-assets.
%CAN message transmission function of the Body Control ECU, and the CAN message transmission function of the Power switch actuator. 
%
Moreover, we create security claims for the identified sub-assets: \emph{C:1.2.2.1}, \emph{C:1.2.2.2}, \emph{C:1.2.2.3}, and \emph{C:1.2.2.4}.
Lastly, strategy \emph{S:1.2.2} is associated with quality claim \emph{QC:1.2.2.1}. 
%, and connected to the claim of the corresponding asset through one of the strategy nodes (S:1.2.1 and S:1.2.2).

At this point, we link the assets to the security goals (i.e., second level). 
%Once we have created the security claims for the sub-assets, 
To do so, we apply an argumentation strategy (e.g., \emph{S:1.3.2}) to decompose the security claims of the sub-assets based on the CIA triad attributes.
%
%We create claims stating that ``the \emph{security property} of the \emph{sub-asset} is secure''. 
As a result, we create claims about the achievement of security goals such as \emph{C:1.3.2.1}: ``\emph{The integrity of CAN message transmission in the body control ECU is preserved}''.%\rodi{update fig 2.}
To make sure that we cover the relevant properties, we create a quality claim \emph{QC:1.3.2.1} and argue (\emph{S:1.4.2}) about possible damage scenarios that could invalidate the claims.
%
%We then
Accordingly, we create quality claims which make sure that these damage scenarios do not happen. 
An example of theses claims is \emph{QC:1.4.2.1}: ``\emph{Unintended turning off of headlamps during night driving is not possible}''. 
%At this point, we reach our security goals for the system. %\todo[inline]{Does that mean that the above claim is a security goal? Then say that:}
At this point, the claim is fine-grained enough and counts as a security goal.
Next, we create the black-hat block.

\subsection{Black-hat Block}
%After identifying the claims that correspond to the security goals of the item in the top claim, we start to 
Here we argue over the threat scenarios that could lead to compromising a security goal.  
%For simplicity, we did not include the whole argument of the white-hat block in Figure \ref{fig:ill_white}, but rather a subset of it. This applies to all Figures depicting the example case.
%Black hat block

\begin{figure*}
\begin{center}
  \includegraphics[width=0.85\linewidth]{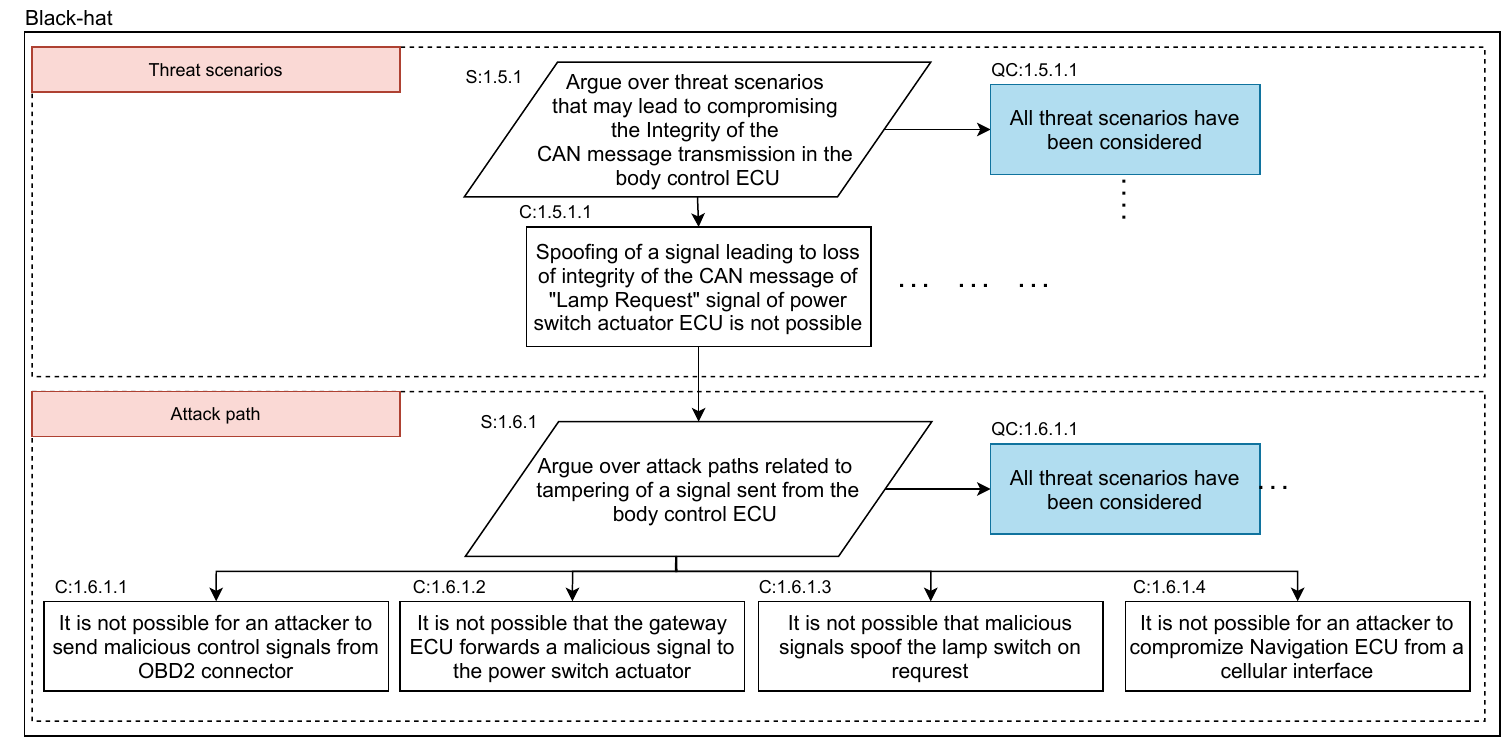}
\end{center}
        \caption{Black-hat block of the headlamp use case} 
        \label{fig:ill_black}
\end{figure*}

%
Figure \ref{fig:ill_black} shows a part of the black-hat block of the headlamp use case. 
This part is associated with the claim about achieving a security goal \emph{C:1.4.2.1} that is shown in Figure \ref{fig:ill_white}.
%Now that we have identified the claims which correspond to the security goals of the item in the top claim, we start to argue over the threat scenarios that could lead to a damage scenario, and hence compromise a security goal.  
We start by creating strategy \emph{S:1.5.1} to argue over the used threat model. 
%that might lead to damaging and compromising the security goal. 
%This strategy depends on the threat model that %the company uses.
%is used. %For example, 
%
If e.g., STRIDE is used as a threat model, % (as it is the case of our example), 
then the strategy would be to create a claim for each %of the 
STRIDE %threat
category.%ies. 
In our example case, %such a claim would be: 
we create claim \emph{C:1.5.1.1}: ``\emph{Spoofing of a signal leading to loss of integrity of the CAN message of \emph{Lamp Request} signal of power switch actuator ECU is not possible}''.
To ensure the completeness of the case, we further associate the strategy \emph{S:1.5.1} %used in the threat scenarios level\todo{Why ``threat scenarios level''?}
with a quality assurance claim (\emph{QC:1.5.1.1}).

At this point, our claims become more concrete as we have a specific item, asset, container component, security property, damage scenario, and threat scenario. 
We use the analysis of attack paths to further decompose and populate the example case.
We apply strategy \emph{S:1.6.1} to argue over the attacks and create attack path claims.
The resulting claims %take the form of the negation of
negate the possibility for an attack path to take place, e.g., \emph{C:1.6.1.4} ``\emph{It is not possible for an attacker to compromise the Navigation ECU from a cellular interface}''. %(C:1.6.1.4). 
As for all strategies in CASCADE, we associate the strategy used in the attack path with a quality assurance claim (\emph{QC:1.6.1.1}) to ensure the completeness of the case.
%(QC:1.5.1.1 and QC:1.6.1.1).

%Resolver block
\afterpage{
\begin{figure}
\begin{center}
  \includegraphics[width=1\linewidth]{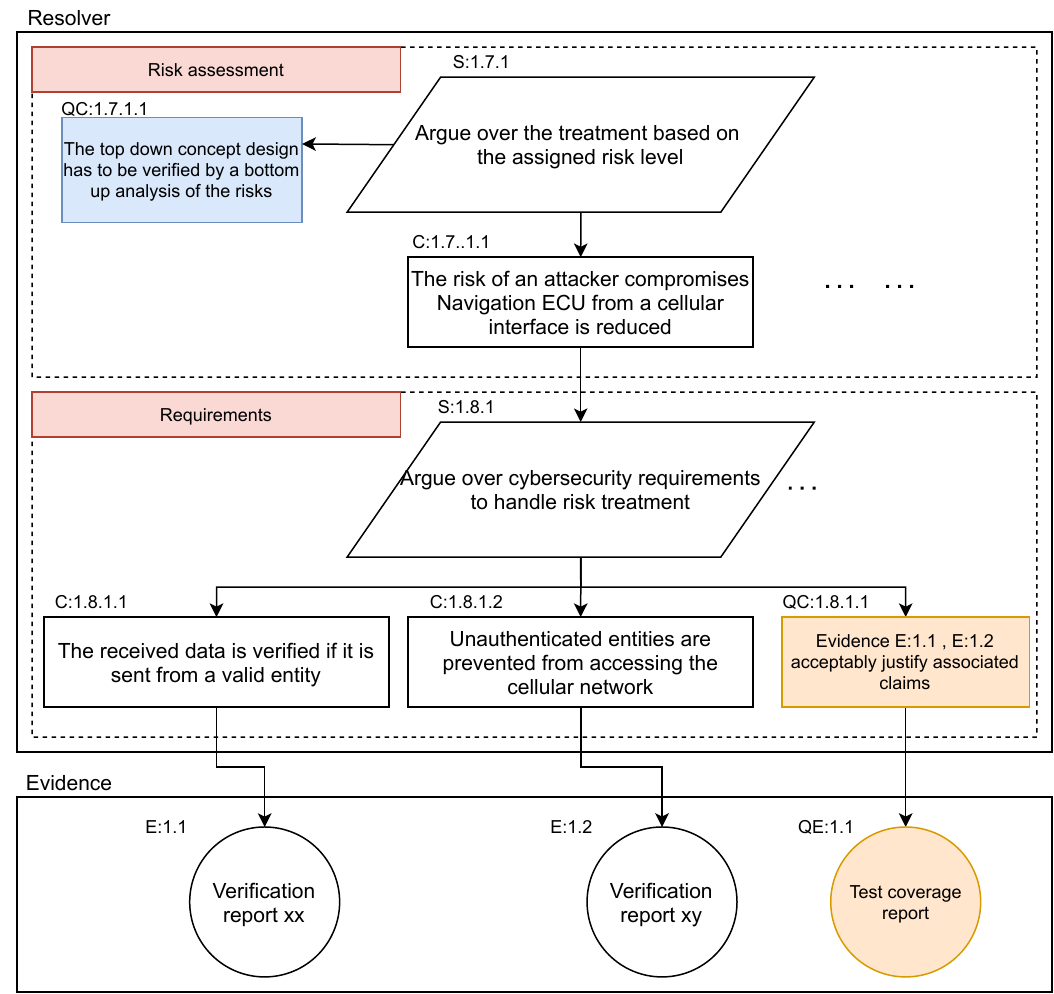}
\end{center}
        \caption{Resolver and evidence blocks of the headlamp use case} 
        \label{fig:ill_resolver}
\end{figure}
}

\subsection{Resolver and Evidence Blocks}
%In the next stage of the case creation
In this stage, we create the resolver block by investigating ways to resolve the attack paths based on a risk %analysis
assessment %done on them,
and creating requirements for the intended risk treatments. 

Figure~\ref{fig:ill_resolver} shows a part of %this
the resolver block %for the headlamp example 
for our example case associated with the attack path \emph{C:1.6.1.4}.
The outcome of the risk %analysis
assessment would be to accept, mitigate, transfer, or solve the risk. 
When a risk is accepted, then there is no need to further decompose the claim.
In the other cases, a strategy (\emph{S:1.7.1}) to decompose the risk of an attack path has to be created. 
In our example, we create claim \emph{C:1.7.1.1} to mitigate the risk %would be written 
as follows: ``\emph{The risk of an attacker compromising the Navigation ECU from a cellular interface is reduced}''. %(C:1.7.1.1)

This leads %then
to the stage where we argue on the requirements in order to specify how the risk has to be reduced or mitigated.
An example of a requirement claim is \emph{C:1.8.1.1}: ``\emph{The received data is verified if it is sent from a valid entity''}.% and ``Unauthenticated entities are prevented from accessing the cellular network'' (C:1.8.1.1 and C:1.8.1.2 respectively).
Figure \ref{fig:ill_resolver} also shows the evidence block which provides examples of evidence to justify the requirement claims. 
The evidence (e.g., \emph{E:1.1}) is supported by quality evidence (e.g., \emph{QE:1.1}) which, in tun, is complemented with requirement quality claims (e.g., \emph{QC:1.8.1.1}) % that it 
to confidently justify the associated requirement claims. %(QC:1.8.1.1), and is supported by a quality evidence (QE:1.1) 

%Generic sub-case
\begin{figure}
\begin{center}
  \includegraphics[width=1\linewidth]{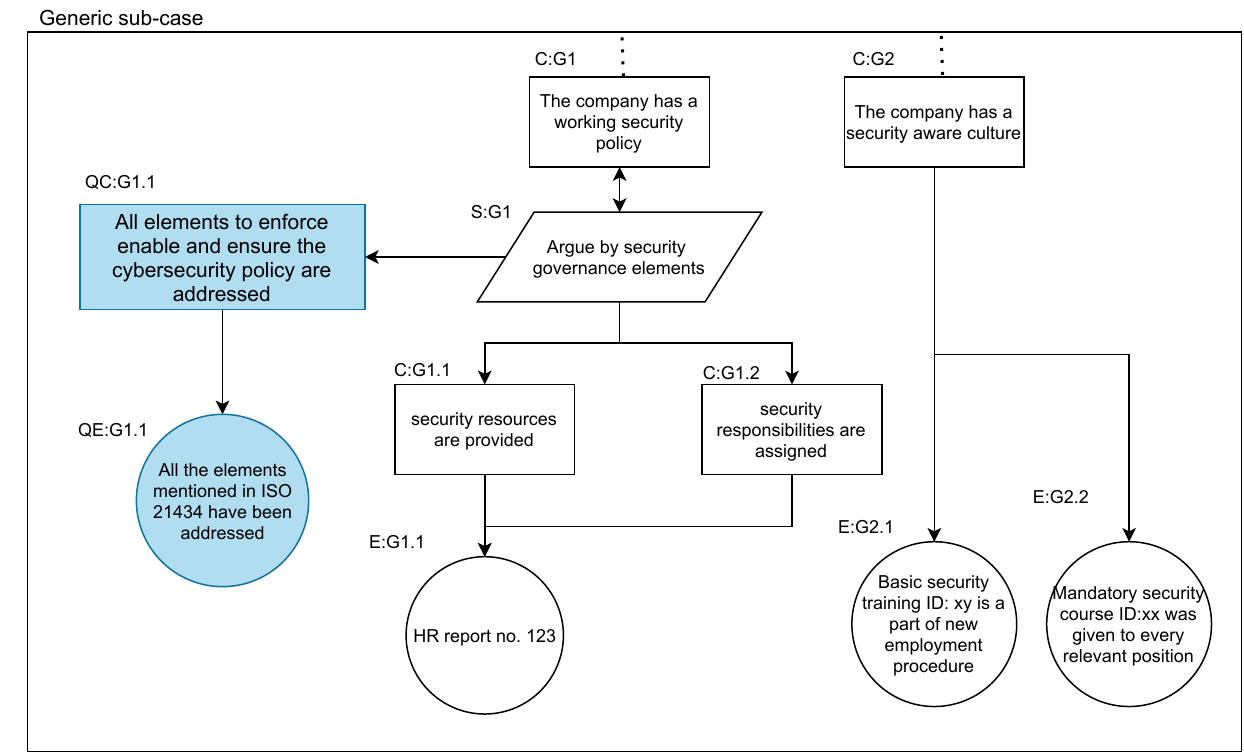}
\end{center}
        \caption{Generic sub-case block of the headlamp use case} 
        \label{fig:ill_generic}
\end{figure}

\subsection{Generic Sub-case Block}
Figure \ref{fig:ill_generic} shows the last block in our example; the generic sub-case.
This block includes claims that are relevant to the example case, but are not specific to it.
For example, claim \emph{C:G2} states that ``\emph{The company has a security aware culture}'', which is supported by two evidence statements; \emph{E:G2.1} and \emph{E:G2.2} to prove that the employees of the company were given a security training. 
Similarly to other blocks, the generic sub-case block might include strategies (e.g., \emph{S:G1}) to break down claims.
%which are in turn 
Moreover, these strategies are associated with quality assurance claims (e.g., \emph{QC:G.1.1}) as shown in %S:1.7.1 and QC:1.7.1.1 in 
Figure \ref{fig:ill_generic}.

%%%%JP and Riccardo's comments:

%- The quality assurance part needs to be done better (the example does not show that)...
%- Mapping of the standards (make more explicit)
%- Section 4. Level 2 titles for each of the blocks.
%- QA: not direct match in the text. Level 2 title.
%- Focus on the evaluation from volvo and put it as a section. 
%- Move methodology to introduction.
%- 1. We followed an iterative process. Initial decision to focus on assets (matches wow of volvo and ISO).
%  2. We created an initial version and iterated by looking at the standard and discussing with the company.
% - Mazen: move 3 to intro. add todo to Riccardo.
% - Section 4 JP will read and give feedback.
% - Example case and validation: Mazen writes. Riccardo -> feedback on example, JP on validation.
% - Mazen -> ping JP before the 23rd 
% - Mazen -> 2.3 make it readable (at the end).
% - 5,6 are the most important parts -> Focus on them!

\section{Validation}\label{sec:validation}
In order to evaluate our approach, we reached out to a security expert from the cybersecurity team at Volvo Trucks, which is a leading OEM that manufactures trucks in Sweden. We conducted several sessions during the development of CASCADE where we discussed the approach, its limitations and possible enhancements. When the approach was fully developed, we conducted a final evaluation session with the expert. 
We first discussed the way of working of the company when it comes to security activities and security assurance. We used the headlamp example from ISO/SAE-21434 as a context for this discussion. We then presented our approach and the example case for the headlamp item. The expert evaluated the approach by discussing how the overall structure of an SAC should look like from the company's perspective in order to satisfy the requirement for security cases in ISO/SAE-21434 and mapping the different elements of the example case to the internal way of working. The expert also provided insights on how to further enhance the approach.   

\begin{figure}
\begin{center}
  \includegraphics[width=1\linewidth]{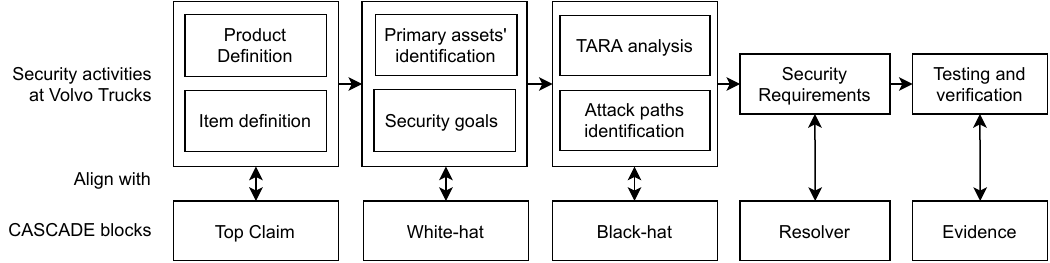}
\end{center}
        \caption{Mapping of the company's security activities to CASCADE blocks} 
        \label{fig:mapping}
\end{figure}

Figure \ref{fig:mapping} shows the different security activities at the company along with the corresponding CASCADE block. A link between an activity and a block indicates that the outcomes of the activity are used to create the SAC elements in the corresponding block.
%\subsection{Asset analysis}
%\subsection{Damage analysis}
%\subsection{AB Volvo perspective}

Software products at Volvo Trucks contain both on-board and off-board parts.  
The off-board parts establish the communication between the vehicles and the back-end systems. For example, the diagnostics services receive data from the vehicle's ECUs and store and use it in a back-end system.
The on-board parts are software components installed in the ECUs of the vehicle, e.g., the engine control and the head-up display unit. 
These parts are divided into items to facilitate the security-related analysis. The items of the off-board systems can be seen as individual services which communicate with the vehicles, whereas the on-board items are end user functionalities, e.g., external lighting and automated parking assistance.
In order to argue about the security of a complete product, both off-board and on-board items have to be considered. Hence, if the company wants to adopt CASCADE to create an SAC for a complete product, then the top claim block would contain claims for the individual items of that product. 
The Generic sub-case block of CASCADE helps to remove redundancy of arguments and evidence applicable to different items.

The assets of a product are identified by considering damage scenarios on the items. In general, these assets can be generalized into the following categories:
\begin{itemize}
    \item Vehicle's functionality (the attackers want to use the vehicle or tamper with the vehicle’s functionality for their own purpose or impede the rightful user from utilizing the vehicle functionality);
    \item Information (the attackers want to gain access to sensitive information); and
    %\item vehicle or parts of the vehicle (the attackers want to steal the vehicle or expensive parts), services. 
    \item Brand (the attackers want to discredit the brand and/or credit themselves).
\end{itemize}

%In the example, from an AB Volvo perspective, we would consider the vehicle with offboard services, to be the product. The product would be divided down into items, in order to make the analysis feasible. For vehicle onboard parts, we would mainly divide them into EndUserFunctions and I do not know the exact name, but I think we could say that the ISO/SAE-21434 Appendix G example would fall into the EnduserFunction "external lighting". 
%That is the EnduserFunction "external lighting" would be our item. The next step would then be to perform damage analysis on this item to find the related assets. The assets related to our product (and all items then) might perhaps be generalized into the following categories:
%\begin{itemize}
%    \item vehicle functionality (the attackers want to use the vehicle or tamper with the vehicle’s functionality for own purpose or impede the rightful user from utilizing the vehicle functionality),
%    \item information (the attackers want gain access to sensitive information),
%    \item vehicle or parts of the vehicle (the attackers want to steal the vehicle or expensive parts), services (similar to vehicle functionality?), 
%    \item brand (the attackers want to discredit the brand and/or credit themselves). In order to simplify the asset identification, we want to divide the vehicle in parts (items).
%\end{itemize}

The identified assets are further categorized into primary and secondary assets in accordance with the definitions in ISO-27005 \cite{iso27005}. Considering the headlamp example case, two possible damage scenarios would be ``Loosing the headlamp will drastically reduce the driver's sight and the vehicle's visibility, which may result in a severe accident'' and ``Applying the headlamp at incorrect times could dazzle other vehicles, which may increase the risk of an accident''. These lead to considering the headlamp functionality as a primary asset.
%When applying damage analysis on this item and checking the generalized asset list above, we would find that "Loosing the headlamp will drastically reduce the driver's sight and the veihcle's visability, which may result in a severe accident". This would mean that headlamp functionality is one of our primary assets (ISO-27005). We would also find that "Applying the headlamp at incorrect times could dazzle other vehicles, which may will increase the risk of an accident".

Then, relevant security attributes for the primary asset are identified and security goals as are derived: 
\begin{itemize}
    \item The integrity of the headlamp control functionality shall be preserved.
    \item The availability of the headlamp control functionality shall be preserved.
\end{itemize}

The identification of assets and security goals corresponds to the white-hat block of CASCADE, as shown in Figure \ref{fig:mapping}.
These goals only take relevant security properties into consideration, i.e., integrity and availability. Hence, other properties such as confidentiality and authenticity are not considered.
%The item description, security goals and other non-functionality requirements will then be used at concept design phase.
During the concept design phase, a Threat Assessment and Remediation Analysis (TARA) is performed on the item's primary assets using STRIDE, which will result in cybersecurity requirements on certain components which are considered as supporting assets. These requirements are converted to claims in the black-hat block of CASCADE, as shown in Figure \ref{fig:mapping}.
%We would also claim (ISO-21434 definition of claim) that authenticity, non-repudiation, confidentiality and authority is not relevant for this primary asset. The item description, security goals and other non-functionality requirements will then be used at concept design. At concept design, a TARA will be performed on the item's primary assets using STRIDE that will result in cybersecurity requirements on certain components, which we will call supporting assets (ISO-27005). 

After that, attack path analyses are performed bottom-up using an attack library, including but not limited to:
\begin{itemize}
    \item Intended over-the-air connection (e.g., 2,5G, 3G, 4G or 5G, Wi-Fi, WPAN, Bluetooth, IrDA, Wireless USB, and UBW);
    \item Intended physical connection points (e.g., OBD, USB, and CD-Rom);
    \item Unintended over-the-air disturbances (e.g., Radar, Laser, electro-magnetic, microwaves, infra-waves, ultrasound, and infra-sound);
    \item Unintended physical connection points (e.g., ECU, network, sensors, and actuators).
\end{itemize}

These attack paths are also expressed as claims in the Black-hat block.
%This analysis verifies that all supporting assets have been identified. 
Then the component design will be started considering all requirements, including cybersecurity. The components and systems are described in ``product descriptions''. These are verified against the requirements, including cybersecurity. 
The cybersecurity requirements in the product description correspond to the requirements of the Resolver block in CASCADE.
The components and systems are then tested against the product descriptions, and the test results are considered as evidence in CASCADE.

Other SAC requirements also emerged from the discussion with the security experts. For instance, they emphasised the need to validate that production, operation, service and decommissioning are all adequately handled. We believe that this would be covered by QA claims in the resolver block. Another requirement is that the product along with the SAC is maintained throughout the life-cycle. This is not covered by CASCADE, and we consider it to be an important complementary aspect for future work. In particular, we will be looking into methods to ensure traceability between the elements of SACs and the corresponding development artefacts. This traceability allows impact analysis for maintaining SACs. 
Lastly, the experts stress that it is important to argue that the performed product work is adequate with respect to cybersecurity policies and practices adopted by the company.%\todo{What does that mean?}. 
We believe that this is covered by the generic sub-case block of CASCADE.

To summarise the validation, we showed that CASCADE aligns well with respect to the way of working at Volvo Trucks. The structure of SAC built with CASCADE follows the structure of work done at the design phase at Volvo to a large extent and the generic sub-case and quality blocks help to serve the abstraction and completeness requirements of the company. 
We have identified some limitations in the approach which will be the basis for future work.
\section{Conclusion and Future work}

We have presented CASCADE, an approach to build security assurance cases driven by assets and geared towards automotive companies that want to conform to the upcoming ISO/SAE-21434. We illustrated the approach using an example case from the standard and validated it at an industrial OEM. We found that the way of working at the company aligns with our approach. 

%\textbf{Future work}.
As a future work, we plan to extend the approach to take into consideration the maintenance of SAC. We will also look into requirements sources other than ISO/SAE-21434 to better cover the needs of the automotive industry. Additionally, we plan to further validate the approach by %setting up an online survey and
including a larger community of automotive companies and automotive security experts.
Additionally, we plan to create a systematic methodology to create SAC in the automotive industry by mapping CASCADE to the requirements and work products of ISO/SAE-21434.

%\section*{Acknowledgement}
%This work is supported by the CASUS research project funded by VINNOVA, a Swedish funding agency.

%%================================================================%
%% Bibliography
%%================================================================%
\cleardoublepage\addcontentsline{toc}{chapter}{Bibliography}
%Different styles depending on whether you want numeric/alphabetic citation key
\bibliographystyle{IEEEtran}
%\bibliographystyle{apalike}

%Your bibtex file
\bibliography{lit}

\begin{thebibliography}{100}
\providecommand{\url}[1]{#1}
\csname url@samestyle\endcsname
\providecommand{\newblock}{\relax}
\providecommand{\bibinfo}[2]{#2}
\providecommand{\BIBentrySTDinterwordspacing}{\spaceskip=0pt\relax}
\providecommand{\BIBentryALTinterwordstretchfactor}{4}
\providecommand{\BIBentryALTinterwordspacing}{\spaceskip=\fontdimen2\font plus
\BIBentryALTinterwordstretchfactor\fontdimen3\font minus
  \fontdimen4\font\relax}
\providecommand{\BIBforeignlanguage}[2]{{%
\expandafter\ifx\csname l@#1\endcsname\relax
\typeout{** WARNING: IEEEtran.bst: No hyphenation pattern has been}%
\typeout{** loaded for the language `#1'. Using the pattern for}%
\typeout{** the default language instead.}%
\else
\language=\csname l@#1\endcsname
\fi
#2}}
\providecommand{\BIBdecl}{\relax}
\BIBdecl

\bibitem{paperA}
M.~Mohamad, A.~{\AA}str{\"o}m, {\"O}.~Askerdal, J.~Borg, and R.~Scandariato,
  ``Security assurance cases for road vehicles: an industry perspective,'' in
  \emph{Proceedings of the 15th International Conference on Availability,
  Reliability and Security}, 2020, pp. 1--6.

\bibitem{paperA_extended}
------, ``Security assurance cases for road vehicles: an industry
  perspective,'' \emph{arXiv preprint arXiv:2003.14106}, 2020.

\bibitem{paperB}
M.~Mohamad, J.-P. Stegh{\"o}fer, and R.~Scandariato, ``Security assurance
  cases—state of the art of an emerging approach,'' \emph{Empirical software
  engineering}, vol.~26, no.~4, p.~70, 2021.

\bibitem{paperC}
M.~Mohamad, {\"O}.~Askerdal, R.~Jolak, J.-P. Stegh{\"o}fer, and R.~Scandariato,
  ``Asset-driven security assurance cases with built-in quality assurance,'' in
  \emph{2021 IEEE/ACM 2nd International Workshop on Engineering and
  Cybersecurity of Critical Systems (EnCyCriS)}, 2021, pp. 1--8.

\bibitem{locococo}
\BIBentryALTinterwordspacing
M.~Mohamad, G.~Liebel, and E.~Knauss, ``Loco coco: Automatically constructing
  coordination and communication networks from model-based systems engineering
  data,'' \emph{Information and Software Technology}, vol.~92, pp. 179--193,
  2017. [Online]. Available:
  \url{https://www.sciencedirect.com/science/article/pii/S0950584916304670}
\BIBentrySTDinterwordspacing

\bibitem{conserve}
\BIBentryALTinterwordspacing
R.~Jolak, T.~Rosenstatter, M.~Mohamad, K.~Strandberg, B.~Sangchoolie,
  N.~Nowdehi, and R.~Scandariato, ``Conserve: A framework for the selection of
  techniques for monitoring containers security,'' \emph{Journal of Systems and
  Software}, vol. 186, p. 111158, 2022. [Online]. Available:
  \url{https://www.sciencedirect.com/science/article/pii/S0164121221002478}
\BIBentrySTDinterwordspacing

\bibitem{designDes}
J.-P. Stegh{\"o}fer, B.~Koopmann, J.~S. Becker, M.~Törnlund, Y.~Ibrahim, and
  M.~Mohamad, ``Design decisions in the construction of traceability
  information models for safe automotive systems,'' in \emph{2021 IEEE 29th
  International Requirements Engineering Conference (RE)}, 2021, pp. 185--196.

\bibitem{iso21434}
{International Organization for Standardization and Society of Automotive
  Engineers}, ``{ISO / SAE 21434 Road vehicles -- Cybersecurity Engineering, CD
  Draft},'' {2018}.

\bibitem{iso26262}
{International Organization for Standardization}, ``{ISO 26262 Road vehicles --
  Functional safety, 1st Edition},'' {Geneva, Switzerland}, {2011}.

\bibitem{GSN_standard}
{GSN Community Standard Working Group}, ``{GSN} community standard,''
  \emph{Available at www.goalstructuringnotation.info/}, 2011.

\bibitem{bloomfield2010}
R.~Bloomfield and P.~Bishop, ``Safety and assurance cases: Past, present and
  possible future--an adelard perspective,'' in \emph{Making Systems
  Safer}.\hskip 1em plus 0.5em minus 0.4em\relax Springer, 2010, pp. 51--67.

\bibitem{palin2011}
R.~Palin, D.~Ward, I.~Habli, and R.~Rivett, ``{ISO} 26262 safety cases:
  Compliance and assurance,'' in \emph{6th IET International Conference on
  System Safety}.\hskip 1em plus 0.5em minus 0.4em\relax IET, 2011.

\bibitem{birch2013}
J.~Birch, R.~Rivett, I.~Habli, B.~Bradshaw, J.~Botham, D.~Higham, P.~Jesty,
  H.~Monkhouse, and R.~Palin, ``Safety cases and their role in {ISO} 26262
  functional safety assessment,'' in \emph{International Conference on Computer
  Safety, Reliability, and Security}.\hskip 1em plus 0.5em minus 0.4em\relax
  Springer, 2013, pp. 154--165.

\bibitem{medicalSafetyExample}
\BIBentryALTinterwordspacing
M.~Sujan, P.~Spurgeon, M.~Cooke, A.~Weale, P.~Debenham, and S.~Cross, ``The
  development of safety cases for healthcare services: Practical experiences,
  opportunities and challenges,'' \emph{Reliability Engineering \& System
  Safety}, vol. 140, pp. 200--207, 2015. [Online]. Available:
  \url{https://www.sciencedirect.com/science/article/pii/S095183201500099X}
\BIBentrySTDinterwordspacing

\bibitem{20_xu2017}
B.~Xu, M.~Lu, and D.~Zhang, ``A layered argument strategy for software security
  case development,'' in \emph{2017 IEEE International Symposium on Software
  Reliability Engineering Workshops (ISSREW)}.\hskip 1em plus 0.5em minus
  0.4em\relax IEEE, 2017, pp. 331--338.

\bibitem{luburic2018}
N.~Luburi{\'c}, G.~Sladi{\'c}, B.~Milosavljevi{\'c}, and A.~Kaplar,
  ``Demonstrating enterprise system security using an asset-centric security
  assurance framework,'' in \emph{8th International Conference on Information
  Society and Technology}, 2018, p.~16.

\bibitem{8_finnegan2014}
A.~Finnegan and F.~McCaffery, ``A security argument pattern for medical device
  assurance cases,'' in \emph{2014 IEEE International Symposium on Software
  Reliability Engineering Workshops}.\hskip 1em plus 0.5em minus 0.4em\relax
  IEEE, 2014, pp. 220--225.

\bibitem{45_finnegan2014}
{A. Finnegan and F. McCaffery}, ``Towards an international security case
  framework for networked medical devices,'' in \emph{International Conference
  on Computer Safety, Reliability, and Security}.\hskip 1em plus 0.5em minus
  0.4em\relax Springer, 2014, pp. 197--209.

\bibitem{9_ankum2005}
T.~S. {Ankrum} and A.~H. {Kromholz}, ``Structured assurance cases: three common
  standards,'' in \emph{Ninth IEEE International Symposium on High-Assurance
  Systems Engineering (HASE'05)}, Oct 2005, pp. 99--108.

\bibitem{iso15408}
{International Organization for Standardization}, ``{Information technology —
  Security techniques — Evaluation criteria for IT security — Part 1:
  Introduction and general model},'' {Geneva, Switzerland}, {1999}.

\bibitem{39_cheah2018}
M.~Cheah, S.~A. Shaikh, J.~Bryans, and P.~Wooderson, ``Building an automotive
  security assurance case using systematic security evaluations,''
  \emph{Computers \& Security}, vol.~77, pp. 360--379, 2018.

\bibitem{cohn2004}
M.~Cohn, \emph{User stories applied: For agile software development}.\hskip 1em
  plus 0.5em minus 0.4em\relax Addison-Wesley Professional, 2004.

\bibitem{kh2007guidelines}
B.~Kitchenham \emph{et~al.}, ``Guidelines for performing systematic literature
  reviews in software engineering,'' Keele University, Tech. Rep.
  EBSE-2007-12007, 2007.

\bibitem{snowballing}
C.~Wohlin, ``Guidelines for snowballing in systematic literature studies and a
  replication in software engineering,'' in \emph{Proceedings of the 18th
  international conference on evaluation and assessment in software
  engineering}.\hskip 1em plus 0.5em minus 0.4em\relax Citeseer, 2014, p.~38.

\bibitem{Hevner2004}
A.~R. Hevner, S.~T. March, J.~Park, and S.~Ram, ``Design science in information
  systems research,'' \emph{MIS Q.}, vol.~28, no.~1, pp. 75--105, Mar. 2004.

\bibitem{vaishnavi2004}
V.~Vaishnavi and B.~Kuechler, ``Design research in information systems.''
  \emph{Association for Information Systems}, 2004.

\bibitem{cocome}
``The common component modeling example - cocome,''
  \url{https://www.cocome.org/downloads/documentation/cocome.pdf}, accessed:
  2020-04-09.

\bibitem{18_othmane2016}
L.~Ben~Othmane and A.~Ali, ``Towards effective security assurance for
  incremental software development the case of zen cart application,'' in
  \emph{2016 11th International Conference on Availability, Reliability and
  Security (ARES)}.\hskip 1em plus 0.5em minus 0.4em\relax IEEE, 2016, pp.
  564--571.

\bibitem{34_vivas2011}
J.~L. Vivas, I.~Agudo, and J.~L{\'o}pez, ``A methodology for security
  assurance-driven system development,'' \emph{Requirements Engineering},
  vol.~16, no.~1, pp. 55--73, 2011.

\bibitem{1_cyra2007}
L.~Cyra and J.~Gorski, ``Supporting compliance with security standards by trust
  case templates,'' in \emph{2nd International Conference on Dependability of
  Computer Systems (DepCoS-RELCOMEX'07)}.\hskip 1em plus 0.5em minus
  0.4em\relax IEEE, 2007, pp. 91--98.

\bibitem{11_chindamaikul2014}
K.~Chindamaikul, T.~Takai, and H.~Iida, ``Retrieving information from a
  document repository for constructing assurance cases,'' in \emph{2014 IEEE
  International Symposium on Software Reliability Engineering Workshops}.\hskip
  1em plus 0.5em minus 0.4em\relax IEEE, 2014, pp. 198--203.

\bibitem{25_rodes2014}
B.~D. Rodes, J.~C. Knight, and K.~S. Wasson, ``A security metric based on
  security arguments,'' in \emph{Proceedings of the 5th International Workshop
  on Emerging Trends in Software Metrics}.\hskip 1em plus 0.5em minus
  0.4em\relax ACM, 2014, pp. 66--72.

\bibitem{secGoals}
C.~{Haley}, R.~{Laney}, J.~{Moffett}, and B.~{Nuseibeh}, ``Security
  requirements engineering: A framework for representation and analysis,''
  \emph{IEEE Transactions on Software Engineering}, vol.~34, no.~1, pp.
  133--153, 2008.

\bibitem{ttv1}
D.~T. Campbell and J.~C. Stanley, \emph{Experimental and quasi-experimental
  designs for research}.\hskip 1em plus 0.5em minus 0.4em\relax Ravenio Books,
  2015.

\bibitem{ttv3}
C.~Wohlin, P.~Runeson, M.~H{\"o}st, M.~C. Ohlsson, B.~Regnell, and
  A.~Wessl{\'e}n, \emph{Experimentation in software engineering}.\hskip 1em
  plus 0.5em minus 0.4em\relax Springer Science \& Business Media, 2012.

\bibitem{goodenough2007}
J.~Goodenough, H.~Lipson, and C.~Weinstock, ``Arguing security: Creating
  security assurance cases,'' Software Engineering Institute CMU, Tech. Rep.,
  2007.

\bibitem{iso26262_1ed}
{International Organization for Standardization}, ``{ISO 26262 Road vehicles --
  Functional safety, 1st Edition},'' {Geneva, Switzerland}, {2011}.

\bibitem{knight2015}
J.~{Knight}, ``The importance of security cases: Proof is good, but not
  enough,'' \emph{IEEE Security Privacy}, vol.~13, no.~4, 2015.

\bibitem{alexander2011}
R.~Alexander, R.~Hawkins, and T.~Kelly, ``Security assurance cases: motivation
  and the state of the art,'' University of York, Tech. Rep., 2011.

\bibitem{gsn}
J.~Spriggs, \emph{GSN-the goal structuring notation: A structured approach to
  presenting arguments}.\hskip 1em plus 0.5em minus 0.4em\relax Springer
  Science \& Business Media, 2012.

\bibitem{CAE}
Adelard, ``The adelard safety case development manual,'' 1998.

\bibitem{26_poreddy2011}
B.~R. Poreddy and S.~Corns, ``Arguing security of generic avionic mission
  control computer system (mcc) using assurance cases,'' \emph{Procedia
  Computer Science}, vol.~6, pp. 499--504, 2011.

\bibitem{16_ray2015}
A.~Ray and R.~Cleaveland, ``Security assurance cases for medical
  cyber--physical systems,'' \emph{IEEE Design \& Test}, vol.~32, no.~5, pp.
  56--65, 2015.

\bibitem{4_he2012}
Y.~He and C.~Johnson, ``Generic security cases for information system security
  in healthcare systems,'' in \emph{7th IET International Conference on System
  Safety, incorporating the Cyber Security Conference}.\hskip 1em plus 0.5em
  minus 0.4em\relax IET, 2012.

\bibitem{15_hawkins2015}
R.~Hawkins, I.~Habli, D.~Kolovos, R.~Paige, and T.~Kelly, ``Weaving an
  assurance case from design: a model-based approach,'' in \emph{2015 IEEE 16th
  International Symposium on High Assurance Systems Engineering}.\hskip 1em
  plus 0.5em minus 0.4em\relax IEEE, 2015, pp. 110--117.

\bibitem{35_fung2018}
N.~L. Fung, S.~Kokaly, A.~Di~Sandro, R.~Salay, and M.~Chechik, ``Mmint-a: a
  tool for automated change impact assessment on assurance cases,'' in
  \emph{International Conference on Computer Safety, Reliability, and
  Security}.\hskip 1em plus 0.5em minus 0.4em\relax Springer, 2018, pp. 60--70.

\bibitem{iso26262_2ed}
{International Organization for Standardization}, ``{ISO 26262 Road vehicles --
  Functional safety, 2nd Edition},'' {Geneva, Switzerland}, {2018}.

\bibitem{SELFDRIVE}
{National Highway Traffic Safety Administration (NHTSA), U.S. Government},
  ``{H.R.3388 - SELF DRIVE Act (Safely Ensuring Lives Future Deployment and
  Research In Vehicle Evolution Act},'' {2017}.

\bibitem{gdpr}
{European Union}, ``{General Data Protection Regulation (GDPR) 2016/679},''
  {2016}.

\bibitem{GBT35273}
{Standardization Administration of China}, ``{GB/T 35273-2017 Information
  security technology - Personal Information Security Specification},'' {2017}.

\bibitem{J3061}
{Society of Automotive Engineers}, ``{J3061 - Cybersecurity Guidebook for
  Cyber-Physical Vehicle Systems},'' {2016}.

\bibitem{ADS2}
{National Highway Traffic Safety Administration (NHTSA), U.S. Government},
  ``{DOT HS 812 442 - ADS 2.0 (Automated Driving Systems)},'' {2017}.

\bibitem{AV3}
{U.S. Government}, ``{DOT 2018-21840 - Preparing for the Future of
  Transportation Automated Vehicles 3.0. National Highway Traffic Safety
  Administration (NHTSA)},'' {2018}.

\bibitem{SPYCar}
{National Highway Traffic Safety Administration (NHTSA), U.S. Government},
  ``{S. 680 - SPY Car Act of 2017 (Security and Privacy in Your Car Act of
  2017)},'' {2017}.

\bibitem{CCPA}
{State of California}, ``{CCPA - Assembly Bill No. 375 - "California Consumer
  Privacy Act - AB-375 Privacy: personal information: businesses"},'' {2018}.

\bibitem{UNECE-Reg}
{United Nations Economic Commission for Europe (UNECE) WP.29 GRVA}, ``{Proposal
  for a new UN Regulation on uniform provisions concerning the approval of
  vehicles with regard to cyber security and of cybersecurity management
  systems},'' {March 2020}.

\bibitem{UNECE-OTA}
------, ``{Proposal for a new UN Regulation on uniform provisions concerning
  the approval of vehicles with regard to software update processes and of
  software update management systems},'' {March 2020}.

\bibitem{ICV}
{Chinese Ministry of Industry and Information Technology (MIIT) and the
  Standardization Administration of China}, ``{Intelligent and Connected
  Vehicles (ICV)},'' {2018}.

\bibitem{CSL}
{National People's Congress (NPC) of the People's Republic of China}, ``{The
  Cyber Security Law},'' {2016}.

\bibitem{nair2013}
S.~Nair, J.~L. de~la Vara, M.~Sabetzadeh, and L.~Briand, ``Classification,
  structuring, and assessment of evidence for safety--a systematic literature
  review,'' in \emph{2013 IEEE Sixth International Conference on Software
  Testing, Verification and Validation}.\hskip 1em plus 0.5em minus 0.4em\relax
  IEEE, 2013, pp. 94--103.

\bibitem{maksimov2}
\BIBentryALTinterwordspacing
M.~Maksimov, S.~Kokaly, and M.~Chechik, ``A survey of tool-supported assurance
  case assessment techniques,'' \emph{ACM Comput. Surv.}, vol.~52, no.~5, Sep.
  2019. [Online]. Available: \url{https://doi.org/10.1145/3342481}
\BIBentrySTDinterwordspacing

\bibitem{gade2015}
D.~Gade and S.~Deshpande, ``A literature review on assurance driven software
  design,'' \emph{International Journal of Advanced Research in Computer and
  Communication Engineering}, vol.~4, no.~9, 2015.

\bibitem{industrialNeeds}
\BIBentryALTinterwordspacing
M.~Mohamad, A.~\r{A}str\"{o}m, O.~Askerdal, J.~Borg, and R.~Scandariato,
  ``Security assurance cases for road vehicles: An industry perspective,'' in
  \emph{Proceedings of the 15th International Conference on Availability,
  Reliability and Security}, ser. ARES '20.\hskip 1em plus 0.5em minus
  0.4em\relax New York, NY, USA: Association for Computing Machinery, 2020.
  [Online]. Available: \url{https://doi.org/10.1145/3407023.3407033}
\BIBentrySTDinterwordspacing

\bibitem{caseStudies}
S.~Easterbrook, J.~Singer, M.-A. Storey, and D.~Damian, ``Selecting empirical
  methods for software engineering research,'' in \emph{Guide to advanced
  empirical software engineering}.\hskip 1em plus 0.5em minus 0.4em\relax
  Springer, 2008, pp. 285--311.

\bibitem{runeson2009}
P.~Runeson and M.~H{\"o}st, ``Guidelines for conducting and reporting case
  study research in software engineering,'' \emph{Empirical software
  engineering}, vol.~14, no.~2, p. 131, 2009.

\bibitem{yin2003}
R.~K. Yin \emph{et~al.}, ``Design and methods,'' \emph{Case study research},
  vol.~3, 2003.

\bibitem{10_othmane2014}
L.~Ben~Othmane, P.~Angin, and B.~Bhargava, ``Using assurance cases to develop
  iteratively security features using scrum,'' in \emph{2014 Ninth
  International Conference on Availability, Reliability and Security}.\hskip
  1em plus 0.5em minus 0.4em\relax IEEE, 2014, pp. 490--497.

\bibitem{2_cockram2007}
T.~Cockram and S.~Lautieri, ``Combining security and safety principles in
  practice,'' in \emph{Proceedings of the 2nd institution of engineering and
  technology international conference on system safety}.\hskip 1em plus 0.5em
  minus 0.4em\relax IET, 2007, pp. 159--164.

\bibitem{3_goodger2012}
A.~Goodger, N.~Caldwell, and J.~Knowles, ``What does the assurance case
  approach deliver for critical information infrastructure protection in
  cybersecurity?'' in \emph{7th IET International Conference on System Safety,
  incorporating the Cyber Security Conference}.\hskip 1em plus 0.5em minus
  0.4em\relax IET, 2012.

\bibitem{14_netkachova2015}
K.~Netkachova, K.~M{\"u}ller, M.~Paulitsch, and R.~Bloomfield, ``Investigation
  into a layered approach to architecting security-informed safety cases,'' in
  \emph{2015 IEEE/AIAA 34th Digital Avionics Systems Conference (DASC)}.\hskip
  1em plus 0.5em minus 0.4em\relax IEEE, 2015, pp. 6B4--1.

\bibitem{19_netkachova2016}
K.~Netkachova and R.~E. Bloomfield, ``Security-informed safety,''
  \emph{Computer}, vol.~49, no.~6, pp. 98--102, 2016.

\bibitem{24_gacek2014}
A.~Gacek, J.~Backes, D.~Cofer, K.~Slind, and M.~Whalen, ``Resolute: an
  assurance case language for architecture models,'' \emph{ACM SIGAda Ada
  Letters}, vol.~34, no.~3, pp. 19--28, 2014.

\bibitem{29_bloomfield2017}
R.~Bloomfield, P.~Bishop, E.~Butler, and K.~Netkachova, ``Using an assurance
  case framework to develop security strategy and policies,'' in
  \emph{International Conference on Computer Safety, Reliability, and
  Security}.\hskip 1em plus 0.5em minus 0.4em\relax Springer, 2017, pp. 27--38.

\bibitem{30_netkachova2014}
K.~Netkachova, R.~Bloomfield, P.~Popov, and O.~Netkachov, ``Using structured
  assurance case approach to analyse security and reliability of critical
  infrastructures,'' in \emph{International Conference on Computer Safety,
  Reliability, and Security}.\hskip 1em plus 0.5em minus 0.4em\relax Springer,
  2014, pp. 345--354.

\bibitem{37_gallo2015}
R.~Gallo and R.~Dahab, ``Assurance cases as a didactic tool for information
  security,'' in \emph{IFIP World Conference on Information Security
  Education}.\hskip 1em plus 0.5em minus 0.4em\relax Springer, 2015, pp.
  15--26.

\bibitem{47_ionita2017}
D.~Ionita, M.~Ford, A.~Vasenev, and R.~Wieringa, ``Graphical modeling of
  security arguments: Current state and future directions,'' in
  \emph{International Workshop on Graphical Models for Security}.\hskip 1em
  plus 0.5em minus 0.4em\relax Springer, 2017, pp. 1--16.

\bibitem{core}
\BIBentryALTinterwordspacing
{Computing Research and Education Association of Australasia}, ``{CORE} ranking
  portal -- computing research and education.'' [Online]. Available:
  \url{https://www.core.edu.au/conference-portal}
\BIBentrySTDinterwordspacing

\bibitem{arc_2018}
\BIBentryALTinterwordspacing
{Australian Research Council}, ``{Excellence in Research for Australia},'' May
  2018. [Online]. Available:
  \url{https://www.arc.gov.au/excellence-research-australia}
\BIBentrySTDinterwordspacing

\bibitem{50_knight2015}
J.~Knight, ``The importance of security cases: Proof is good, but not enough.''
  \emph{IEEE Security \& Privacy}, vol.~13, no.~4, pp. 73--75, 2015.

\bibitem{53_alexander2011}
R.~Alexander, R.~Hawkins, and T.~Kelly, ``Security assurance cases: motivation
  and the state of the art,'' \emph{High Integrity Systems Engineering
  Department of Computer Science University of York Deramore Lane York YO10
  5GH}, 2011.

\bibitem{55_calinescu2017}
R.~Calinescu, D.~Weyns, S.~Gerasimou, M.~U. Iftikhar, I.~Habli, and T.~Kelly,
  ``Engineering trustworthy self-adaptive software with dynamic assurance
  cases,'' \emph{IEEE Transactions on Software Engineering}, vol.~44, no.~11,
  pp. 1039--1069, 2017.

\bibitem{51_mohammadi2018}
N.~G. Mohammadi, N.~Ulfat-Bunyadi, and M.~Heisel, ``Trustworthiness
  cases--toward preparation for the trustworthiness certification,'' in
  \emph{International Conference on Trust and Privacy in Digital
  Business}.\hskip 1em plus 0.5em minus 0.4em\relax Springer, 2018, pp.
  244--259.

\bibitem{21_sklyar2017}
V.~Sklyar and V.~Kharchenko, ``Challenges in assurance case application for
  industrial iot,'' in \emph{2017 9th IEEE International Conference on
  Intelligent Data Acquisition and Advanced Computing Systems: Technology and
  Applications (IDAACS)}, vol.~2.\hskip 1em plus 0.5em minus 0.4em\relax IEEE,
  2017, pp. 736--739.

\bibitem{28_sklyar2017}
V.~V. Sklyar and V.~S. Kharchenko, ``Assurance case driven design based on the
  harmonized framework of safety and security requirements.'' in \emph{ICTERI},
  2017, pp. 670--685.

\bibitem{31_sklyar2019}
V.~Sklyar and V.~Kharchenko, ``Green assurance case: Applications for internet
  of things,'' in \emph{Green IT Engineering: Social, Business and Industrial
  Applications}.\hskip 1em plus 0.5em minus 0.4em\relax Springer, 2019, pp.
  351--371.

\bibitem{23_sljivo2016}
I.~Sljivo and B.~Gallina, ``Building multiple-viewpoint assurance cases using
  assumption/guarantee contracts,'' in \emph{Proccedings of the 10th European
  Conference on Software Architecture Workshops}.\hskip 1em plus 0.5em minus
  0.4em\relax ACM, 2016, p.~39.

\bibitem{22_strielkina2018}
A.~Strielkina, O.~Illiashenko, M.~Zhydenko, and D.~Uzun, ``Cybersecurity of
  healthcare iot-based systems: Regulation and case-oriented assessment,'' in
  \emph{2018 IEEE 9th International Conference on Dependable Systems, Services
  and Technologies (DESSERT)}.\hskip 1em plus 0.5em minus 0.4em\relax IEEE,
  2018, pp. 67--73.

\bibitem{17_sklyar2016}
V.~Sklyar and V.~Kharchenko, ``Assurance case driven design for computer
  systems: graphical notations versus mathematical methods,'' in \emph{2016
  Third International Conference on Mathematics and Computers in Sciences and
  in Industry (MCSI)}.\hskip 1em plus 0.5em minus 0.4em\relax IEEE, 2016, pp.
  308--312.

\bibitem{56_lipson2008}
H.~Lipson and C.~Weinstock, ``Evidence of assurance: Laying the foundation for
  a credible security case,'' Carnegie Mellon University, Tech. Rep., 2008.

\bibitem{52_weinstock2007}
\BIBentryALTinterwordspacing
C.~B. Weinstock, J.~B. Goodenough, and H.~F. Lipson, ``Arguing
  security-creating security assurance cases,'' Software Engineering Institute
  -- Carnegie Mellon University, Tech. Rep., 2007, part of the collection
  ``Resources for Assurance Cases''. [Online]. Available:
  \url{https://resources.sei.cmu.edu/library/asset-view.cfm?assetid=293629}
\BIBentrySTDinterwordspacing

\bibitem{38_agudo2009}
I.~Agudo, J.~L. Vivas, and J.~L{\'o}pez, ``Security assurance during the
  software development cycle,'' in \emph{Proceedings of the International
  Conference on Computer Systems and Technologies and Workshop for PhD Students
  in Computing}.\hskip 1em plus 0.5em minus 0.4em\relax ACM, 2009, p.~20.

\bibitem{43_finnegan2013}
A.~Finnegan, F.~McCaffery, and G.~Coleman, ``A process assessment model for
  security assurance of networked medical devices,'' in \emph{International
  Conference on Software Process Improvement and Capability
  Determination}.\hskip 1em plus 0.5em minus 0.4em\relax Springer, 2013, pp.
  25--36.

\bibitem{5_patu2013}
V.~Patu and S.~Yamamoto, ``Identifying and implementing security patterns for a
  dependable security case--from security patterns to {D-Case},'' in \emph{2013
  IEEE 16th International Conference on Computational Science and
  Engineering}.\hskip 1em plus 0.5em minus 0.4em\relax IEEE, 2013, pp.
  138--142.

\bibitem{46_graydon2013}
P.~J. Graydon and T.~P. Kelly, ``Using argumentation to evaluate software
  assurance standards,'' \emph{Information and Software Technology}, vol.~55,
  no.~9, pp. 1551--1562, 2013.

\bibitem{54_haley2005}
C.~B. Haley, J.~D. Moffett, R.~Laney, and B.~Nuseibeh, ``Arguing security:
  Validating security requirements using structured argumentation,'' in
  \emph{Proceedings of the 3rd Symposium on Requirements Engineering for
  Information Security (SREIS'05)}, 2005.

\bibitem{6_masumoto2013}
M.~Masumoto, T.~Tokuno, and S.~Yanamoto, ``A method for assuring service grade
  with assurance case: An experiment on a portal service,'' in \emph{2013 IEEE
  International Symposium on Software Reliability Engineering Workshops
  (ISSREW)}.\hskip 1em plus 0.5em minus 0.4em\relax IEEE, 2013, pp. 311--314.

\bibitem{49_yamamoto2015}
S.~Yamamoto, ``Assuring security through attribute {GSN},'' in \emph{2015 5th
  International Conference on IT Convergence and Security (ICITCS)}.\hskip 1em
  plus 0.5em minus 0.4em\relax IEEE, 2015, pp. 1--5.

\bibitem{42_gorski2012}
J.~G{\'o}rski, A.~Jarz{\k{e}}bowicz, J.~Miler, M.~Witkowicz,
  J.~Czy{\.z}nikiewicz, and P.~Jar, ``Supporting assurance by evidence-based
  argument services,'' in \emph{International Conference on Computer Safety,
  Reliability, and Security}.\hskip 1em plus 0.5em minus 0.4em\relax Springer,
  2012, pp. 417--426.

\bibitem{41_patu2013}
V.~Patu and S.~Yamamoto, ``How to develop security case by combining real life
  security experiences (evidence) with {D-Case},'' \emph{Procedia Computer
  Science}, vol.~22, pp. 954--959, 2013.

\bibitem{40_coffey2014}
J.~W. Coffey, D.~Snider, T.~Reichherzer, and N.~Wilde, ``Concept mapping for
  the efficient generation and communication of security assurance cases,''
  \emph{Proceedings of IMCIC}, vol.~14, pp. 173--177, 2014.

\bibitem{7_tippenhauer2014}
N.~O. Tippenhauer, W.~G. Temple, A.~H. Vu, B.~Chen, D.~M. Nicol, Z.~Kalbarczyk,
  and W.~H. Sanders, ``Automatic generation of security argument graphs,'' in
  \emph{2014 IEEE 20th Pacific Rim International Symposium on Dependable
  Computing}.\hskip 1em plus 0.5em minus 0.4em\relax IEEE, 2014, pp. 33--42.

\bibitem{36_shortt2015}
C.~Shortt and J.~Weber, ``Hermes: A targeted fuzz testing framework,'' in
  \emph{International Conference on Intelligent Software Methodologies, Tools,
  and Techniques}.\hskip 1em plus 0.5em minus 0.4em\relax Springer, 2015, pp.
  453--468.

\bibitem{toulmin2003}
S.~E. Toulmin, \emph{The uses of argument}.\hskip 1em plus 0.5em minus
  0.4em\relax Cambridge university press, 2003.

\bibitem{tool_adrelard}
Adelard, ``The adelard safety case editor - {ASCE},'' 2003, product description
  available at: http://adelard.co.uk/software/asce/.

\bibitem{tool_turboac}
GessNet, ``Turboac™ assurance cases,'' \url{https://www.gessnet.com//},
  2011--2016.

\bibitem{d-case}
Y.~Matsuno, H.~Takamura, and Y.~Ishikawa, ``A dependability case editor with
  pattern library,'' in \emph{2010 IEEE 12th International Symposium on High
  Assurance Systems Engineering}.\hskip 1em plus 0.5em minus 0.4em\relax IEEE,
  2010, pp. 170--171.

\bibitem{tool_openArgue}
Y.~{Yu}, T.~T. {Tun}, A.~{Tedeschi}, V.~N.~L. {Franqueira}, and B.~{Nuseibeh},
  ``Openargue: Supporting argumentation to evolve secure software systems,'' in
  \emph{2011 IEEE 19th International Requirements Engineering Conference}, Aug
  2011, pp. 351--352.

\bibitem{tool_arg_sec}
D.~{Ionita}, R.~{Kegel}, A.~{Baltuta}, and R.~{Wieringa}, ``Arguesecure:
  Out-of-the-box security risk assessment,'' in \emph{2016 IEEE 24th
  International Requirements Engineering Conference Workshops (REW)}, Sep.
  2016, pp. 74--79.

\bibitem{tool_CyberSAGE}
{Advanced Digital Sciences Center - Singapore}, ``{CyberSAGE},''
  \url{https://www.illinois.adsc.com.sg/cybersage/index.html/}, 2015.

\bibitem{tool_uppaal}
G.~Behrmann, A.~David, K.~G. Larsen, J.~H{\aa}kansson, P.~Pettersson, W.~Yi,
  and M.~Hendriks, ``Uppaal 4.0,'' in \emph{Behrmann, G., et al. "Uppaal 4.0."
  Third International Conference on the Quantitative Evaluation of SysTems
  (QEST 2006)}.\hskip 1em plus 0.5em minus 0.4em\relax Los Alamitos, CA: IEEE
  Computer Society, 2006.

\bibitem{tool_meld}
K.~Willadsen, ``Meld,'' \url{https://meldmerge.org/}, 2011--2012.

\bibitem{tool_norsta}
{Gdansk University of Technology}, ``{NOR-STA},''
  \url{https://www.nor-sta.eu/en/}, 2010--2019.

\bibitem{omg}
{Object Management Group (OMG)}, ``{Structured Assurance Case Metamodel (SACM),
  Version 2.1},'' OMG Document Number formal/20-04-01
  (\url{https://www.omg.org/spec/SACM/2.1/PDF}), 2020.

\bibitem{aadl}
P.~H. Feiler and D.~P. Gluch, \emph{Model-based engineering with AADL: an
  introduction to the SAE architecture analysis \& design language}.\hskip 1em
  plus 0.5em minus 0.4em\relax Addison-Wesley, 2012.

\bibitem{48_taguchi2014}
K.~Taguchi, D.~Souma, and H.~Nishihara, ``Safe \& sec case patterns,'' in
  \emph{International Conference on Computer Safety, Reliability, and
  Security}.\hskip 1em plus 0.5em minus 0.4em\relax Springer, 2014, pp. 27--37.

\bibitem{ADR}
M.~Sein, O.~Henfridsson, S.~Purao, M.~Rossi, and R.~Lindgren, ``Action design
  research,'' \emph{MIS Quarterly}, vol.~35, pp. 37--56, 03 2011.

\bibitem{safetyvssecurity}
\BIBentryALTinterwordspacing
L.~Piètre-Cambacédès and M.~Bouissou, ``Cross-fertilization between safety
  and security engineering,'' \emph{Reliability Engineering \& System Safety},
  vol. 110, pp. 110 -- 126, 2013. [Online]. Available:
  \url{http://www.sciencedirect.com/science/article/pii/S0951832012001913}
\BIBentrySTDinterwordspacing

\bibitem{maksimov2018}
M.~Maksimov, N.~L. Fung, S.~Kokaly, and M.~Chechik, ``Two decades of assurance
  case tools: a survey,'' in \emph{International Conference on Computer Safety,
  Reliability, and Security}.\hskip 1em plus 0.5em minus 0.4em\relax Springer,
  2018, pp. 49--59.

\bibitem{rosenstatter2020remind}
T.~Rosenstatter, K.~Strandberg, R.~Jolak, R.~Scandariato, and T.~Olovsson,
  ``Remind: A framework for the resilient design of automotive systems,'' in
  \emph{2020 IEEE Secure Development (SecDev)}.\hskip 1em plus 0.5em minus
  0.4em\relax IEEE, 2020, pp. 81--95.

\bibitem{hawkins2011}
R.~Hawkins, T.~Kelly, J.~Knight, and P.~Graydon, ``A new approach to creating
  clear safety arguments,'' in \emph{Advances in systems safety}.\hskip 1em
  plus 0.5em minus 0.4em\relax Springer, 2011, pp. 3--23.

\bibitem{howard2006security}
M.~Howard and S.~Lipner, \emph{The security development lifecycle}.\hskip 1em
  plus 0.5em minus 0.4em\relax Microsoft Press Redmond, 2006, vol.~8.

\bibitem{iso27005}
{International Organization for Standardization}, ``{ISO 26262 Information
  technology — Security techniques — Information security risk
  management},'' {Geneva, Switzerland}, {2018}.

\end{thebibliography}

\end{document}